\pdfoutput=1
\documentclass[11pt,twoside,a4paper,cmspaper,final,collab]{cms-tdr}
\def\svnVersion{21fd5f5}\def\svnDate{2025/11/12}\def\cmsCernNoTag{CERN-EP-2025-161}\def\cmsCernDate{\today}\def\cmsMessage{Published in the Journal of Instrumentation as \href{http://dx.doi.org/10.1088/1748-0221/20/11/P11006}{\doi{10.1088/1748-0221/20/11/P11006}.}}
\begin{document}\cmsNoteHeader{BTV-22-001}

\newcommand{\hbb}{\ensuremath{\PH \to \PQb \PAQb}\xspace}
\newcommand{\hcc}{\ensuremath{\PH \to \PQc \PAQc}\xspace}
\newcommand{\zbb}{\ensuremath{\PZ \to \PQb \PAQb}\xspace}
\newcommand{\zcc}{\ensuremath{\PZ \to \PQc \PAQc}\xspace}
\newcommand{\gbb}{\ensuremath{\Pg\to \PQb \PAQb}\xspace}
\newcommand{\bb}{\ensuremath{\PQb \PAQb}\xspace}
\newcommand{\cc}{\ensuremath{\PQc \PAQc}\xspace}
\newcommand{\xbb}{\ensuremath{\PX \to \PQb \PAQb}\xspace}
\newcommand{\xcc}{\ensuremath{\PX \to \PQc \PAQc}\xspace}
\newcommand{\xqq}{\ensuremath{\PX \to \Pq \overline{\Pq}}\xspace}
\newcommand{\Zjets}{{\PZ}+jets\xspace}
\newcommand{\Wjets}{{\PW}+jets\xspace}
\newcommand{\Vjets}{{\PV}+jets\xspace}
\newcommand{\Gjets}{{\PGg}+jets\xspace}
\newcommand{\msd}{\ensuremath{m_{\textrm{SD}}}\xspace}
\newcommand{\mpnet}{\ensuremath{m_{\textrm{PNet}}}\xspace}
\newcommand{\pp}{\ensuremath{\Pp\Pp}\xspace}
\newcommand{\HJMINLO}{{\textsc{HJ-MiNLO}}\xspace}
\newcommand{\MINLO}{{\textsc{MiNLO}}\xspace}
\newcommand{\larger}{large-\ensuremath{R}\xspace}
\newcommand{\sigmaobs}{\ensuremath{\sigma_{\text{obs}}}\xspace}
\newcommand{\sigmaexp}{\ensuremath{\sigma_{\text{exp}}}\xspace}

\cmsNoteHeader{BTV-22-001}

\title{Performance of heavy-flavour jet identification in Lorentz-boosted topologies in proton-proton collisions at \texorpdfstring{$\sqrt{s} = 13\TeV$}{sqrt(s) = 13 TeV}}

\date{\today}

\abstract{
  Measurements in the highly Lorentz-boosted regime provoke increased interest in probing the Higgs boson properties and in searching for particles beyond the standard model at the LHC. In the CMS Collaboration, various boosted-object tagging algorithms, designed to identify hadronic jets originating from a massive particle decaying to \bbbar or \ccbar, have been developed and deployed across a range of physics analyses.  This paper highlights their performance on simulated events, and summarizes novel calibration techniques using proton-proton collision data collected at $\sqrt{s} = 13\TeV$ during the 2016--2018 LHC data-taking period. Three dedicated methods are used for the calibration in multijet events, leveraging either machine learning techniques, the presence of muons within energetic boosted jets, or the reconstruction of hadronically decaying high-energy \PZ bosons. The calibration results, obtained through a combination of these approaches, are presented and discussed.
}

\hypersetup{%
pdfauthor={CMS Collaboration},%
pdftitle={Performance of heavy-flavour jet identification in boosted topologies in proton-proton collisions at sqrt(s) = 13 TeV},%
pdfsubject={CMS},%
pdfkeywords={CMS, boosted topology, heavy-flavour jet tagging, tagging performance, calibration}}

\maketitle 

\tableofcontents

\section{Introduction}

Heavy particles produced in proton-proton (\pp) collisions at a centre-of-mass energy of 13\TeV at the CERN LHC, such as the Higgs boson (\PH) and beyond-the-standard model (BSM) particles, can have high energies reaching up to the TeV scale.
These highly Lorentz-boosted resonances can undergo hadronic decay into quarks, followed by a hadronization process, resulting in the generation of a collimated spray of particles.
These final-state particles can be clustered within a single jet using a large distance parameter $R$. The collection of those clustered particles is commonly referred to as a \larger jet.

Identifying the origin of a \larger jet is crucial to exploring boosted topologies at the LHC~\cite{CMS:BTVFlvTagger,ATLAS:bbTaggerCalibPaper,ATLAS:bbTagger,CMS:JMETagger,ATLAS:TopTagger,ATLAS:WZTagger}. Heavy-flavour tagging of \larger jets aims to identify a boosted resonance (denoted by \PX) decaying to a bottom (\PQb) or charm (\PQc) quark-antiquark pair.
In the CMS experiment, a variety of tagging algorithms (``taggers'') using modern machine learning methods~\cite{CMS:BTVFlvTagger,CMS:JMETagger,CMS-DP-2020-002,CMS-DP-2022-041} such as deep neural networks (DNNs) or boosted decision trees (BDTs) have been developed to distinguish the \xbb or \xcc jets from the background. The latter is mainly composed of multijet events from quantum chromodynamics (QCD) processes.
The main features that distinguish the jets from a heavy boosted object versus jets from QCD multijet events are the invariant mass of the jet and the distribution of particles within the jet.
Most of the tagging algorithms are designed to produce an output score that is uncorrelated with the invariant mass, and are referred to as mass-decorrelated.
Mass decorrelation ensures that selecting events based on the tagger output does not introduce artificial mass peaks in the background distributions associated with jet mass. This property is important when applying these taggers over a wide mass range and is crucial when the mass variable is used to evaluate the background.
The \xbb and \xcc tagging techniques have been applied to various studies by the CMS Collaboration, including searches for boosted standard model (SM) Higgs bosons decaying to \bb~\cite{CMS:Hbb2016,CMS:HbbFullRun2,CMS:HH4b} and to \cc~\cite{CMS:VHcc2016,CMS:VHcc,CMS:ggHcc}, and for a BSM resonance decaying to \bb~\cite{CMS:XY4b}.

It is essential that the tagging efficiency for the \xbb or \xcc jets is calibrated using observed events.
In simulated events, the prediction of the shower and hadronization of jets has large uncertainties since it partially relies on phenomenological models; the tagging efficiency obtained from simulated signal jets is not necessarily equal to that in data.
The calibration is performed by measuring the ratio of the efficiencies of selected events in data to that in simulation.
This ratio is referred to as the scale factor (SF).
It is challenging to obtain a pure sample of \xbb (\cc) jets in data.
In practice, a different sample of jets with characteristics similar to that of signal, referred to as ``proxy jets'', is used to derive the SFs.
Selecting the proxy jets that closely match the characteristics of signal jets is mandatory. References~\cite{CMS:BTVFlvTagger,ATLAS:bbTaggerCalibPaper} investigate the use of gluon-splitting \bb jets as proxies, employing a dedicated reweighting procedure to adjust the jet phase space and acquire a good proxy to \hbb jets.
Since the newer \xbb (\cc) tagging algorithms developed within the CMS experiment have stronger discrimination power between \bb (\cc) jets from a resonance decay and those produced via gluon splitting, it becomes more challenging to select \gbb (\cc) jets as a signal proxy. Another approach is to rely on \zbb jets to calibrate \hbb jets, studied in Ref.~\cite{ATLAS:bbTaggerCalibPaper}.

In this paper, we summarize three methods to calibrate the mass-decorrelated \xbb or \cc jet taggers that were adopted by the CMS Collaboration for the 2016--2018 data-taking period (LHC Run~2). The first method selects dedicated regions of phase space using a BDT selection from gluon-splitting \bb (\cc) jets in QCD multijet events as a proxy to \xbb (\cc) jets; the second method uses \gbb (\cc) jets including a reconstructed muon with low transverse momentum (\pt); the third method uses jets from Lorentz-boosted \PZ boson decays to \bb.
The data used in the derivation of these SFs are largely independent, so the methods can be used as cross-validation for each other.
In addition, a combined measurement of the SFs is performed with the three methods.

The paper is organized as follows.
Sections~\ref{sec:detector}--\ref{sec:object} detail the CMS detector, the simulated samples, and the event reconstruction. Section~\ref{sec:hrtalgo} summarizes and compares the heavy-flavour boosted object tagging algorithms developed by the CMS Collaboration during Run~2. Section~\ref{sec:perfdata} describes the three calibration methods and presents the measured SFs and their combination, and Section~\ref{sec:summary} summarizes the results.

\section{The CMS detector}\label{sec:detector}

The central feature of the CMS apparatus is a superconducting solenoid with an internal dia\-meter of 6\unit{m}, providing a magnetic field of 3.8\unit{T}. Within the solenoid volume are a silicon pixel and strip tracker, a lead tungstate crystal electromagnetic calorimeter (ECAL), and a brass and scintillator hadron calorimeter (HCAL), each composed of a barrel and two endcap sections.
Forward calorimeters, made of steel and quartz fibres, extend the pseudorapidity ($\eta$) coverage provided by the barrel and endcap detectors.
Muons are detected in gas-ionization chambers embedded in the steel flux-return yoke outside the solenoid.
A more detailed description of the CMS detector, together with a definition of the coordinate system used and the relevant kinematic variables, is reported in Refs.~\cite{Chatrchyan:2008zzk, CMS:2023gfb}.

Events of interest are selected using a two-tiered trigger system~\cite{Khachatryan:2016bia}. The first level~\cite{Sirunyan:2020zal}, composed of custom hardware processors, uses information from the calorimeters and muon detectors to select events at a rate of around 100\unit{kHz} within a fixed latency of about 4\mus. The second level, known as the high-level trigger (HLT), consists of a farm of processors running a version of the full event reconstruction software optimized for fast processing that reduces the event rate to around 1\unit{kHz} before data storage.

\section{Simulated events} \label{sec:datamc}

Multiple Monte Carlo (MC) event generators are used to simulate \pp collision events at $\sqrt{s}=13\TeV$.
The dominant MC contributions in the methods used to measure the SFs are: (i) the QCD multijet process, when selecting gluon-splitting \bb or \cc jets; (ii) the \Zjets process when selecting \zbb decays.
Additional simulated processes we used include top quark-antiquark pair (\ttbar), single top quark, and \Wjets production.

The main QCD multijet and \Vjets (\PV = \PZ,~\PW) processes are modelled at leading order (LO) accuracy using the \MGvATNLO v2.6.5 generator~\cite{Alwall:2014hca}.
For the matrix element (ME) calculation, the QCD multijet process includes up to four partons, whereas the \Vjets process accounts for up to three partons.
The \PZ (\PW) boson is required to decay to a quark-antiquark pair at the ME level of the \Zjets (\Wjets) event.
The ME generation of the \ttbar simulation is performed with \POWHEG v2~\cite{Nason:2004rx,Frixione:2007vw,Alioli:2010xd, Campbell:2014kua} at next-to-LO (NLO) accuracy in QCD, and its cross section is scaled to a theoretical prediction at next-to-NLO (NNLO) in QCD, including resummation of next-to-next-to-leading logarithmic soft-gluon terms~\cite{Czakon:2011xx}.
The single top quark production in the $t$-channel (tW channel) is simulated using \POWHEG in the 4-flavour (5-flavour) scheme~\cite{Frixione:2008yi,Alioli:2009je,Re:2010bp,Frederix:2012dh}, with its cross section normalized to the NLO calculations from Ref.~\cite{Kidonakis:2012rm}.

For all processes, the parton shower is simulated with \PYTHIA v8.230~\cite{SJOSTRAND2015159}, using the CP5 underlying event tune~\cite{Sirunyan:2019dfx} with the NNPDF3.1 NNLO parton distribution function (PDF) set~\cite{Ball:2017nwa}. The matching of jets from ME calculations and those from parton showers is done with the MLM~\cite{Alwall:2007fs} technique for LO samples.
The \PYTHIA generator is used for parton showering the simulated \xbb (\cc) signal events in searches described in Refs.~\cite{CMS:Hbb2016,CMS:HbbFullRun2,CMS:HH4b,CMS:VHcc2016,CMS:VHcc,CMS:ggHcc,CMS:XY4b}. Since the performance of a DNN-based jet identification algorithm on simulated events is affected by the parton shower patterns, it is important to ensure that the proxy jet samples use the same generator software for parton showering so that the resulting SFs are applicable to signal jets.

For the method using $\mu$-tagged jets, detailed in Section~\ref{sec:mutagged}, a $\mu$-enriched QCD multijet process is simulated with \PYTHIA by forcing the decay of charged pions and kaons into muons and requiring the presence of at least one generated muon with $\pt > 5\GeV$ inside the jet.
It increases the number of jets in the simulated sample that have an associated low-energy muon.
For the method using \zbb jets, detailed in Section~\ref{sec:boostedzbb}, the differential cross sections of the \Zjets and \Wjets processes are corrected, as a function of boson \pt, for NLO QCD effects. The cross sections are reweighted to NLO using \Zjets and \Wjets events with up to two additional partons, simulated at NLO with \MGvATNLO and using FxFx matching~\cite{Frederix:2012ps}.
Additional corrections are applied to the cross section originating from NLO electroweak effects~\cite{Lindert:2017olm}.

To study the tagging performance on simulated signal events, the gluon-fusion Higgs boson production process is simulated using the \HJMINLO~\cite{Frixione:2007vw,Hamilton:2012rf,Luisoni:2013kna} event generator with the Higgs boson mass $m_{\PH} = 125\GeV$, interfaced with \PYTHIA v8.230 for Higgs boson decays to \bb or \cc and event hadronization. The \hbb and \hcc signal jets are selected from these events; this is discussed in Section~\ref{sec:hrtalgo} and used for deriving the selection thresholds in the \xbb and \xcc identification algorithms.

For all processes, the effect of additional $\Pp\Pp$ interactions within the same or nearby bunch
crossings (pileup) on top of the hard scattering processes is modelled by minimum bias collisions generated with \PYTHIA.
The events are then reweighted to match the pileup profile observed in data.
The interactions between particles and the material of the CMS detector are simulated using \GEANTfour~\cite{AGOSTINELLI2003250}.

The events are simulated separately for four data-taking eras during Run~2 with their corresponding conditions, denoted as the 2016 pre-VFP, 2016 post-VFP, 2017, and 2018 eras, where VFP stands for feedback preamplifier bias voltage~\cite{CMS-DP-2020-045}.
The 2016 pre-VFP and post-VFP eras are treated separately because of the substantial change in the strip tracker conditions between them.
The selection thresholds of each tagging discriminant, referred to as working points (WPs), are determined separately for each era. 
The performance of the tagging algorithms, discussed in Sections~\ref{sec:hrtalgo} and \ref{sec:perfdata}, is evaluated separately for each era, using the corresponding simulated events and data collected during that period. Although the 2018 data-taking conditions are primarily used for illustration in these sections, the resulting SFs for all eras are summarized at the end of Section~\ref{sec:perfdata}.

\section{Event reconstruction and physics objects} \label{sec:object}

The global event reconstruction with the particle-flow (PF) algorithm~\cite{CMS-PRF-14-001} reconstructs and identifies each individual particle in an event, with an optimized combination of all subdetector information.
In this process, particles are identified exclusively as charged or neutral hadrons, photons, electrons, or muons.
Photons (\eg coming from neutral pion decays or from electron bremsstrahlung) are identified as ECAL energy clusters not linked to the extrapolation of any charged-particle trajectory to the ECAL. Electrons (\eg coming from photon conversions in the tracker material or from semileptonic decays of \PQb hadrons) are identified as a primary charged-particle track and potentially as ECAL energy clusters corresponding to this track extrapolation to the ECAL and to possible bremsstrahlung photons emitted along the way through the tracker material. Muons are identified as tracks in the central tracker consistent with either tracks or several hits in the muon system, and associated with calorimeter deposits compatible with the muon hypothesis. Charged hadrons are identified as charged particle tracks neither identified as electrons, nor as muons. Finally, neutral hadrons are identified as HCAL energy clusters not linked to any charged-hadron trajectory, or as a combined ECAL and HCAL energy excess with respect to the expected charged-hadron energy deposit.

Events are required to have at least one reconstructed vertex.
The primary vertex (PV) is taken to be the vertex corresponding to the hardest scattering in the event, evaluated using tracking information alone, as described in Section~9.4.1 of Ref.~\cite{CMS-TDR-15-02}.
The displaced secondary vertices (SVs) used to probe the decays of {\PQb} or {\PQc} hadrons are reconstructed by the inclusive SV-finding algorithm~\cite{Khachatryan:2011wq,CMS-PRF-14-001}, taking reconstructed tracks in an event as input.

A collection of reconstructed low-energy (soft) nonprompt muons is used in the $\mu$-tagged calibration.
These soft muons arise from semileptonic decay modes of hadrons; they typically have low momentum and are surrounded by hadronic activity of the underlying jet in which these hadrons are created. The relative isolation, $I_{\text{rel}}$, is defined as the scalar \pt sum of the PF candidates within a cone of radius $\Delta R = \sqrt{\smash[b]{(\Delta\eta)^2 + (\Delta\phi)^2}} = 0.4$ around the muon candidate (where $\phi$ is the azimuthal angle in radians) divided by the muon \pt. It is corrected for contributions of neutral particles originating from pileup interactions~\cite{Khachatryan:2015hwa,Sirunyan:2018fpa}.
The soft muons are required to satisfy $I_{\text{rel}} > 0.15$ and a set of kinematic criteria based on the track reconstruction quality, hit multiplicities in the tracking and muon subdetector layers, and the displacement of these particles with respect to the PV.

Jets are clustered from PF candidates using the anti-\kt algorithm \cite{Cacciari:2008gp} with a distance parameter of $R=0.4$ (AK4 jets) or 0.8 (AK8 jets). The latter forms the \larger jet collection and is the primary object studied in this paper.
The effect of particles from pileup is mitigated through the charged-hadron subtraction~\cite{Khachatryan:2016kdb} and pileup per particle identification (PUPPI)~\cite{Bertolini:2014bba,CMS:2020ebo} algorithms for AK4 and AK8 jets, respectively.
In PUPPI, charged particles identified as originating from pileup vertices are discarded, and a weight between zero and one is assigned to each neutral particle as the probability for the particle to have originated at the PV.
The resulting list of PF candidates, with each particle four-momentum scaled by its corresponding weight, is input to cluster for AK8 jets. Jet energy corrections are derived from simulation studies so that the average measured energy of jets matches that of particle-level jets.
In situ measurements of the momentum balance in \Gjets, \Zjets, and QCD multijet events are used to determine any residual differences between the jet energy scale in data and in simulation, and appropriate corrections are made~\cite{Khachatryan:2016kdb}.

Algorithms for heavy-flavour tagging of AK8 jets constitute the primary focus of this study.
These algorithms include seven \xbb and \xcc jet taggers developed by the CMS Collaboration for the analysis of Run~2 data, which are discussed in detail in Section~\ref{sec:hrtalgo}.
Various jet observables are used in the tagging performance studies presented in Section~\ref{sec:perfdata}.
Among these, the $N$-subjettiness variable $\tau_{N}$~\cite{Thaler:2010tr} is used to quantify the compatibility of a jet's energy distribution with a hypothesis of having $N$ subjets, where each subjet represents a localized region of energy corresponding to potential partonic activity inside the jet.
A smaller value of $\tau_{N}$ indicates greater compatibility with having $N$ or fewer subjets.
The $N$-subjettiness ratio $\tau_{21} = \tau_2 / \tau_1$ is used to identify jets with a two-prong characteristic that may originate from a resonance or a gluon splitting to \bb or \cc.
The ``soft-drop mass'' \msd of a jet is obtained from the soft-drop (SD) algorithm~\cite{Larkoski:2014wba}.
This algorithm removes wide-angle soft radiation from the jet through a recursive declustering process, removing soft branches from the original structure. In the final step of declustering, two subjet axes are identified.
The SD algorithm, as applied in CMS analyses~\cite{CMS:JMETagger}, uses the parameters $z = 0.1$ and $\beta = 0$, where $z$ is the soft threshold parameter controlling the minimum energy sharing between subjets, and $\beta$ controls the angular exponent in the grooming condition.
In addition to \msd, a regression algorithm~\cite{CMS-DP-2021-017} is developed to reconstruct the AK8 jet mass.
This method exploits properties of the PF candidates and SVs associated with the jet 
using the ``ParticleNet'' graph neural network~\cite{Qu:2019gqs}. The resulting regressed output, \mpnet, has an improved resolution of reconstructing the mass of the two-prong jet initiated by a resonance decay.

In simulation, generator-level variables are used to determine the origin of jets.
Associating the flavour of the generator-level hadron that gave rise to a reconstructed jet is a crucial step in defining jet samples used for calibration. 
Jets are labelled using ghost association~\cite{Cacciari:2007fd}, a widely used approach albeit not guaranteed to be infrared- and collinear-safe.
The reconstructed final-state particles in the jet are reclustered with the generated {\PQb} or {\PQc} hadrons.
Only {\PQb} ({\PQc}) hadrons that are the last {\PQb} ({\PQc}) hadrons in their decay chains are included.
The four-momenta of these hadrons are rescaled to a very small value to ensure that they do not affect the reconstructed jet momentum and that only their directional information is kept. 
The label is determined from the number of {\PQb} or {\PQc} hadrons ghost-associated with a jet.
In addition, for jets originating from the signal events, specifically the gluon-fusion Higgs boson production process with \hbb (\cc) decays, the selection criteria for signal jets require that the direction of the resonance, as well as those of both daughter \PQb (\PQc) quarks, lie within the jet cone defined by $\Delta R < 0.8$.

\section{Overview of tagging algorithms} \label{sec:hrtalgo}

This section reviews and compares various boosted heavy-flavour jet identification algorithms developed by the CMS Collaboration for the analysis of Run~2 data.
The ParticleNet-MD tagger (where MD stands for mass-decorrelated)~\cite{CMS-DP-2020-002}, the DeepDoubleX tagger~\cite{CMS-DP-2022-041}, and the DeepAK8-MD tagger~\cite{CMS:JMETagger} provide discriminants for both \xbb and \xcc identification, whereas the double-b tagger~\cite{CMS:BTVFlvTagger} aims at \xbb identification only.
The performance of these algorithms is evaluated using simulated samples corresponding to 2018 detector conditions, which are used as a representative benchmark in this section.

\subsection{ParticleNet-MD}

The ParticleNet-MD jet tagging algorithm~\cite{CMS-DP-2020-002} provides two discriminants, the ParticleNet-MD bbvsQCD and ccvsQCD, used for \xbb and \xcc identification, respectively.
The ParticleNet-MD algorithm is a DNN-based algorithm designed to identify two-prong hadronic decays (\bb, \cc, and \qqbar, where \PQq represents \PQu, \PQd, and \PQs quarks) of a highly Lorentz-boosted particle across a wide range of resonance mass and has been used in a number of CMS analyses~\cite{CMS:HH4b,CMS:XY4b,CMS:VHcc}.
It takes the particle-level features as input, including a list of PF candidates and SVs associated with the jet.
Input variables for a PF candidate include kinematic features such as its \pt, energy, the differences in $\eta$ and $\phi$ between the particle and the jet axis, and its charge. For charged PF candidates, additional properties measured by the tracking detector are included, such as the track displacement and quality. Variables for an SV include kinematic and displacement features, as well as quality criteria.
At the core of the algorithm is the ``ParticleNet'' neural network architecture.
For networks of this kind, the input PF candidates and SVs are processed in a permutation-invariant way; a convolution operation is performed on each particle, 
grouping it with its nearest neighbours in the geometric $\eta$--$\phi$ space to facilitate information exchange between particles to extract local features.

Two versions of the algorithm with the ParticleNet architecture exist: an MD version (\ie the ParticleNet-MD algorithm) and a non-MD version (the ParticleNet algorithm)~\cite{CMS-DP-2020-002}.
This paper studies the former algorithm; the latter aims at explicitly utilizing the jet mass to identify hadronic decays of Lorentz-boosted SM particles (\PQt, \PW, \PZ, and \PH) and is beyond the scope of this paper.

The ParticleNet-MD algorithm is trained on a set of signal jets with $\pt > 200$\GeV and $30 < \msd < 260$\GeV, including \xbb, \xcc, \xqq, and background QCD jets, where \PX is a variable-mass spin-0 neutral particle.
Jets from both signal and background samples are reweighted to yield flat distributions in both $\pt$ and $\msd$ so as to decorrelate the trained tagger outputs with the jet mass.
The output of the algorithm provides four probability-like scores: $p(\xbb)$, $p(\xcc)$, $p(\xqq)$, and $p(\text{QCD})$.
The discriminants used to separate \xbb and \xcc jets from the dominant QCD multijet background are the binary classification scores:
\begin{equation}\begin{aligned}\label{eq:particlenet_discriminant}
    \text{ParticleNet-MD bbvsQCD disc.} & = \frac{p(\xbb)}{p(\xbb) + p(\text{QCD})},\\
    \text{ParticleNet-MD ccvsQCD disc.} & = \frac{p(\xcc)}{p(\xcc) + p(\text{QCD})}.
\end{aligned}\end{equation}
These MD discriminants have been found to provide consistent responses to heavy resonance decays (\eg \PZ or \PH), regardless of the resonance spin.

\begin{figure}[htbp]
    \centering
    \includegraphics[width=0.48\textwidth]{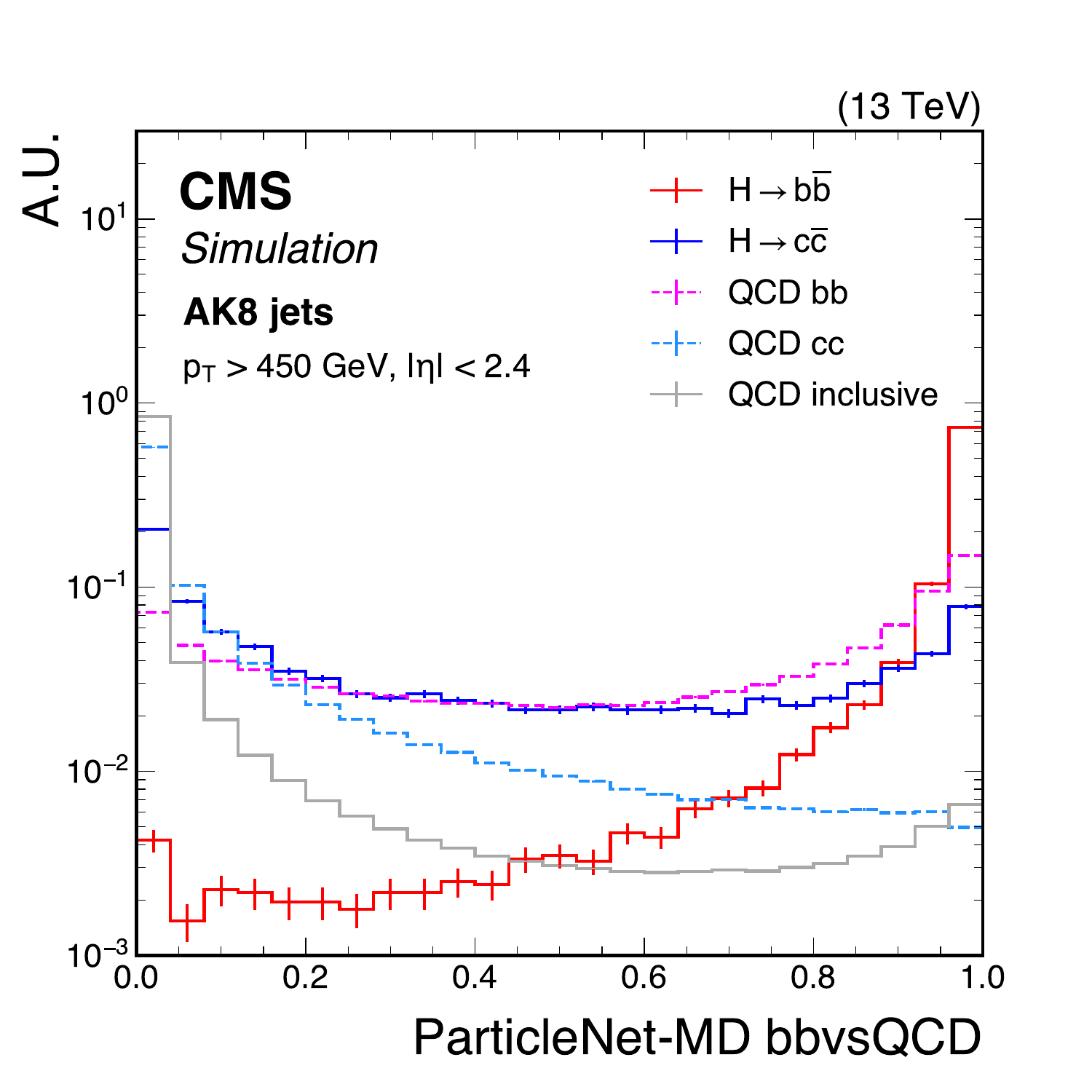}
    \includegraphics[width=0.48\textwidth]{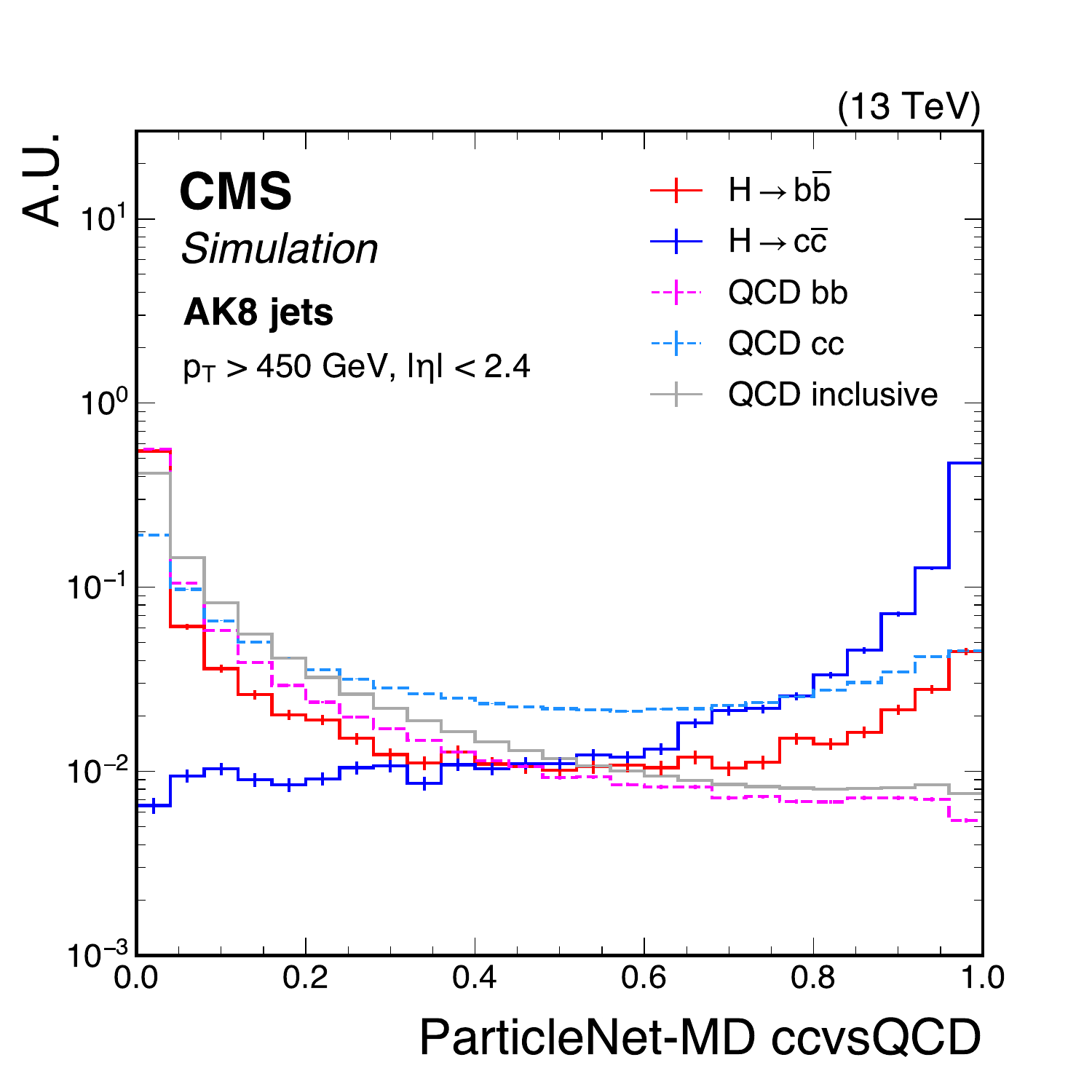}
    \caption{Shape comparison of the ParticleNet-MD bbvsQCD (left) and ParticleNet-MD ccvsQCD (right) discriminants for the simulated standard model \hbb and \hcc jets, the bb and cc components of QCD multijet background jets, and inclusive QCD jets (without flavour-specific selection), using simulated events corresponding to the 2018 data-taking conditions for jets with $\pt > 450\GeV$ and $\abs{\eta} < 2.4$. The error bars represent the statistical uncertainties due to the limited number of simulated events.}
    \label{fig:shape_pnet}
\end{figure}

The discriminants for signal and background jets are shown in Fig.~\ref{fig:shape_pnet}.
The QCD jets matched with at least two ghost \PQb hadrons are designated ``QCD bb''; jets with no ghost-matched \PQb hadrons but including at least two ghost-matched \PQc hadrons are designated ``QCD cc''.
The figure shows that \hbb (\hcc) jets are well separated from the other processes, typically exhibiting high ParticleNet-MD bbvsQCD (ccvsQCD) discriminant scores.

\subsection{DeepDoubleX}\label{sec:ddx}

The DeepDoubleX tagging algorithm~\cite{CMS-DP-2022-041} is a DNN-based algorithm designed to identify \xbb and \xcc in the boosted topology.
The algorithm is employed in the search for boosted Higgs boson decays to \cc~\cite{CMS:ggHcc}.
It is an updated version of the algorithm used in the boosted \hbb search~\cite{CMS:HbbFullRun2}, denoted V1 in Ref.~\cite{CMS-DP-2018-046}.
DeepDoubleX, inspired by the DeepJet model for AK4 jet flavour tagging~\cite{bols2020jet}, combines one-dimensional (1D) convolutional layers and gated recurrent units.
The algorithm is developed for AK8 jets with $\pt > 300\GeV$ and $\abs{\eta} < 2.4$.
The input to the algorithm includes jet-level observables and three groups of low-level input features: charged PF candidates, neutral PF candidates, and SVs.
The jet-level variables include properties of the selected tracks and SVs within the jet, as well as information related to the two-SV system. The low-level variables for PF candidates and SVs are similar to the inputs for the ParticleNet-MD algorithms.
Irrelevant input features from the initial set described above are pruned using the layer-wise relevance propagation technique.
Each group of low-level inputs is organized into an ordered list, where the ordering is determined by specific features: the impact parameter for charged PF candidates, the distance to the nearest SV for neutral PF candidates, and the transverse flight distance for SVs.
These groups are then scaled with a batch normalization layer, and then passed through separate convolutional and gated recurrent units layers successively.
The global jet-level features are passed through a batch normalization layer and combined with the three processed low-level feature groups in a dense layer.

\begin{figure}[htbp]
    \centering
    \includegraphics[width=0.48\textwidth]{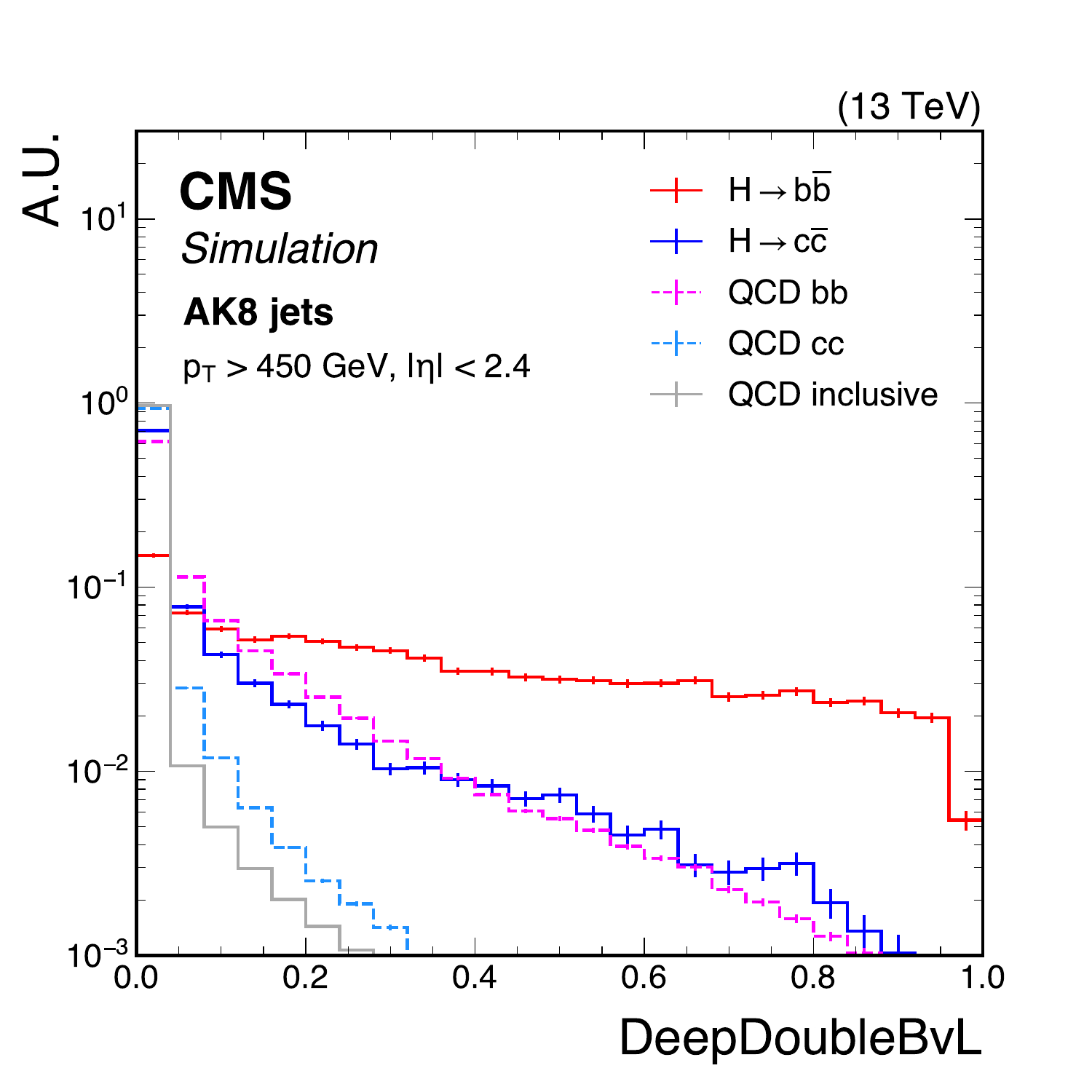}
    \includegraphics[width=0.48\textwidth]{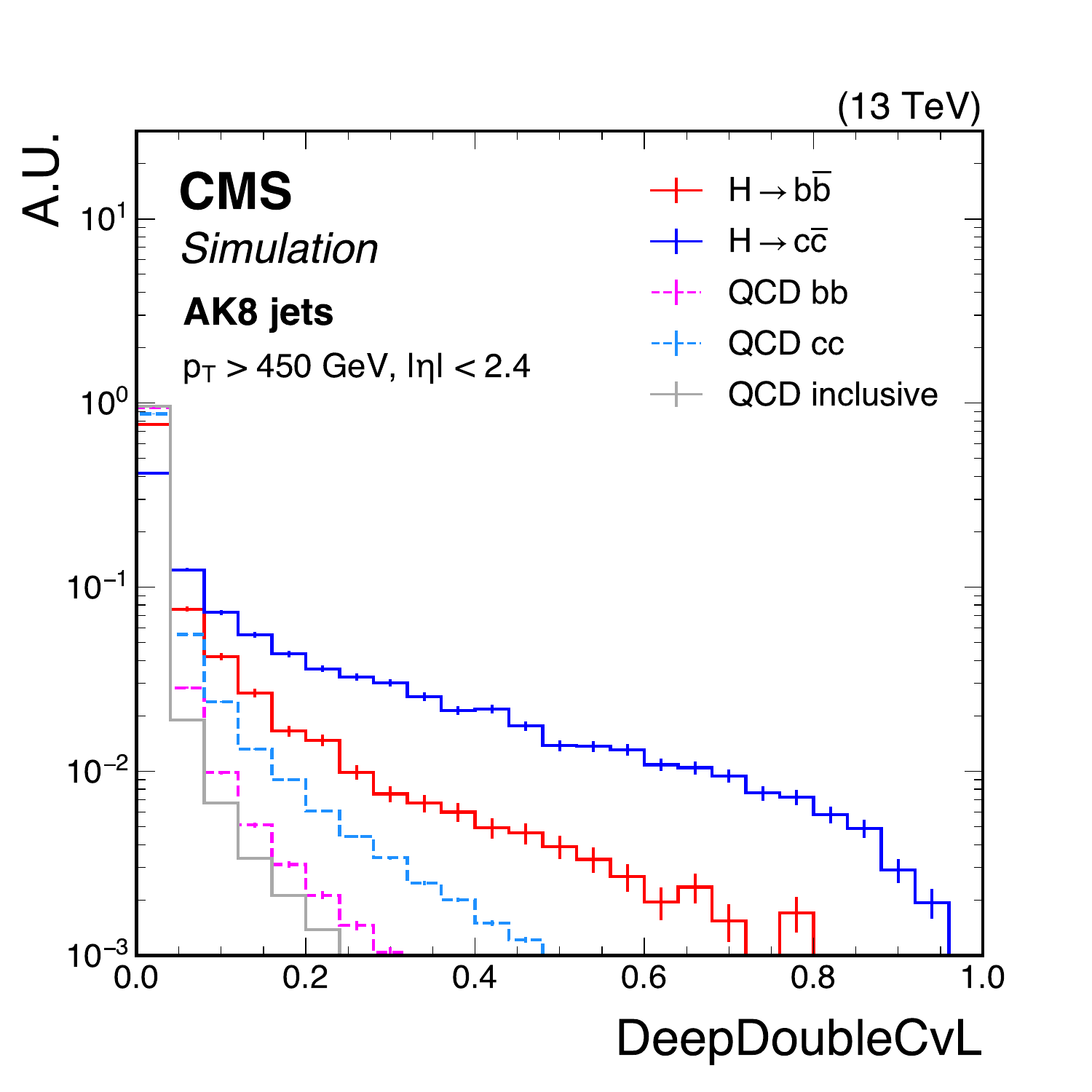}
    \caption{Shape comparison of the DeepDoubleBvL (left) and DeepDoubleCvL (right) discriminants for the simulated standard model \hbb and \hcc jets, the bb and cc components of QCD multijet background jets, and inclusive QCD jets, using simulated events corresponding to the 2018 data-taking conditions for jets with $\pt > 450\GeV$ and $\abs{\eta} < 2.4$. The error bars represent the statistical uncertainties due to the limited number of simulated events.}
    \label{fig:shape_ddx}
\end{figure}

The algorithm is trained for three binary jet classification tasks: distinguishing $\xbb$ from QCD jets, $\xcc$ from QCD jets, and $\xcc$ from $\xbb$ jets.
The signal jets originate from the decay of a spin-0 resonance \PX into \bb or \cc, with the mass of \PX ranging from 15 to 250\GeV. Mass decorrelation is achieved by reweighting the signal jets to match the $\msd$ distribution of the QCD background jets.
This study focuses on the models trained for the first two classification tasks, referred to as DeepDoubleBvL (DDBvL) and DeepDoubleCvL (DDCvL), respectively, as outlined in Ref.~\cite{CMS-DP-2022-041}.
The distributions of DDBvL and DDCvL discriminants for signal and QCD multijet background are shown in Fig.~\ref{fig:shape_ddx}.

\subsection{DeepAK8-MD}

The DeepAK8-MD algorithm~\cite{CMS:JMETagger} is a DNN-based jet tagging algorithm developed for identifying resonance decays to \bb or \cc.
It was developed early in Run~2 and used in several CMS searches, such as Ref.~\cite{CMS:VHcc2016}. 
DeepAK8-MD uses 1D residual convolutional layers~\cite{he2016deep} and is trained with the same low-level input features as in the ParticleNet-MD algorithm, focusing on AK8 jets with $\pt > 200$\GeV.
It functions as a multiclass classifier, with the output classes comprising five main categories (\PQt, \PW, \PZ, \PH, and QCD). Each of these categories is further subdivided; for instance, specific decay modes of a resonance, such as \bb or \cc, are distinguished.
The training dataset includes hadronic jets from SM top quarks and \PW, \PZ, and \PH boson decays, as well as QCD jets.
Mass decorrelation is achieved using adversarial training.
A mass prediction network is added to the classification network.
The training target is modified to include the accuracy of the mass prediction as a penalty in the loss.
After training, the algorithm outputs probability scores that are largely independent of jet mass.

The discriminants used to identify a resonance decay to \bb or \cc are:
\begin{linenomath*}
\begin{equation}\begin{aligned}\label{eq:deepak8_discriminant}
    \text{DeepAK8-MD bbvsQCD disc.} & = \frac{p(\hbb) + p(\zbb)}{p(\hbb) + p(\zbb)+ p(\text{QCD})},\\
    \text{DeepAK8-MD ccvsQCD disc.} & = \frac{p(\hcc) + p(\zcc)}{p(\hcc) + p(\zcc)+ p(\text{QCD})}.
\end{aligned}\end{equation}
\end{linenomath*}

\begin{figure}[htbp]
    \centering
    \includegraphics[width=0.48\textwidth]{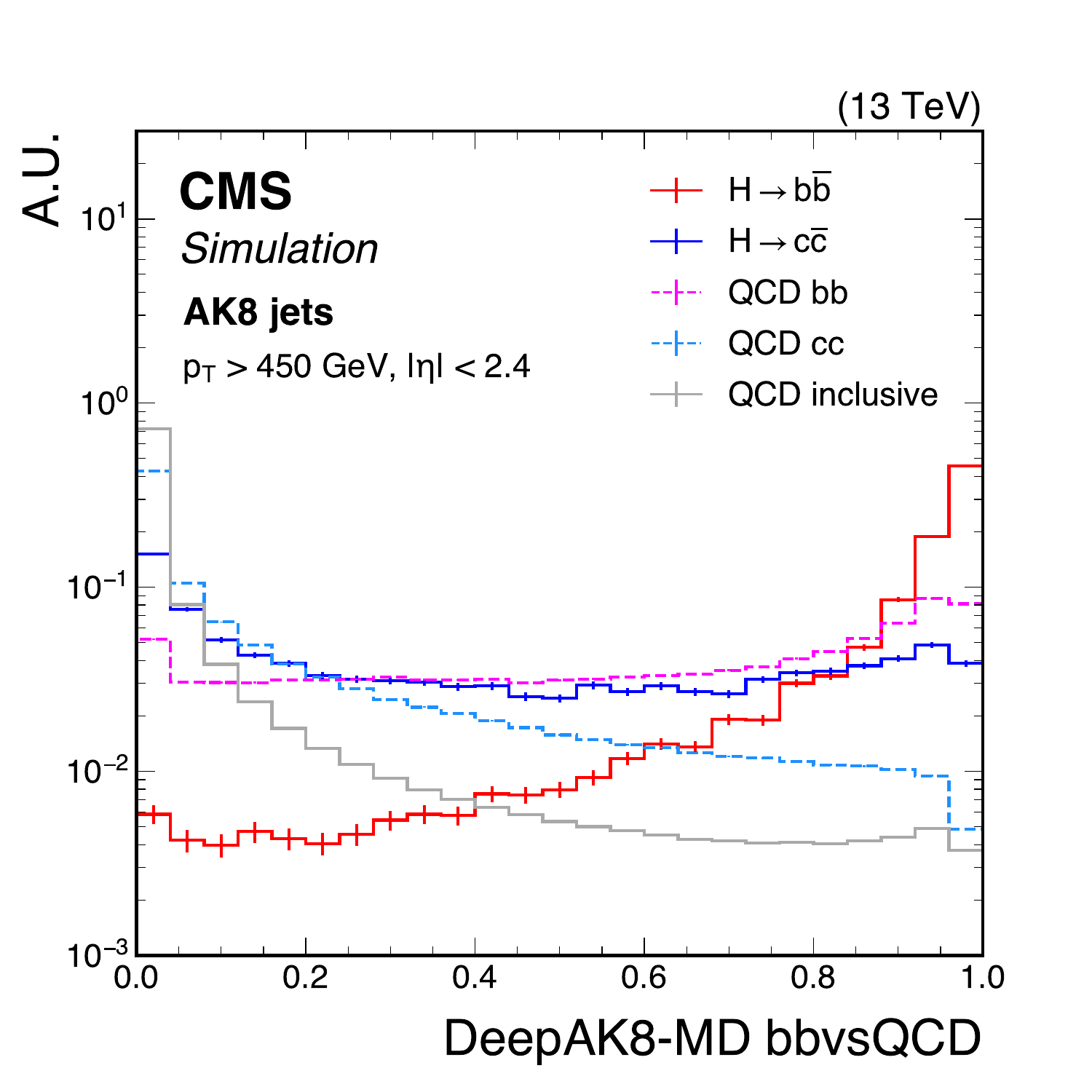}
    \includegraphics[width=0.48\textwidth]{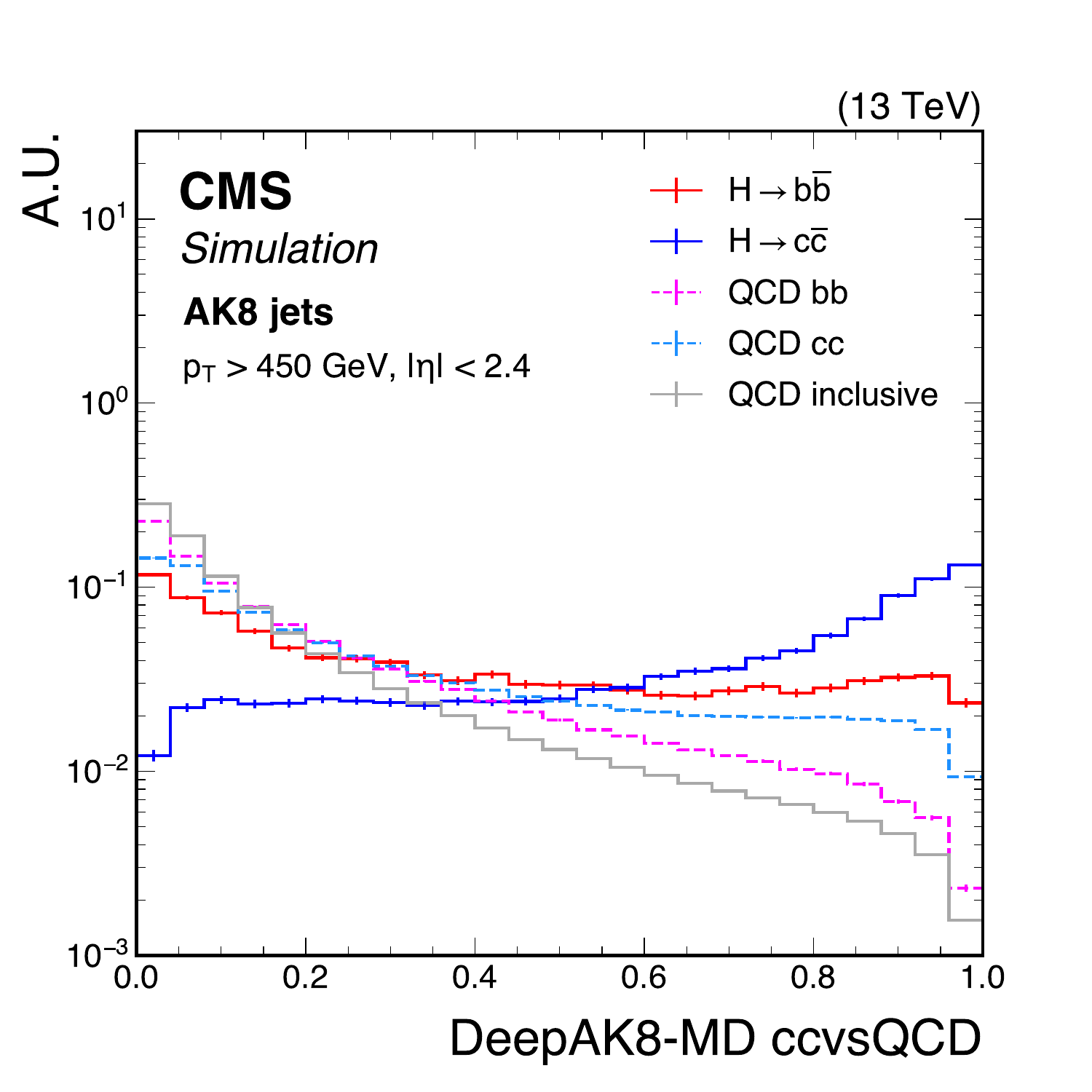}
    \caption{Shape comparison of the DeepAK8-MD bbvsQCD (left) and DeepAK8-MD ccvsQCD (right) discriminants for the simulated standard model \hbb and \hcc jets, the bb and cc components of QCD multijet background jets, and inclusive QCD jets, using simulated events corresponding to the 2018 data-taking conditions for jets with $\pt > 450\GeV$ and $\abs{\eta} < 2.4$. The error bars represent the statistical uncertainties due to the limited number of simulated events.}
    \label{fig:shape_da}
\end{figure}

The distribution of the discriminants on the signal and background jets is shown in Fig.~\ref{fig:shape_da}.

\subsection{The double-b tagger}\label{sec:double-b}

The double-b tagger, detailed in Ref.~\cite{CMS:BTVFlvTagger}, is a BDT algorithm trained to distinguish $\hbb$ jets from the QCD multijet background.
It was developed early in Run~2 and used in the search for boosted \hbb decays~\cite{CMS:Hbb2016}.
The input to the tagger includes high-level variables constructed from tracks and SVs associated with the jet, as introduced in Section~\ref{sec:ddx}.
For training, \hbb jets are used directly as the signal.
The input variables are chosen to have a weak dependence on jet \pt and mass, thereby ensuring a stable performance across a wide kinematic range.

\begin{figure}[htbp]
    \centering
    \includegraphics[width=0.48\textwidth]{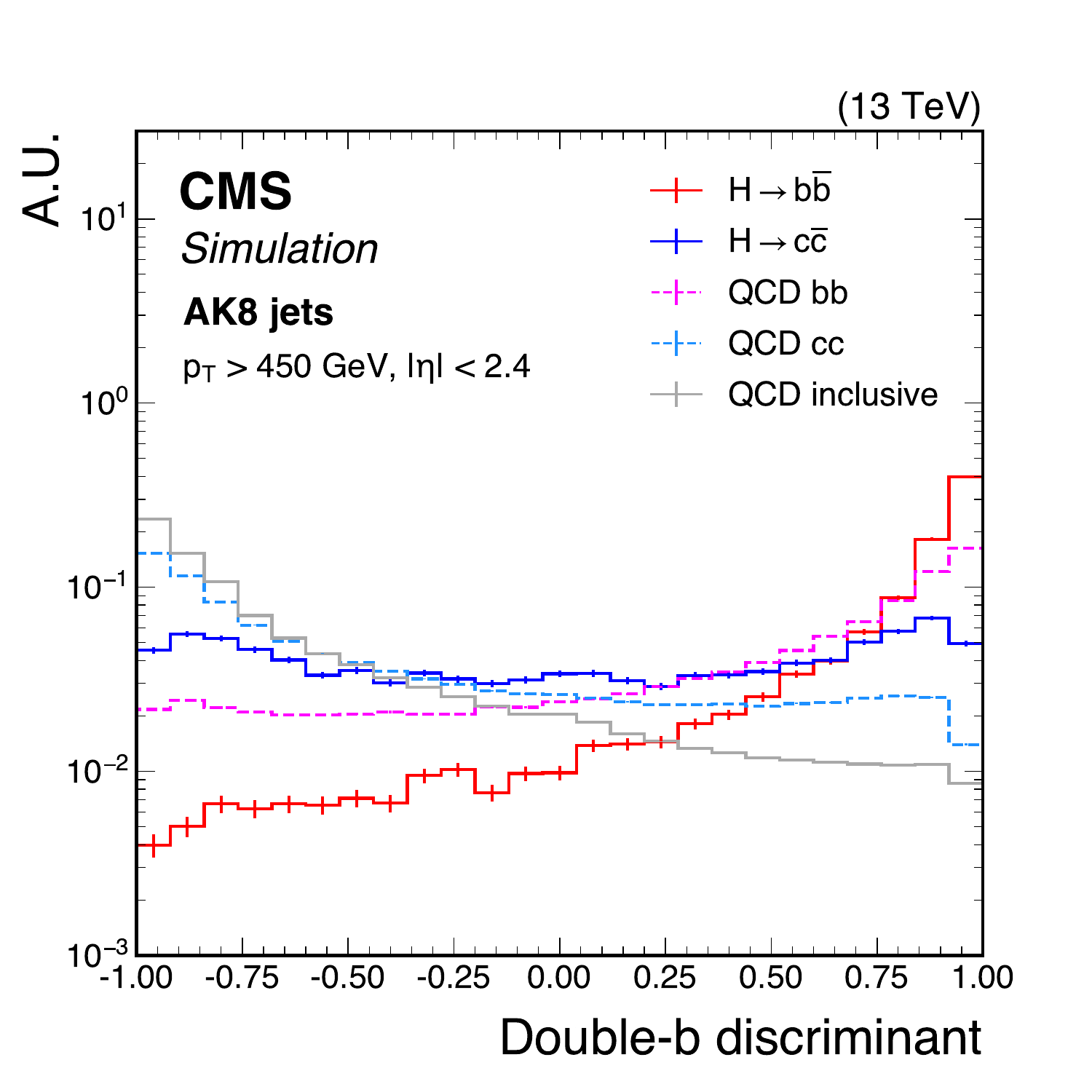}
    \caption{Shape comparison of the double-b discriminant for the simulated standard model \hbb and \hcc jets, the bb and cc components of QCD multijet background jets, and inclusive QCD jets, using simulated events corresponding to the 2018 data-taking conditions for jets with $\pt > 450\GeV$ and $\abs{\eta} < 2.4$. The error bars represent the statistical uncertainties due to the limited number of simulated events.}
    \label{fig:shape_doubleb}
\end{figure}

The discriminant distribution for the signal jets and QCD multijet background is shown in Fig.~\ref{fig:shape_doubleb}.

\subsection{Working points}

Three WPs are determined for each of the discriminants described above.
They are referred to as high-purity (HP), medium-purity (MP), and low-purity (LP) WPs and are defined to result in \hbb (\cc) selection efficiencies of 40\% (15\%), 60\% (30\%), and 80\% (50\%), respectively, based on simulated events.
The WPs are determined separately in simulation corresponding to each of the four data-taking eras.

\subsection{Performance comparison}

\begin{figure}[htbp]
    \centering
    \includegraphics[width=0.48\textwidth]{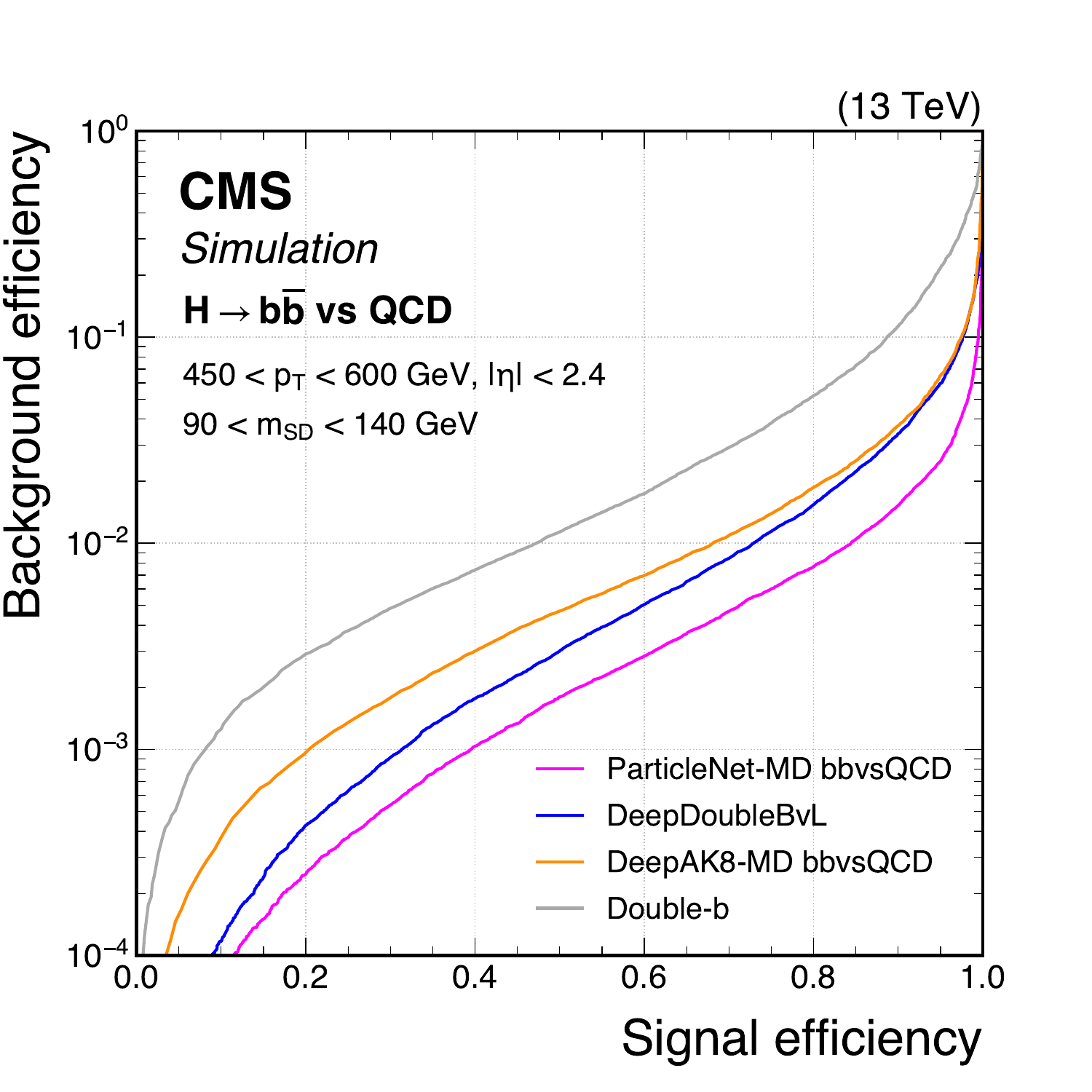}
    \includegraphics[width=0.48\textwidth]{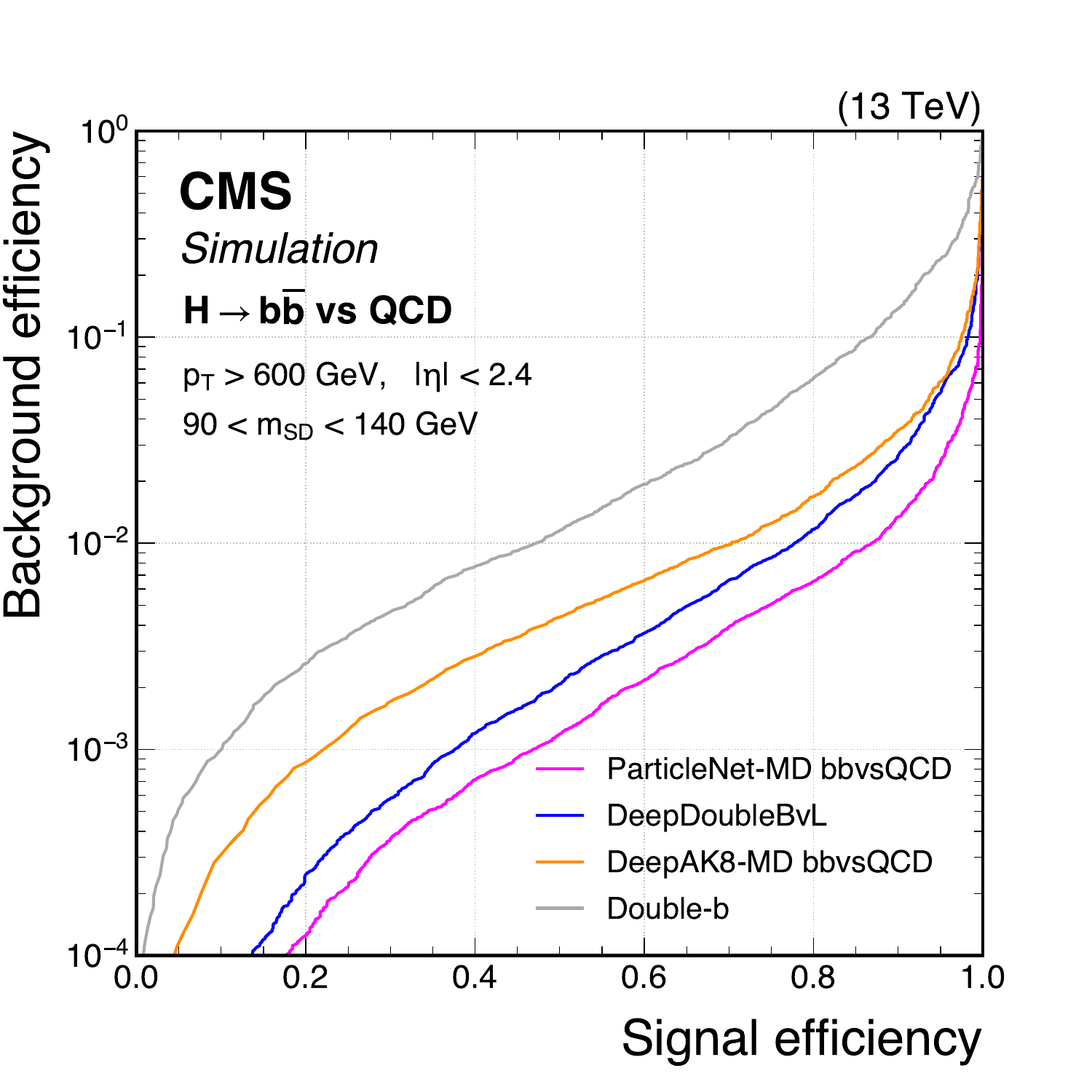}
    \caption{Comparison of the performance of the \xbb identification algorithms in terms of receiver operating characteristic (ROC) curves for \hbb signal jets versus the inclusive QCD jets as background, using simulated events with the 2018 data-taking conditions. Performance is shown in the $450 < \pt < 600\GeV$ (left) and $\pt > 600\GeV$ (right) regions. Additional selection criteria applied to the jets are displayed on the plots.}
    \label{fig:roc_bb}
\end{figure}

\begin{figure}[htbp]
    \centering
    \includegraphics[width=0.48\textwidth]{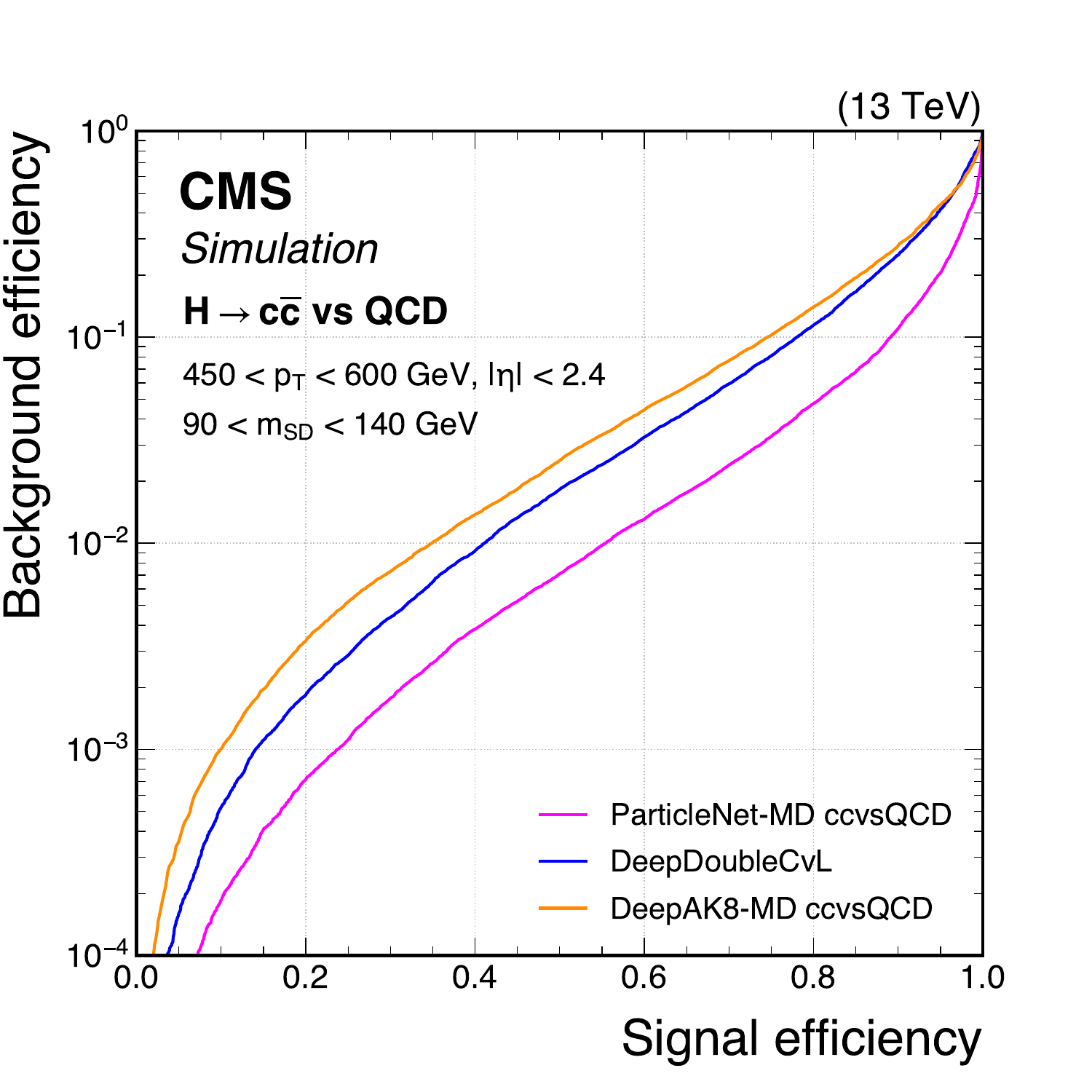}
    \includegraphics[width=0.48\textwidth]{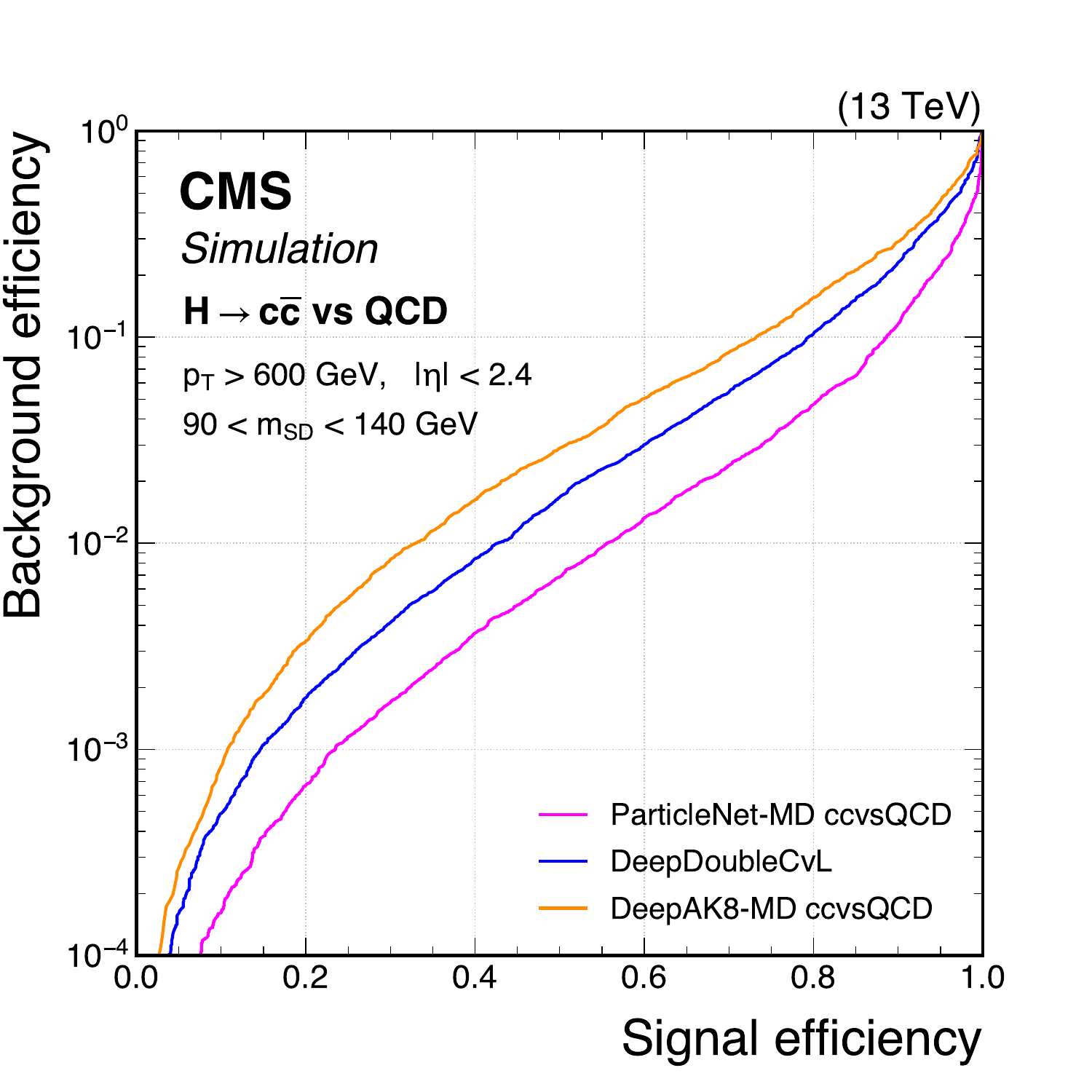}
    \caption{Comparison of the performance of the \xcc identification algorithms in terms of receiver operating characteristic (ROC) curves for \hcc signal jets versus the inclusive QCD jets as background, using simulated events with the 2018 data-taking conditions. Performance is shown in the $450 < \pt < 600\GeV$ (left) and $\pt > 600\GeV$ (right) regions. Additional selection criteria applied to the jets are displayed on the plots.}
    \label{fig:roc_cc}
\end{figure}

\begin{figure}[htbp]
    \centering
    \includegraphics[width=0.48\textwidth]{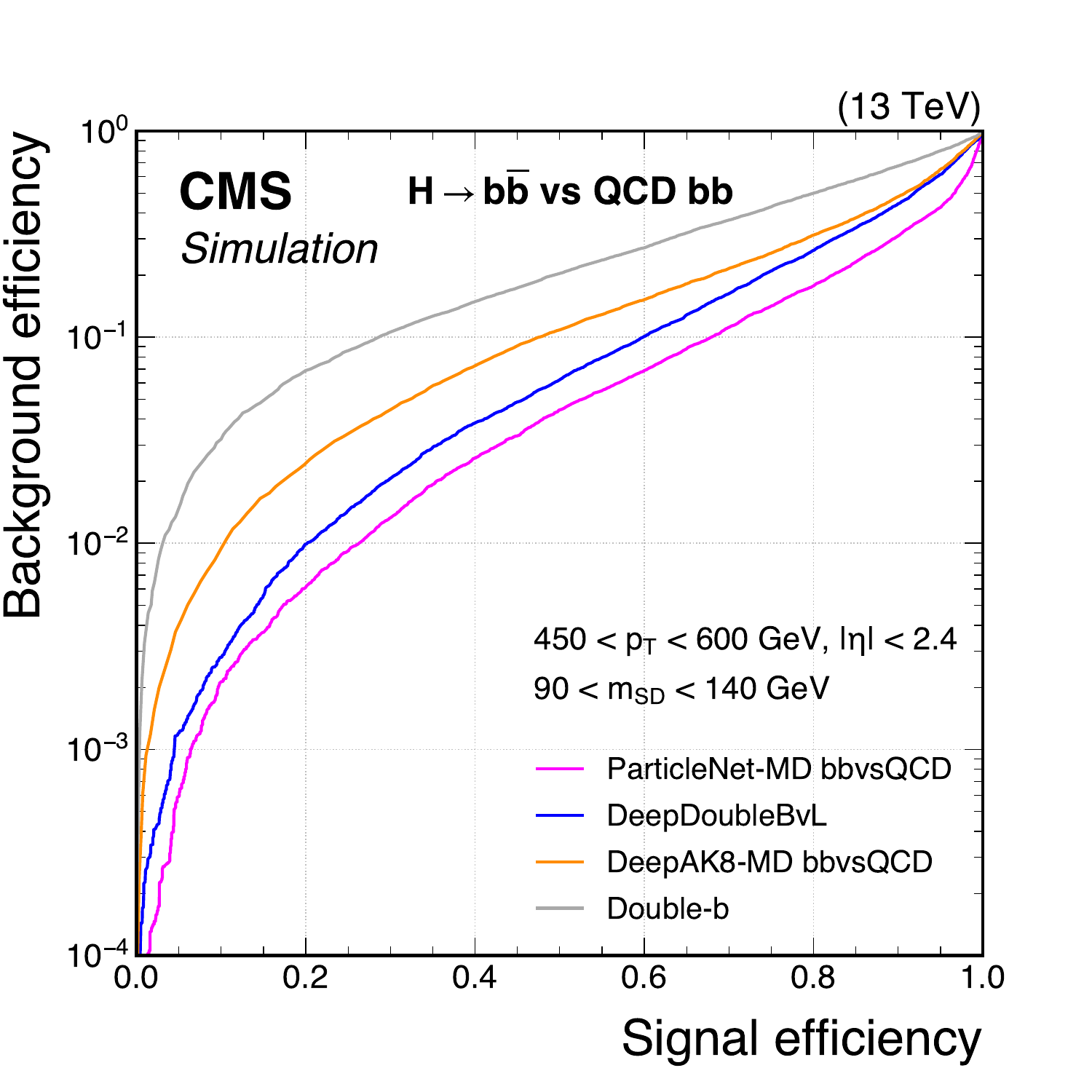}
    \includegraphics[width=0.48\textwidth]{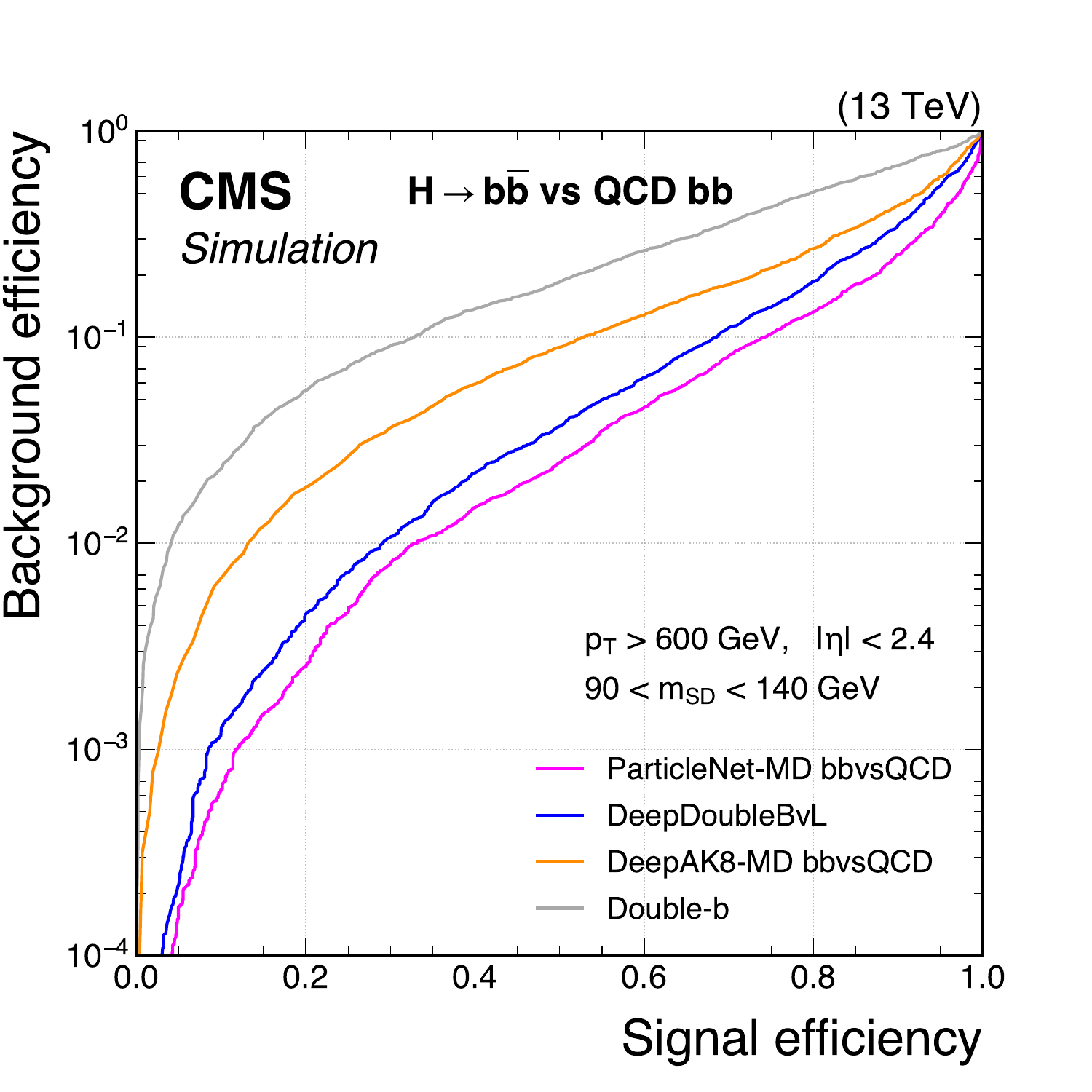}
    \caption{Comparison of the performance of the \xbb identification algorithms in terms of receiver operating characteristic (ROC) curves for \hbb signal jets versus the bb component of the QCD jets as background, using simulated events with the 2018 data-taking conditions. Performance is shown in the $450 < \pt < 600\GeV$ (left) and $\pt > 600\GeV$ (right) regions. Additional selection criteria applied to the jets are displayed on the plots.}
    \label{fig:roc_bbvsgbb}
\end{figure}

\begin{figure}[htbp]
    \centering
    \includegraphics[width=0.48\textwidth]{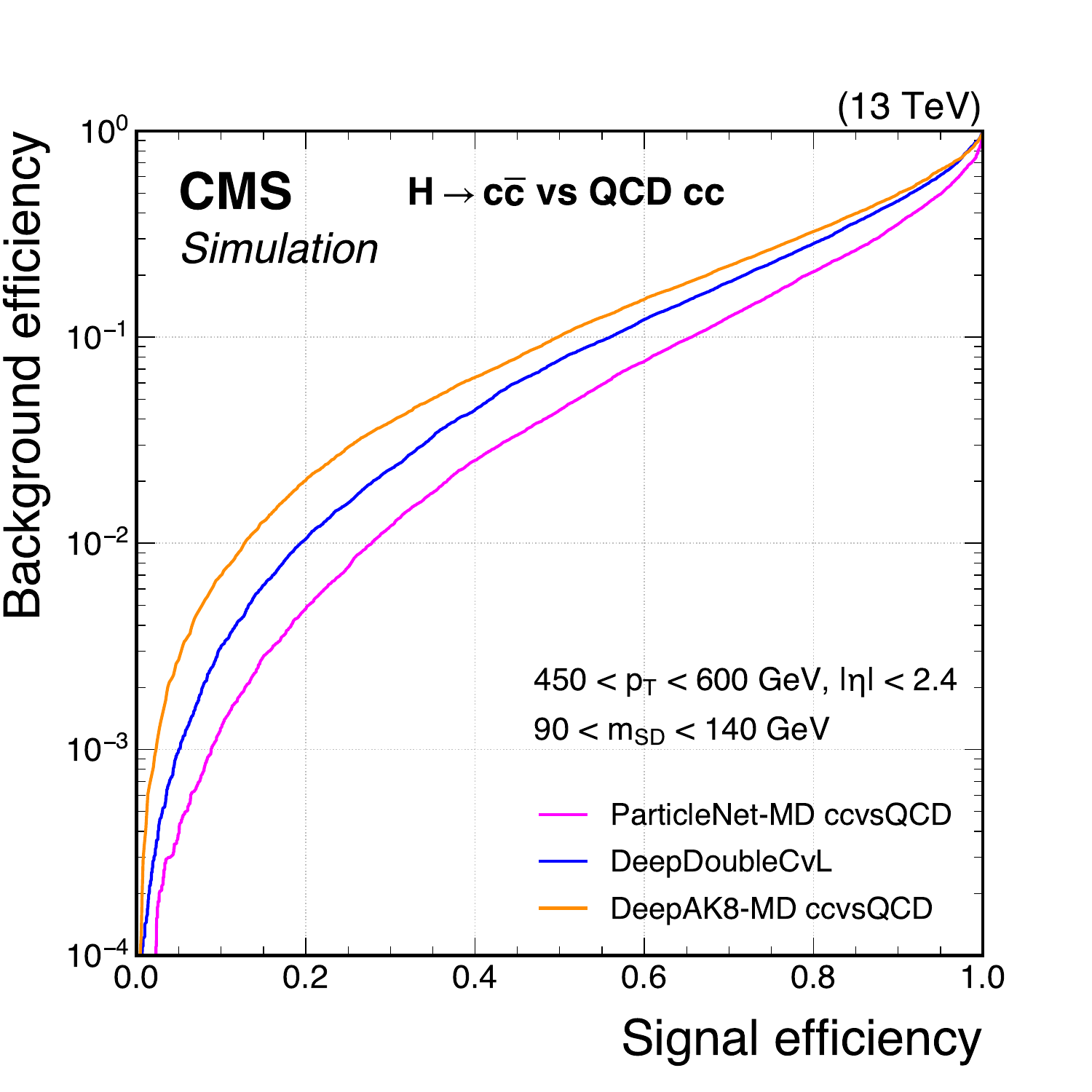}
    \includegraphics[width=0.48\textwidth]{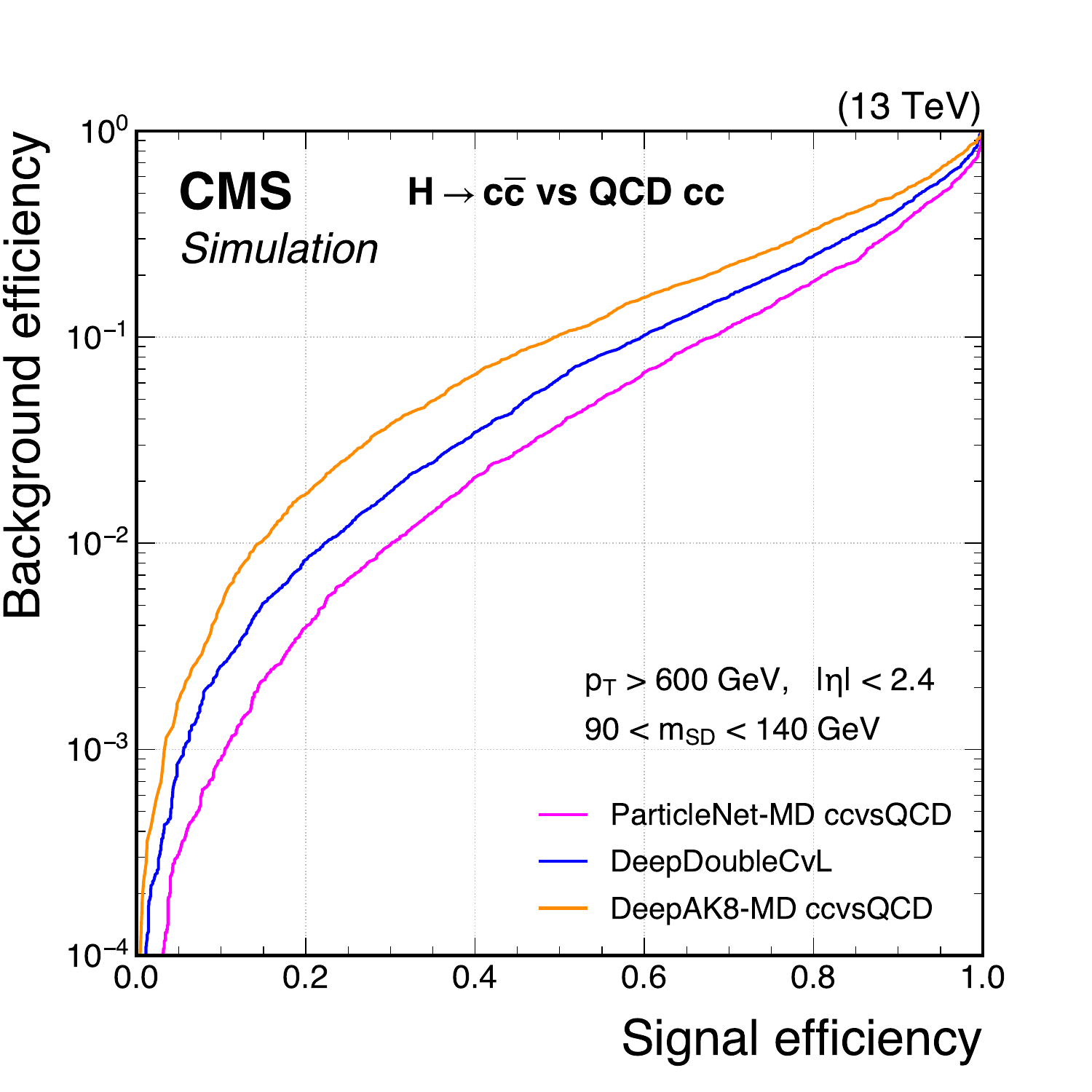}
    \caption{Comparison of the performance of the \xcc identification algorithms in terms of receiver operating characteristic (ROC) curves for \hcc signal jets versus the cc component of the QCD jets as background, using simulated events with the 2018 data-taking conditions. Performance is shown in the $450 < \pt < 600\GeV$ (left) and $\pt > 600\GeV$ (right) regions. Additional selection criteria applied to the jets are displayed on the plots.}
    \label{fig:roc_ccvsgcc}
\end{figure}

The performance of the tagging algorithms is compared in Figs.~\ref{fig:roc_bb}--\ref{fig:roc_ccvsgcc}, within the \pt ranges of 450--600 and $>$600\GeV.
The figures show the selection efficiency of \hbb (\cc) signal jets as a function of the background selection efficiency, in terms of the receiver operating characteristic (ROC) curves.
The performance is shown with respect to both inclusive QCD multijet background and separately for the QCD-bb and QCD-cc components.
Notably, since QCD-bb (cc) is a background component that closely resembles \hbb (\cc), the performance of separating \hbb (\cc) from QCD-bb (cc) background jets is significantly worse than separating them from inclusive QCD jets.

For \xbb tagging, the three neural-network-based taggers significantly outperform the double-b tagger. This improvement can be attributed to the utilization of low-level PF candidates and SV inputs, along with the capability of neural networks to effectively process such detailed information.
Similarly, when considering both \xbb and \xcc tagging, the improvements from DeepAK8-MD to DeepDoubleX, and subsequently to ParticleNet-MD, reflect the advancements made in neural network architecture.
These findings demonstrate the effectiveness of neural-network-based approaches in enhancing tagger performance for both \xbb and \xcc tagging tasks.

\begin{figure}[htbp]
    \centering
    \includegraphics[width=0.48\textwidth]{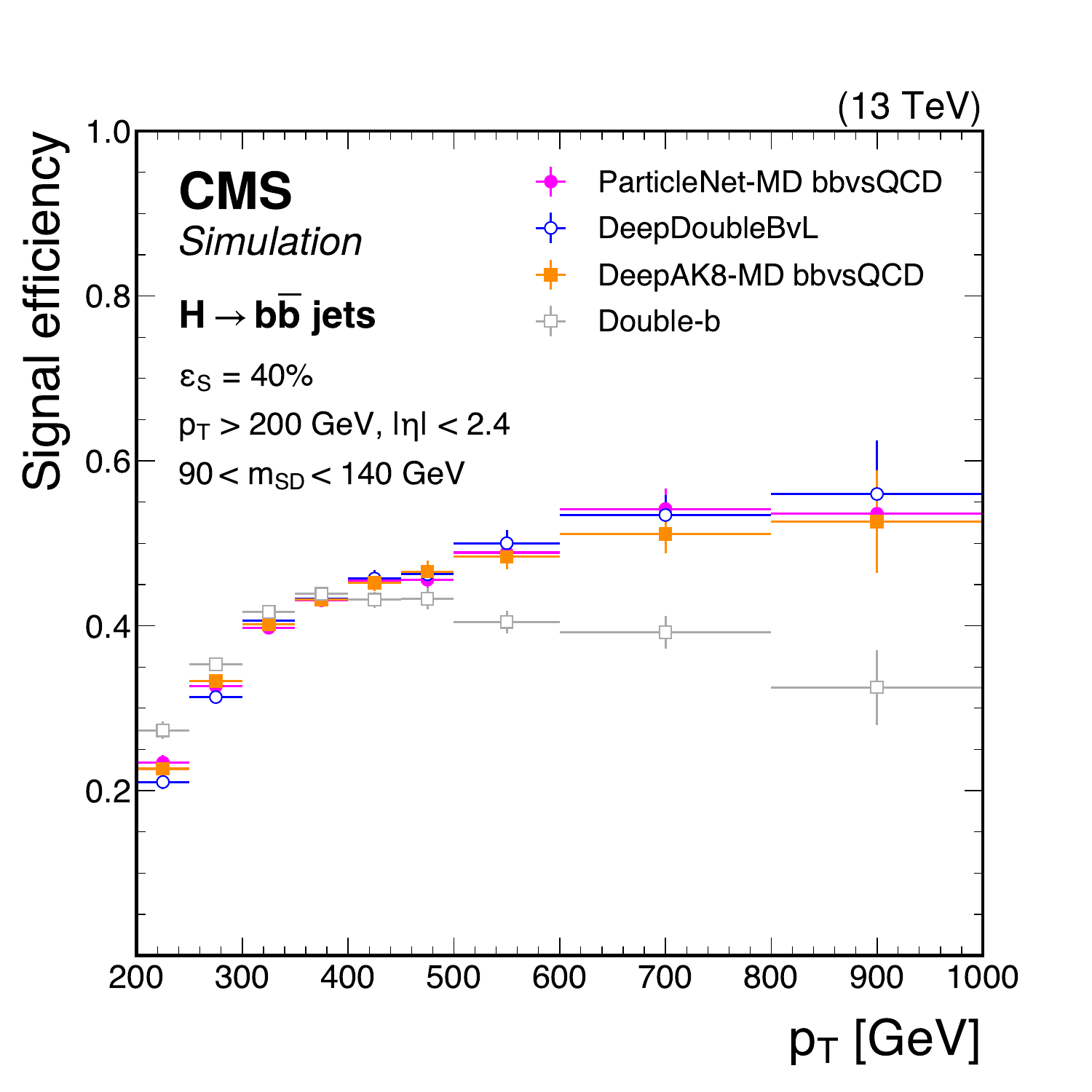}
    \includegraphics[width=0.48\textwidth]{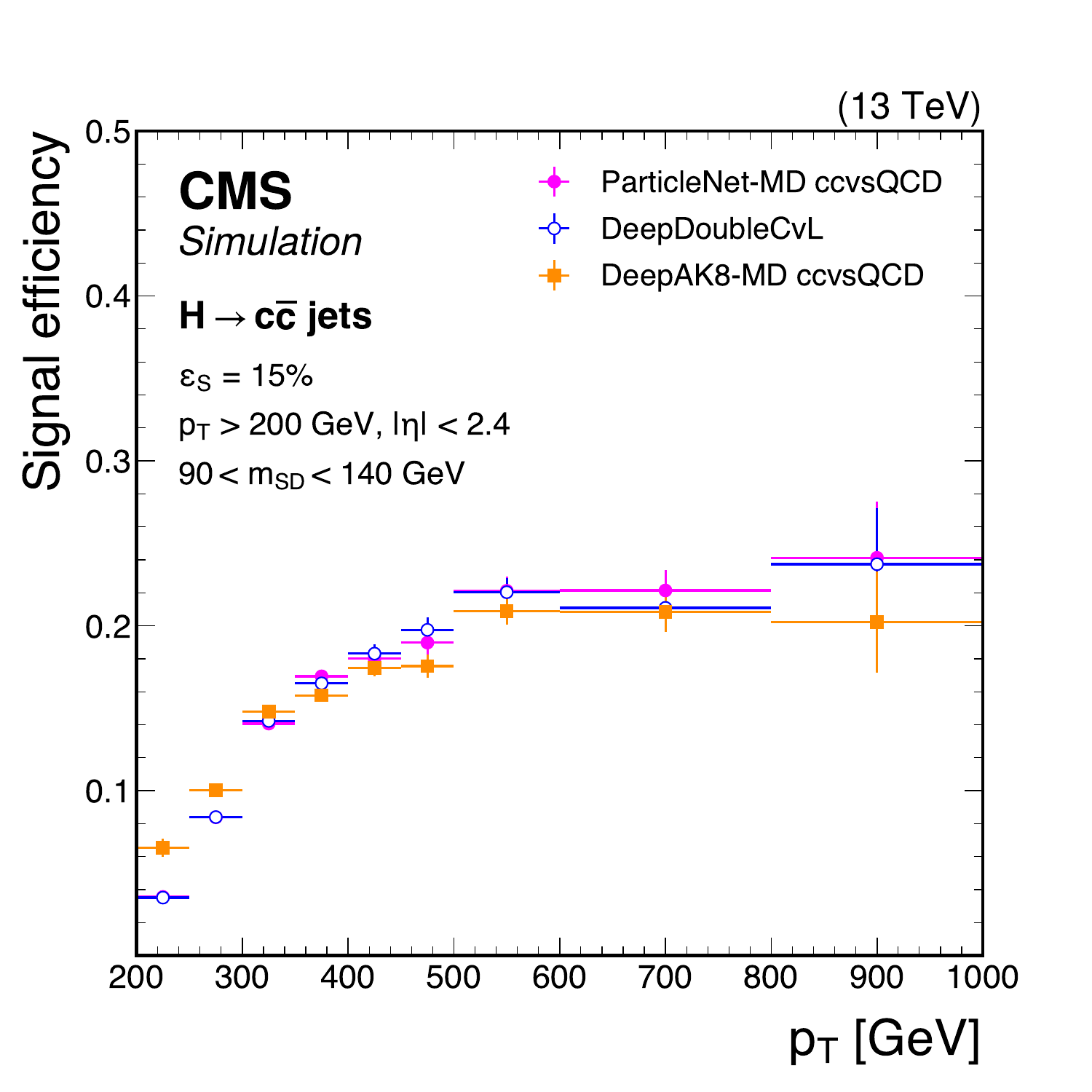}
    \caption{Signal efficiency $\epsilon_\mathrm{S}$ as a function of jet \pt for a working point corresponding to overall selection efficiencies of 40\% in \hbb and 15\% in \hcc jets. The left and right plots compare the performance of various \xbb and \xcc tagging algorithms, respectively. The error bars represent the statistical uncertainties due to the limited number of simulated events. Additional selection criteria applied to the jets are displayed on the plots.}
    \label{fig:effsig_vs_pt}
\end{figure}

The dependence of signal efficiency on jet \pt is evaluated for all algorithms, as shown in Fig.~\ref{fig:effsig_vs_pt}. In the low-\pt region, the efficiency rises rapidly up to $\pt \approx 500\GeV$. At higher \pt, the DeepAK8-MD, DeepDoubleX, and ParticleNet-MD algorithms exhibit comparable performance, with signal efficiency remaining stable. In contrast, the double-b algorithm shows a decline in performance at high \pt.

\section{Measurements of the tagging efficiency in data} \label{sec:perfdata}

This section introduces three methods to measure the efficiency of \xbb and \xcc taggers in data.
Results are presented for each of the tagger WPs in three exclusive \pt\ bins, \ie 450--500, 500--600, and $>$600\GeV, in terms of the SFs:
\begin{equation}
    \text{SF} = \epsilon_{\text{data}}(\pt) \big/ \epsilon_{\text{sim}}(\pt),
\end{equation}
where $\epsilon_{\text{data}}(\pt)$ and $\epsilon_{\text{sim}}(\pt)$ represent the \pt-dependent tagging efficiency in data and for simulated jets.
In the first method, referred to as the sfBDT method, a BDT is trained to identify a sample of QCD \bb (\cc) jets arising from gluon splitting with characteristics similar to the \xbb (\cc) signal jets.
These gluon-splitting jets are used as proxies for the signal.
In the second method, QCD \bb (\cc) jets are selected using requirements on soft muons inside the jets and the $N$-subjettiness feature. 
The third method uses \zbb jets in data as a proxy for \xbb signal jets.

A detailed description of each method, followed by the calibration results, is given below. A combination of the measured SFs is presented at the end of the section.

\subsection{The sfBDT method} \label{sec:sfbdt_method}

\subsubsection{Method description} \label{sec:sfbdt_method_description}

The sfBDT method, first introduced in Ref.~\cite{CMS-DP-2022-005}, calibrates the \xbb (\cc) signal jets using gluon-splitting jets as a proxy.
At the core of the method is a BDT, a multivariate technique employed to integrate various jet observables and construct a discriminant to identify regions of phase space of \gbb (\cc) jets that closely resemble corresponding regions of \xbb (\cc) signal jet phase space. As such, it serves as an enhanced version of early calibration methods based on gluon-splitting proxies with manually constructed selection variables, as detailed in Section~9.3 of Ref.~\cite{CMS:BTVFlvTagger}.
This section provides a self-contained summary of the method, with particular emphasis on the design of the BDT, the validation of the similarity between proxy and signal jets, and the study of the dependence of the SFs on the sfBDT selection criteria.

Events are selected online using a logical OR of HLT algorithms with different \HT (scalar \pt sum of all AK4 jets) thresholds, starting from 125 (180)\GeV for the 2016 (2017--2018) era.
Events are required to have at least one AK8 jet.
To ensure a sufficient number of selected jets, for each event, the leading AK8 jet and the subleading one (if it exists), ordered by \pt, are selected if they satisfy $\pt > 200\GeV$, $\abs{\eta} < 2.4$, and $50 < \msd < 200\GeV$.
Light-flavour jets are suppressed by requiring preselected jets to have at least two SVs within the cone of the jet, with each matched to one of the two subjets produced by the SD algorithm.
Since HLTs with low \HT thresholds are prescaled triggers, a reweighting procedure is applied to the simulated events to align them with the data. This is performed on a 2D binned histogram defined by the event \HT and jet \pt, for the leading and subleading jets separately.

In the simulation, each selected jet is classified as bottom (\PQb), charm (\PQc), or light flavour (\Pl) depending on the number of ghost-matched \PQb and \PQc hadrons.
Jets with at least one matched \PQb hadron are assigned to the ``\PQb'' category; jets with no matched \PQb hadrons but including at least one matched \PQc hadron go to the ``\PQc'' category; the remaining jets are labelled as ``\Pl'' type.
In multijet QCD events, the \PQb (\PQc) category mainly comprises \gbb (\cc) jets.
These jets differ from the \xbb (\cc) signal jets, especially in the tagging discriminant scores.
The aim of the sfBDT method is to obtain a more representative, signal-like sample of jets using a BDT discriminant. The sfBDT is specifically trained to distinguish between two groups of jets, both originating from QCD multijet events and selected based on generator-level jet information. One group consists of jets that closely resemble \xbb (\cc) jets, whereas the other group comprises jets that are less similar to the signal.

The determination of suitable generator-level variables is therefore critical for the performance of the sfBDT method. Dedicated studies are performed to characterize the differences between \gbb (\cc) jets and the signal \xbb (\cc) jets at the generator level.
A notable distinction is that \gbb (\cc) jets are more frequently contaminated with additional gluons.
Consequently, the previous iteration of the method, detailed in Ref.~\cite{CMS-DP-2022-005}, defined the training samples using the parton-level variable,
\begin{equation}\label{eq:kappa_g}
    \kappa_\Pg= \frac{\sum_{i\in \{\Pg\}}p_{\mathrm{T},i}}{\sum_{i\in \{\Pg,\,\Pq\}}p_{\mathrm{T},i}},
\end{equation}
which measures the fraction of energy inside a jet due to gluons.
Smaller $\kappa_\Pg$ corresponds to QCD jets that exhibit closer resemblance to the resonance \bb or \cc jets.
The updated version of the method presented in this paper uses a new variable based on the generator-level hadrons, instead of the partons.
It has been observed that \gbb (\cc) jets tend to contain extra radiations, which arise from either gluon or quark emissions.
From the perspective of the distribution of generator-level hadrons within the jet, such emissions often result in a distinct multiprong structure.
To characterize this behaviour, we define the $N$-subjettiness using first-generation hadrons---those directly produced from partons prior to any sequential decay. This approach provides a measure of the multiprongness of jets at the generator level.
The $\tau_{MN}$ of the hadrons, denoted by $\tau_{MN}^{\text{h}}$, is defined analogously to the standard $N$-subjettiness ratio~\cite{Thaler:2010tr},
\begin{equation}
    \tau_{MN}^{\text{h}} = \frac{\sum_{i\in \{\text{had.}\}}p_{\mathrm{T},i}\, \min_{j=1}^{M} \{\Delta R_{i,\,\hat{n}_{M,j}}\}}{\sum_{i\in \{\text{had.}\}}p_{\mathrm{T},i}\, \min_{j=1}^{N} \{\Delta R_{i,\,\hat{n}_{N,j}}\}},
\end{equation}
where $\hat{n}_{N,j}$ $(j=1,\,2,\cdots,\,N)$ are the $N$ subjet axes of the hadrons, obtained by performing the exclusive $k_\mathrm{T}$ algorithm~\cite{Catani:1993hr,Ellis:1993tq} on the hadron list.
The $\tau_{31}^{\text{h}}$ variable, with the signal (background) for training of sfBDT defined as $\tau_{31}^{\text{h}} < 0.1$ ($> 0.1$), yields the sfBDT with the strongest ability to select \gbb (\cc) jets that resemble \xbb (\cc) jets.
This choice also demonstrates superior performance compared with the earlier variable described in Eq.~(\ref{eq:kappa_g}).

\begin{figure}[htbp]
    \centering
    \includegraphics[width=0.48\textwidth]{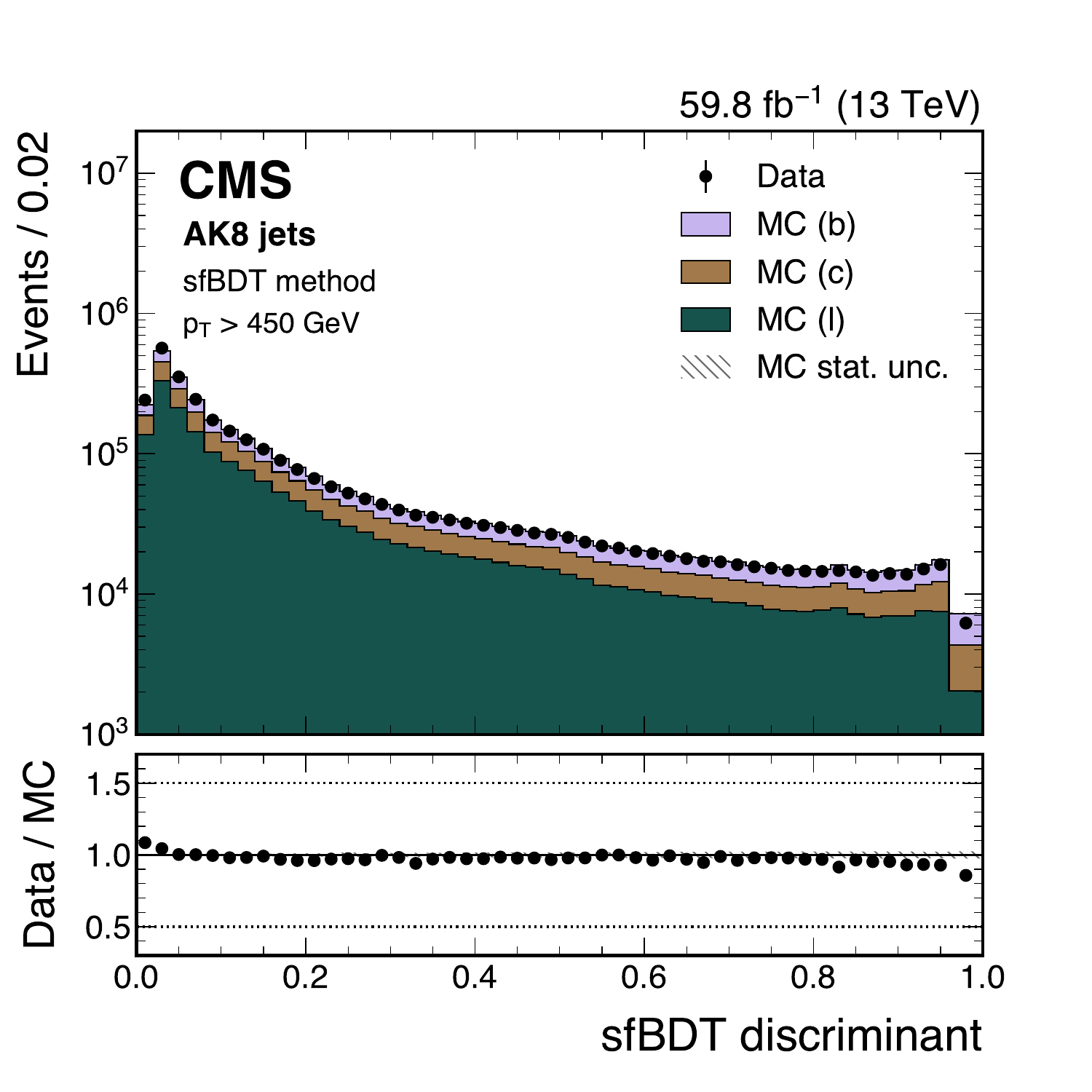}
    \caption{Distributions of the sfBDT discriminant for data and simulation, illustrated using the 2018 data-taking conditions, for jets with $\pt > 450\GeV$. The error bars indicate statistical uncertainties in observed data, which may be too small to be visible.
    }
    \label{fig:sfbdt_dist}
\end{figure}

The sfBDT is trained on the simulated QCD multijet events enriched with \PQb and \PQc partons. The same preselection and jet categories (``\PQb'', ``\PQc'', and ``\Pl'') are applied. Signal and background jets are selected from the combined ``\PQb'' and ``\PQc'' categories based on the criterion involving $\tau_{31}^{\text{h}}$. This allows the same sfBDT discriminant to be used in calibrating both \xbb and \xcc jets.
The input to the sfBDT includes six jet-level variables: the constituent-based $N$-subjettiness ratio $\tau_{21}$, the masses of the two subjets obtained from the SD algorithm, the \pt of the two SVs matched to each subjet, and the total number of tracks associated with the two SVs.
Figure~\ref{fig:sfbdt_dist} shows the distribution of the trained sfBDT discriminant for both data and simulated events. Overall, good agreement is observed between data and simulation in the sfBDT discriminant. Residual differences are a source of systematic uncertainty, as discussed in detail in Section~\ref{sec:sfbdt_method_uncertainty}.

The proxy jets are defined as jets passing the preselection, belonging to ``\PQb'' (``\PQc'') category in the calibration of \xbb (\cc) jets, and satisfying a dedicated selection on the sfBDT discriminant. The choices of sfBDT selections are detailed below.

\begin{figure}[htbp]
    \centering
    \includegraphics[width=0.48\textwidth]{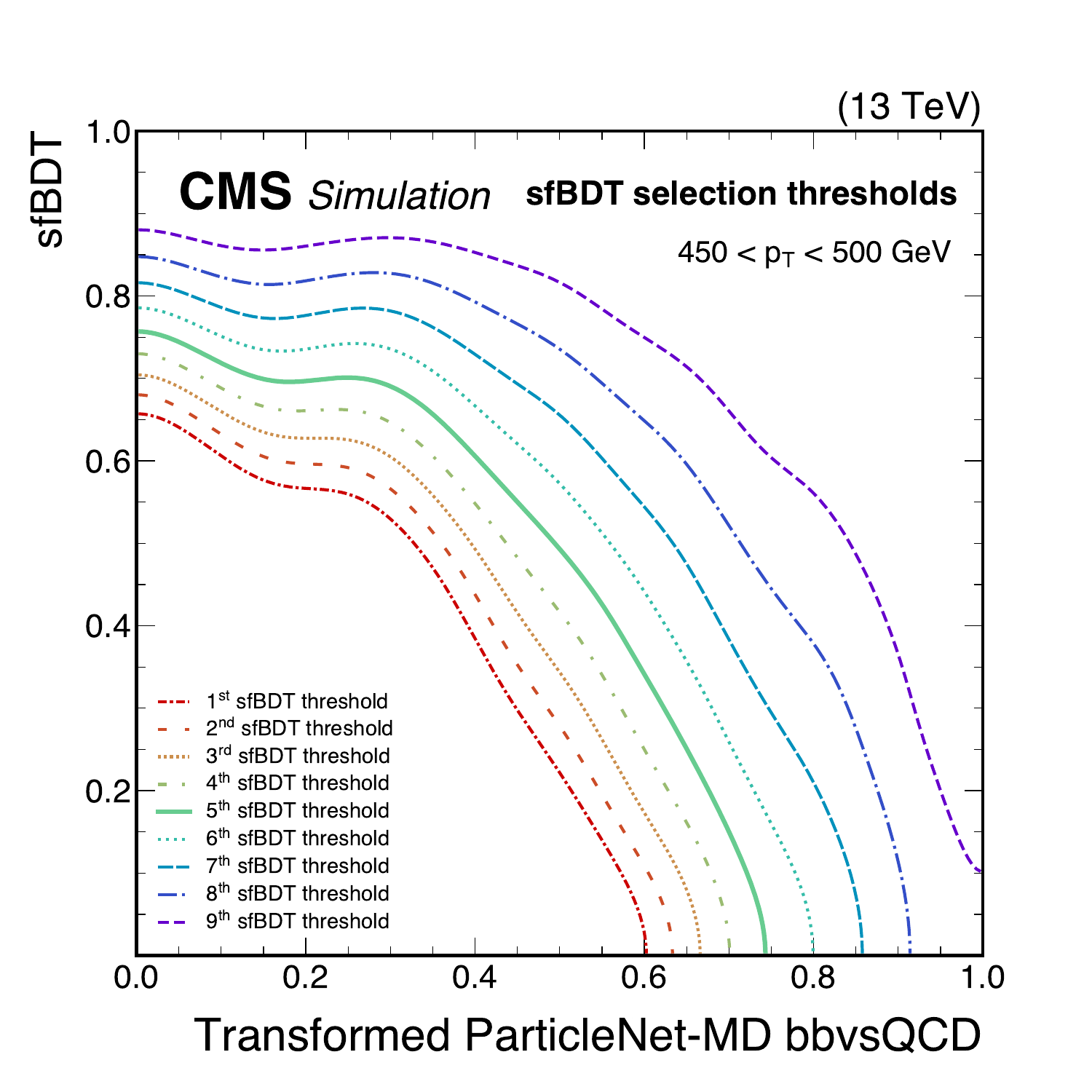}
    \includegraphics[width=0.48\textwidth]{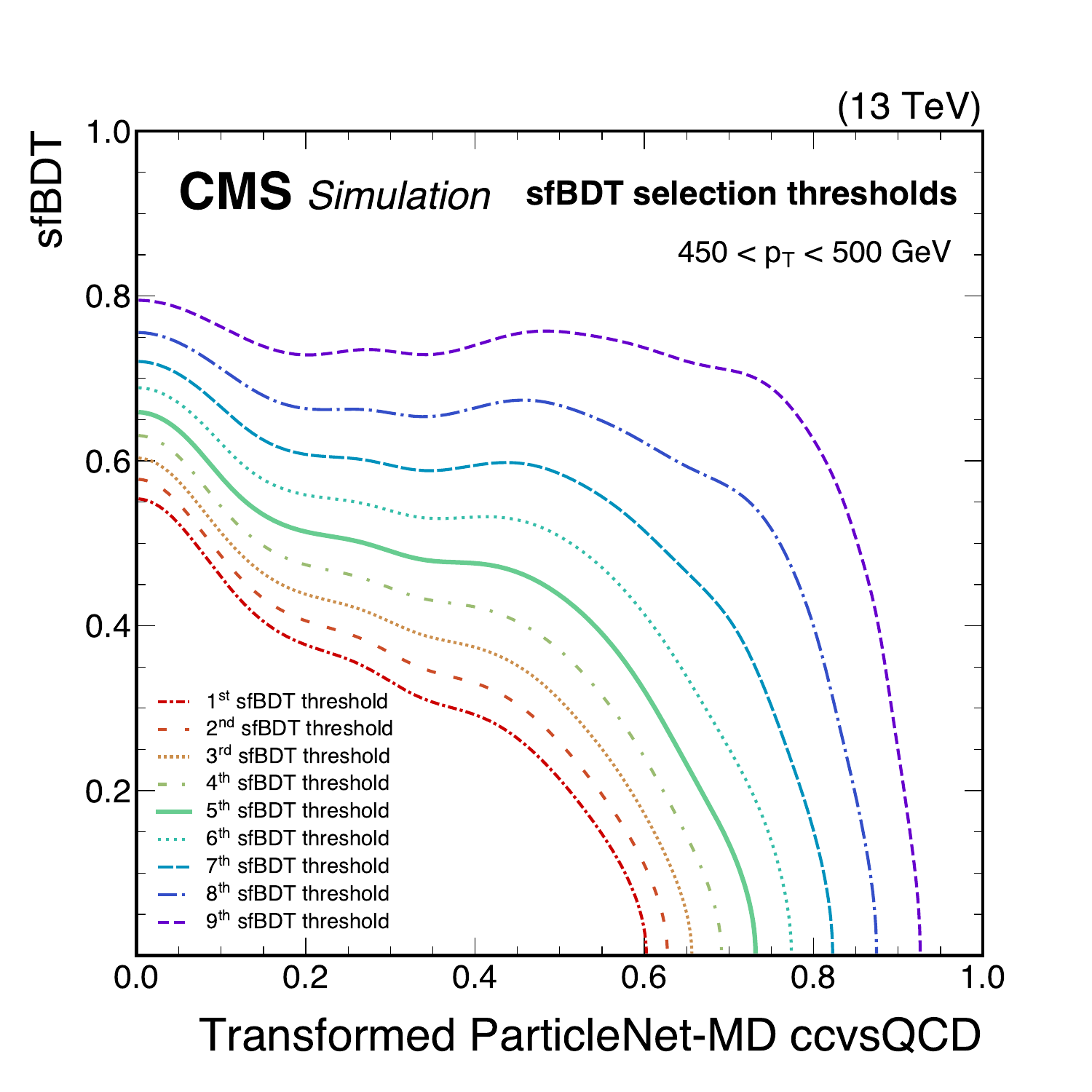}
    \caption{
    Illustration of nine predefined ``reference selection thresholds'' visualized on the two-dimensional plane spanned by the sfBDT score and the transformed tagger discriminant scores.
    Selections based on these thresholds can be interpreted as sfBDT selections with thresholds as a function of the tagger discriminant score. Each selection aims to match the tagger discriminant distribution of the proxy jet to that of the signal.
    The examples shown correspond to the calibration of the ParticleNet-MD \xbb (left) and ParticleNet-MD \xcc (right) discriminants, using simulated events under 2018 data-taking conditions in the jet \pt range of $(450,\,500)\GeV$.}
    \label{fig:sfbdt}
\end{figure}

To define a jet selection using the sfBDT discriminant, the updated method introduces an automated and more sophisticated procedure, improving upon the earlier approach described in Ref.~\cite{CMS-DP-2022-005}.
The method introduces nine predefined ``reference selection thresholds'', which are selections on the sfBDT discriminant as a function of the tagger discriminant score. Each reference selection threshold is chosen to align the tagger discriminant distributions of proxy and signal jets.
They can be visualized on a 2D plane of the sfBDT score versus the transformed tagger discriminant score, as shown in Fig.~\ref{fig:sfbdt}, with the threshold index increasing from the loosest to the tightest selection. The tagger discriminant is transformed to $X\in (0,\,1)$ such that a selection of $X>X_0$ corresponds to the signal jet selection efficiency of $1-X_0$. Events that pass the selection thresholds are located above the corresponding curves.
As observed, the reference selection thresholds apply looser constraints on the sfBDT score in regions with higher tagger discriminant scores.
Since the predefined thresholds generate a set of references where the proxy jet phase space matches that of the signal jets, variations around these thresholds are introduced to produce conditions with differing signal-to-proxy similarity levels, resulting in a total of 81 selection choices. Additional details are provided in Section~\ref{sec:sfbdt_method_uncertainty}. Specifically, each selection yields a corresponding SF, and the spread among these SFs is used to quantify an uncertainty term related to the dependence of the SF on the choice of sfBDT selections.

\begin{figure}[htbp]
    \centering
    \includegraphics[width=0.48\textwidth]{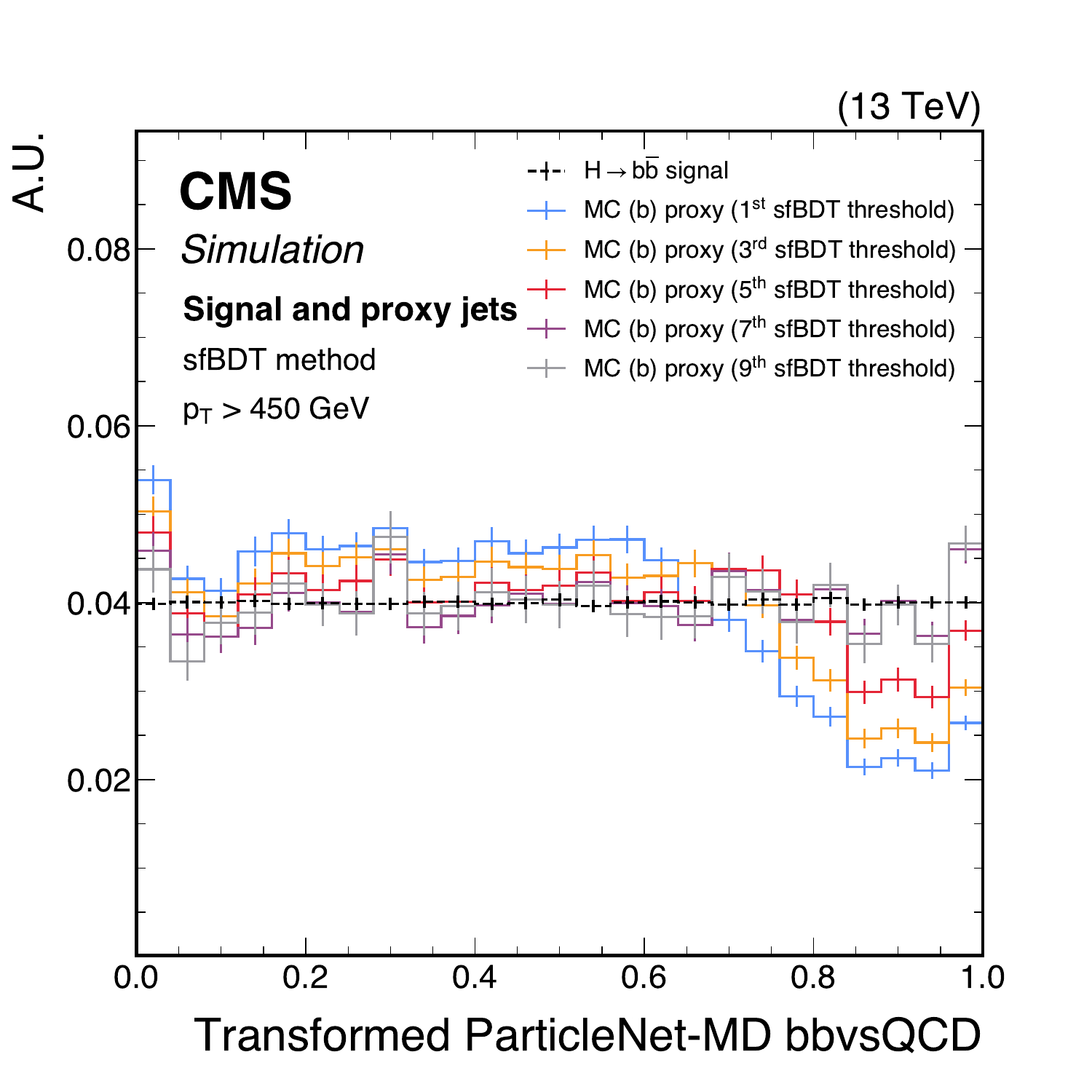}
    \includegraphics[width=0.48\textwidth]{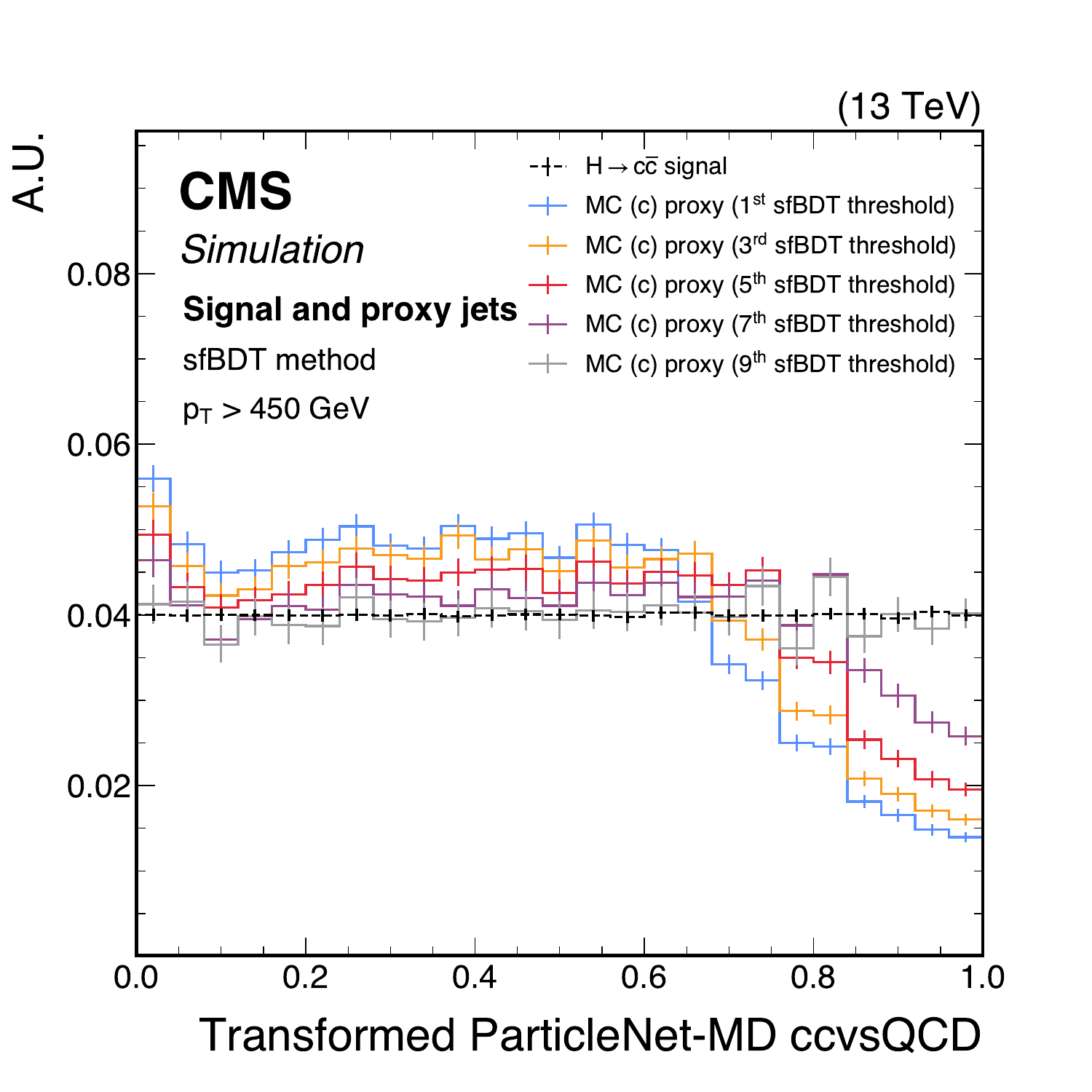}
    \caption{Shapes of the transformed ParticleNet-MD \xbb (left) and \xcc (right) discriminants for SM \hbb (\cc) signal jets and proxy jets selected with different sfBDT selection thresholds. The examples correspond to the calibration of the ParticleNet-MD \xbb and \xcc discriminants with the sfBDT method, using simulated events under 2018 data-taking conditions for jets with $\pt > 450\GeV$. The error bars represent the statistical uncertainties due to the limited number of simulated events.
    }
    \label{fig:sigvsproxy_sfbdt}
\end{figure}

Figure~\ref{fig:sigvsproxy_sfbdt} demonstrates the closure of proxy and signal jets on the transformed tagger discriminant after applying selections based on the ``reference selections''.
The closure is also evaluated across various jet observables, including the kinematic properties of subjets, SV kinematics and impact parameters, and the number of tracks associated with the SVs. The sfBDT selection substantially improves the agreement between proxy and signal jets.

\begin{figure}[htbp]
    \centering
    \includegraphics[width=0.48\textwidth]{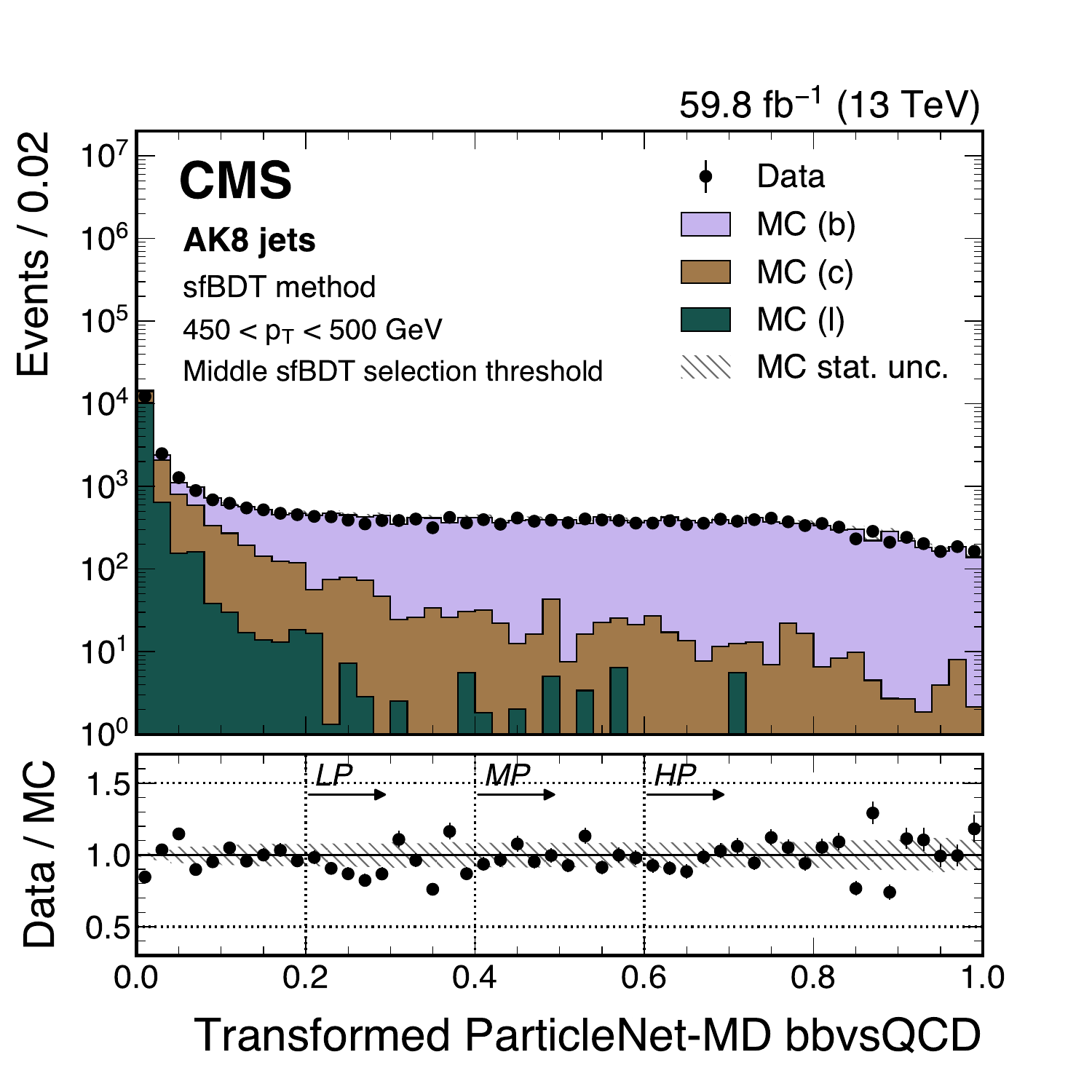}
    \includegraphics[width=0.48\textwidth]{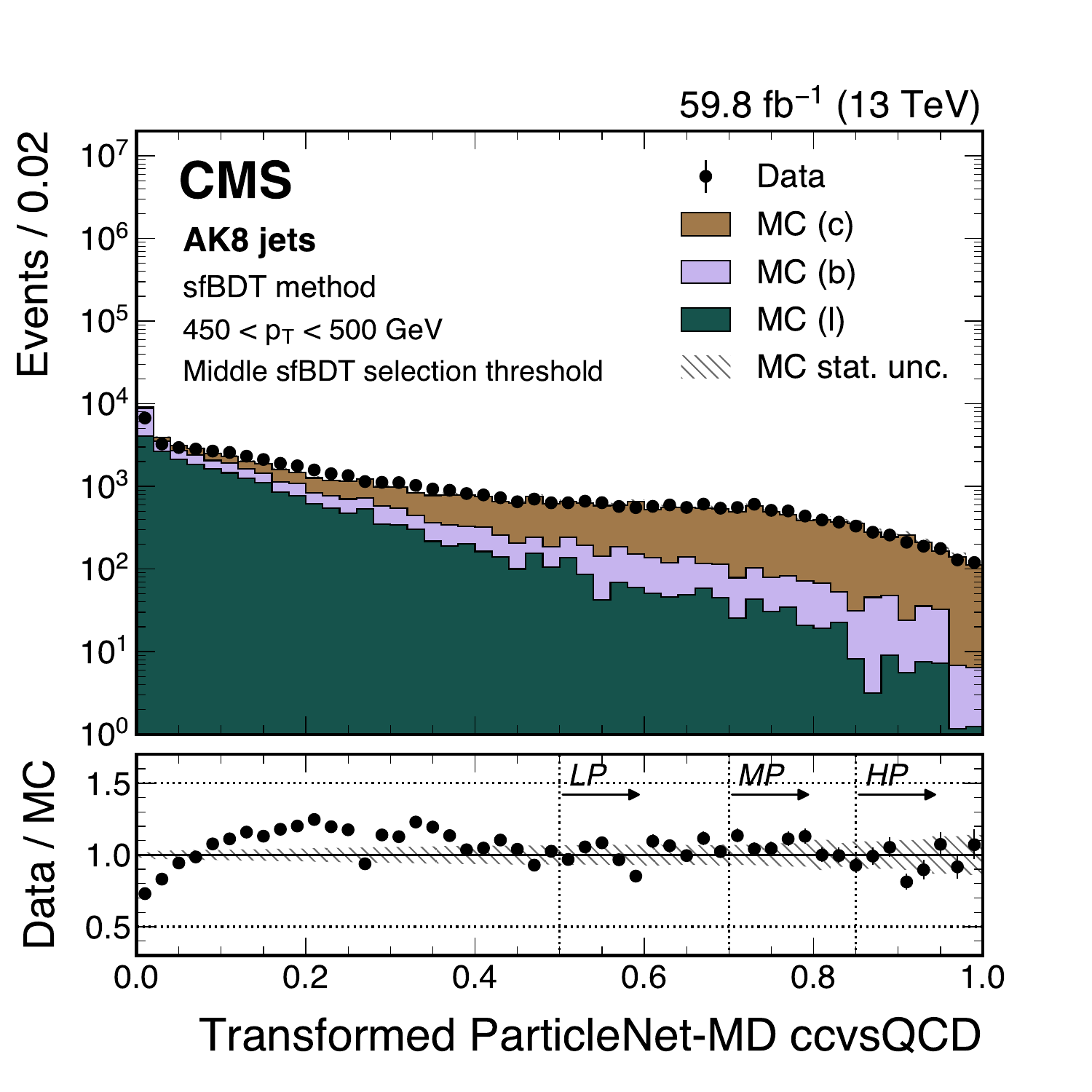}
    \caption{An example of the transformed ParticleNet-MD \xbb (left) and \xcc (right) distribution in data and simulated events, after applying the preselection and the middle sfBDT selection threshold in the sfBDT method. The high-purity (HP), medium-purity (MP), and low-purity (LP) working points for the left (right) plot correspond to selections of $X>0.6,\,0.4,\,0.2$ $(0.85,\,0.7,\,0.5)$ on the transformed tagger discriminant. The error bars represent the statistical uncertainties in observed data. The lower panels display the ratio of data to simulation, with the hatched bands representing the normalized statistical uncertainty of simulated events for each bin. The distributions are based on data and simulated events with the 2018 data-taking conditions, in the jet \pt range of $(450,\,500)\GeV$.}
    \label{fig:datamc_xtagger_pnet}
\end{figure}

\begin{figure}[htbp]
    \centering
    \includegraphics[width=0.48\textwidth]{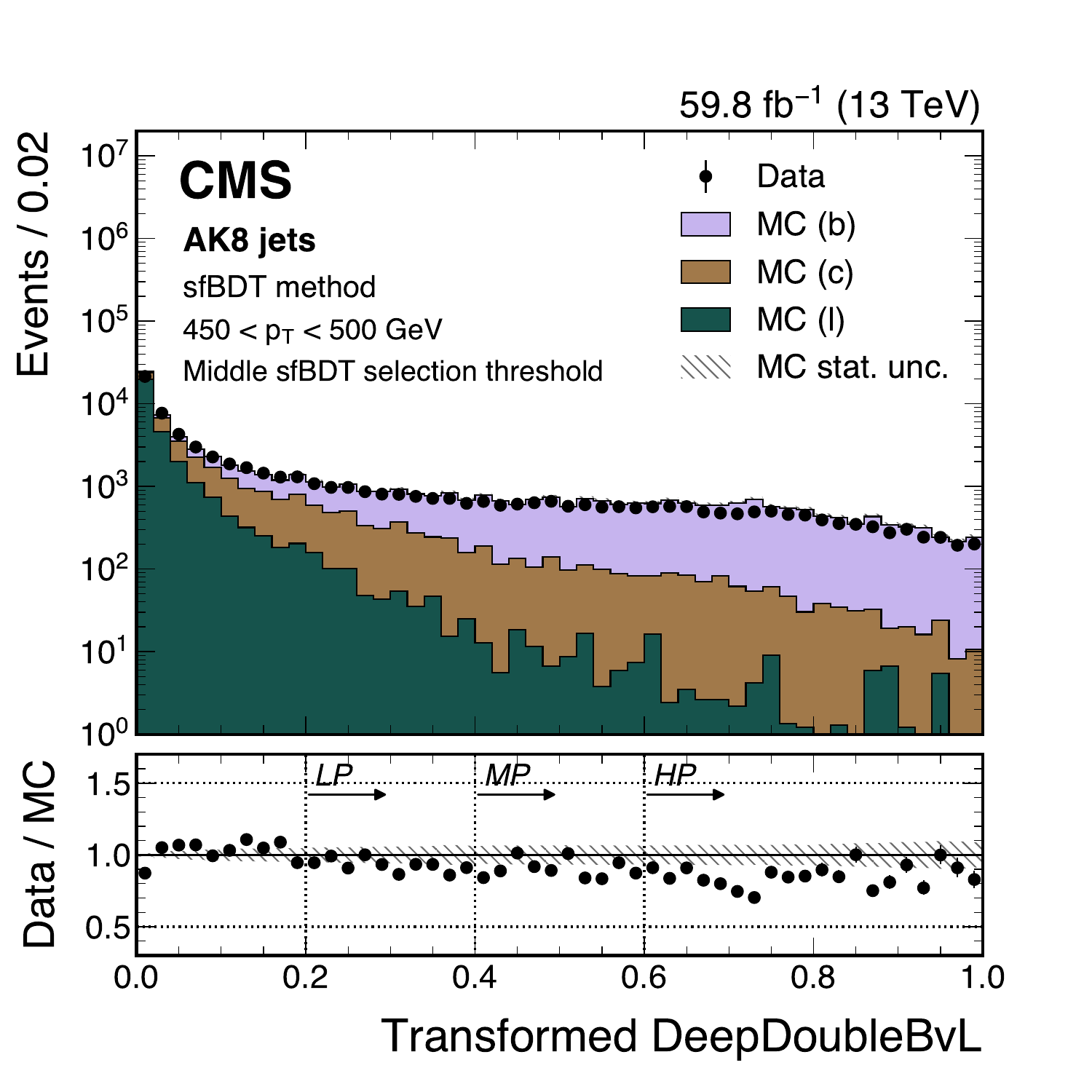}
    \includegraphics[width=0.48\textwidth]{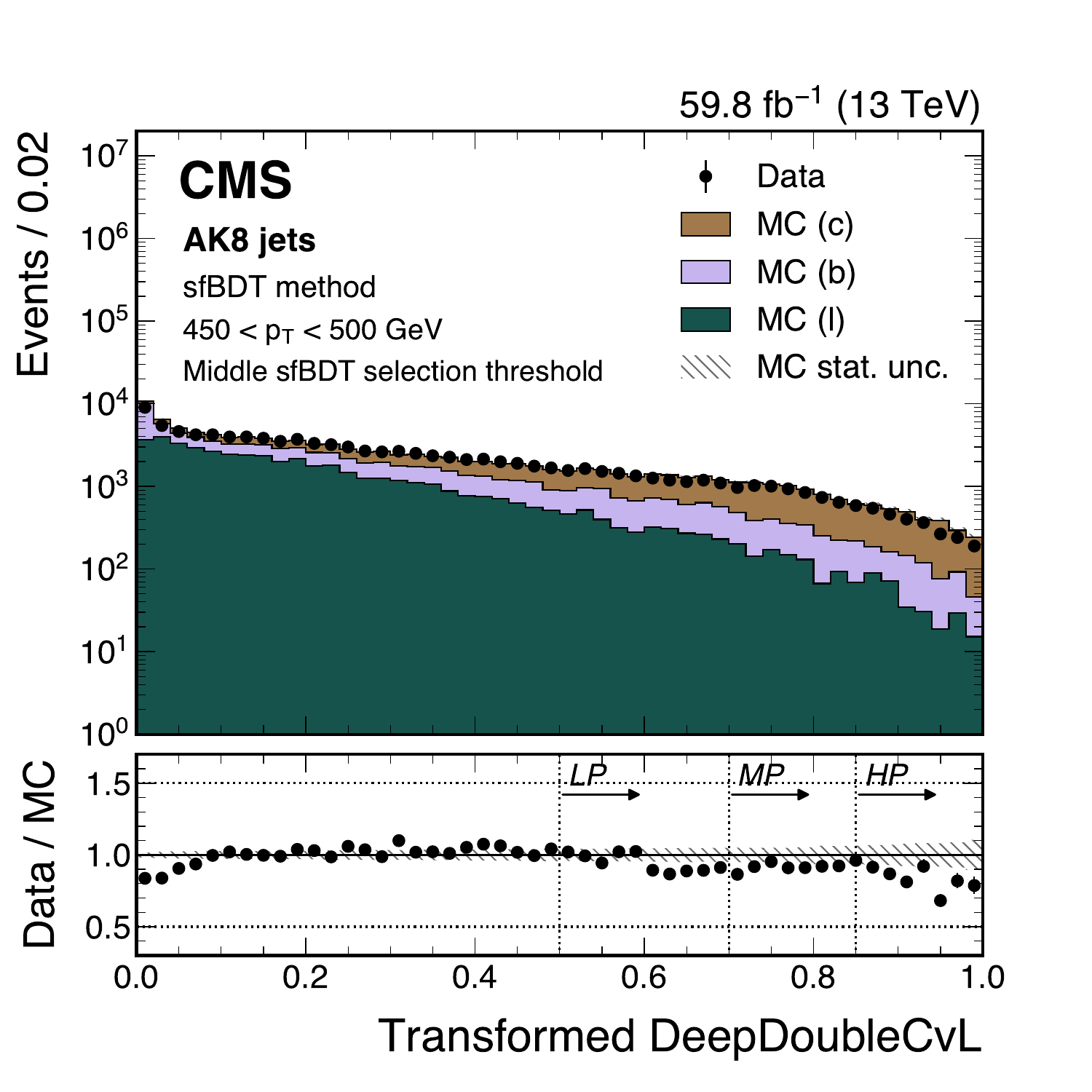}
    \caption{An example of the transformed DeepDoubleX \xbb (left) and \xcc (right) distribution in data and simulated events, after applying the preselection and the middle sfBDT selection threshold in the sfBDT method. The high-purity (HP), medium-purity (MP), and low-purity (LP) working points for the left (right) plot correspond to selections of $X>0.6,\,0.4,\,0.2$ $(0.85,\,0.7,\,0.5)$ on the transformed tagger discriminant. The error bars represent the statistical uncertainties in observed data. The lower panels display the ratio of data to simulation, with the hatched bands representing the normalized statistical uncertainty of simulated events for each bin. The distributions are based on data and simulated events with the 2018 data-taking conditions, in the jet \pt range of $(450,\,500)\GeV$.}
    \label{fig:datamc_xtagger_ddx}
\end{figure}

\begin{figure}[htbp]
    \centering
    \includegraphics[width=0.48\textwidth]{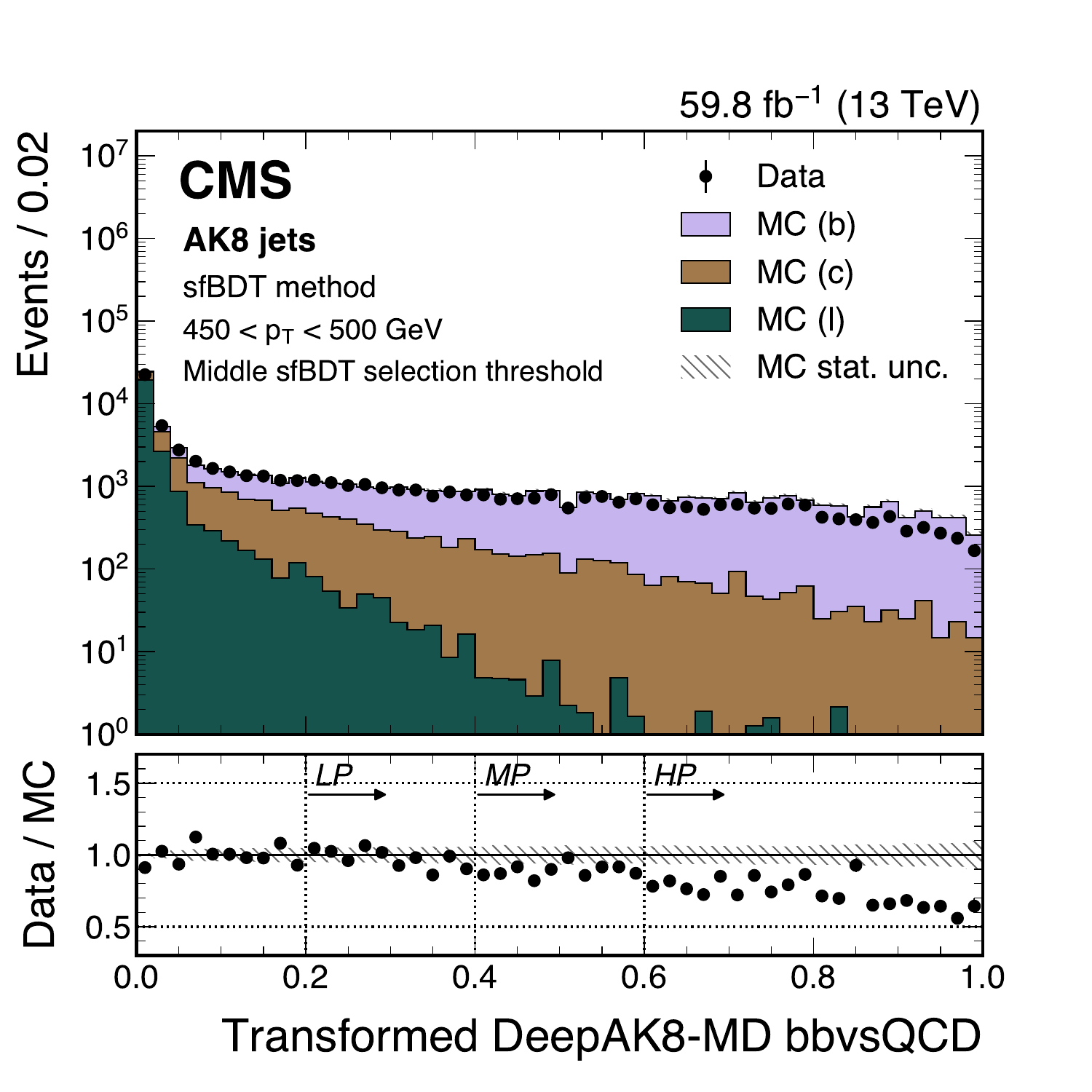}
    \includegraphics[width=0.48\textwidth]{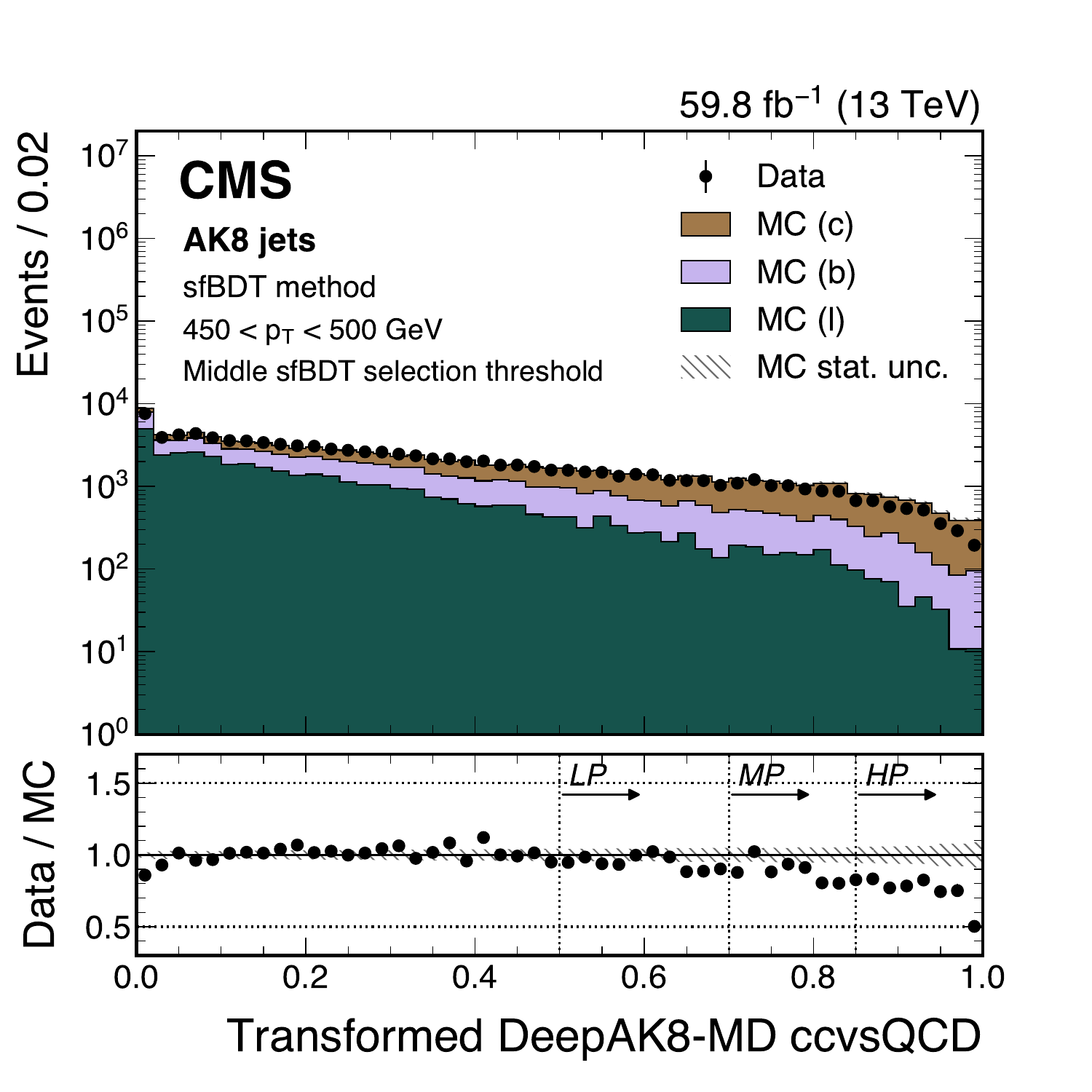}
    \caption{An example of the transformed DeepAK8-MD \xbb (left) and \xcc (right) distribution in data and simulated events, after applying the preselection and the middle sfBDT selection threshold in the sfBDT method. The high-purity (HP), medium-purity (MP), and low-purity (LP) working points for the left (right) plot correspond to selections of $X>0.6,\,0.4,\,0.2$ $(0.85,\,0.7,\,0.5)$ on the transformed tagger discriminant. The error bars represent the statistical uncertainties in observed data, which may be too small to be visible. The lower panels display the ratio of data to simulation, with the hatched bands representing the normalized statistical uncertainty of simulated events for each bin. The distributions are based on data and simulated events with the 2018 data-taking conditions, in the jet \pt range of $(450,\,500)\GeV$.}
    \label{fig:datamc_xtagger_da}
\end{figure}

\begin{figure}[htbp]
    \centering
    \includegraphics[width=0.48\textwidth]{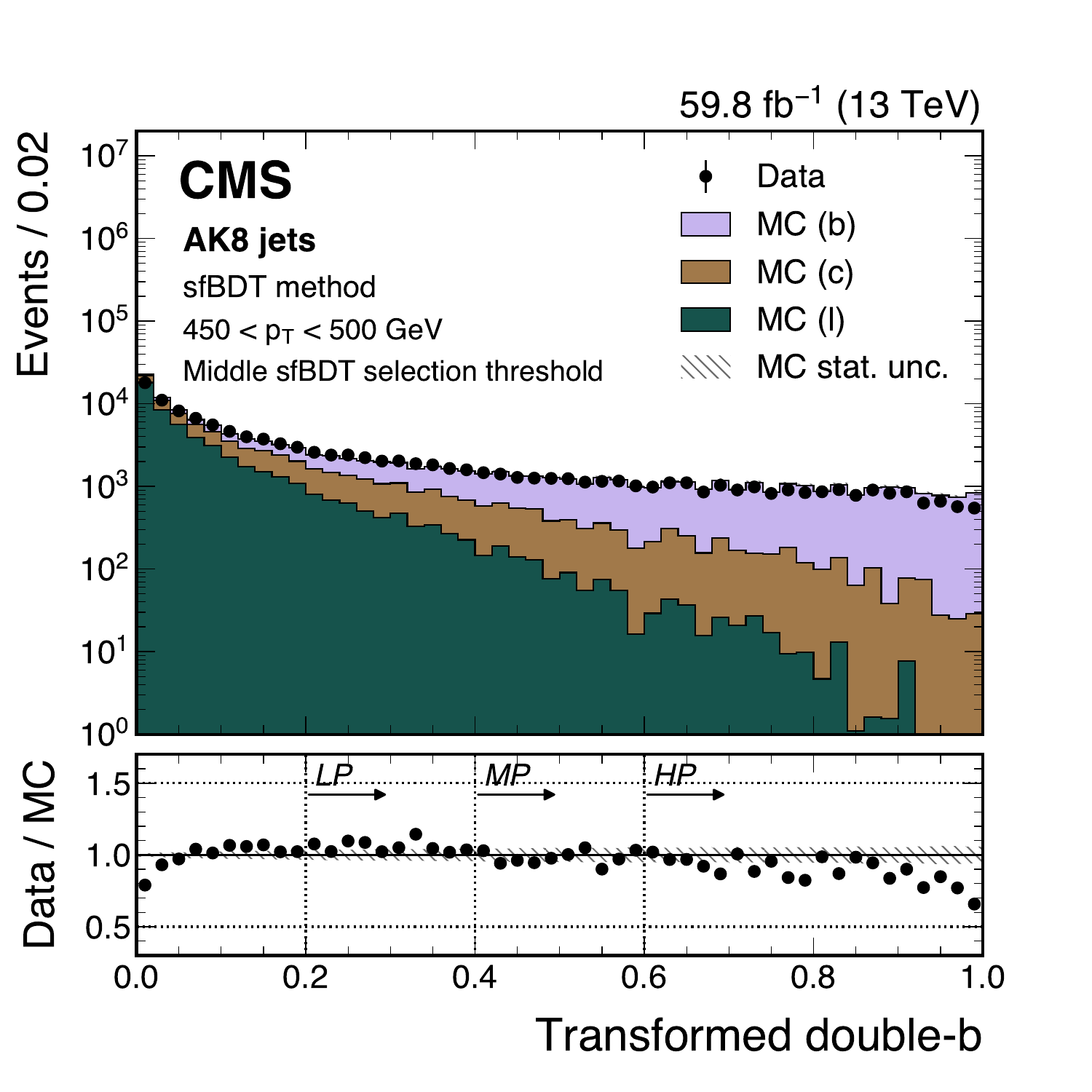}
    \caption{An example of the transformed double-b distribution in data and simulated events, after applying the preselection and the middle sfBDT selection threshold in the sfBDT method. The high-purity (HP), medium-purity (MP), and low-purity (LP) working points correspond to selections of $X>0.6,\,0.4,\,0.2$ on the transformed tagger discriminant. The error bars represent the statistical uncertainties in observed data, which may be too small to be visible. The lower panel displays the ratio of data to simulation, with the hatched bands representing the normalized statistical uncertainty of simulated events for each bin. The distribution is based on data and simulated events with the 2018 data-taking conditions, in the jet \pt range of $(450,\,500)\GeV$.}
    \label{fig:datamc_xtagger_doubleb}
\end{figure}

Figures~\ref{fig:datamc_xtagger_pnet}--\ref{fig:datamc_xtagger_doubleb} show the tagger discriminant distribution in data and simulation, after applying the middle sfBDT selection among the nine options illustrated in Fig.~\ref{fig:sfbdt} (solid curve).
The level of agreement between data and simulation varies depending on the tagger type and the tagging WP. Discrepancies are more pronounced for higher purity WPs, highlighting the need to calibrate the selection efficiencies. Overall, the ParticleNet-MD and DeepDoubleX discriminants exhibit better agreement between data and simulation in the high-discriminant-score region, whereas the DeepAK8-MD and double-b discriminants show larger discrepancies. These differences will be further discussed in Section~\ref{sec:combination}.

The yields of ``\PQb'', ``\PQc'', and ``\Pl'' categories in data are determined from a template of fit to the variable $\ln(m_{\text{SV\textsubscript{1}}}^{\text{corr}}/\GeV)$,
where SV\textsubscript{1} is the SV with maximum $d_{xy}$ significance and $m_{\text{SV}}^{\text{corr}}$ is the corrected SV mass. It is defined as
\begin{equation} \label{eq:svmass}
    m_{\text{SV}}^{\text{corr}} = \sqrt{{m_{\text{SV}}^2 + p^2\sin^2\theta}} + p\sin\theta, 
\end{equation}
where $m_{\text{SV}}$ is the invariant mass of the tracks associated with the SV, $p$ is the SV momentum obtained from associated tracks, and $\theta$ the angle between the SV momentum and the vector pointing from the PV to the SV.
This correction to the SV mass accounts for the difference between the SV's flight direction and its momentum, considering the effects of potential particles that were either not reconstructed or failed to be associated with the SV.
The fit variable is constructed to enhance separation among the three categories. For the ``\PQb'' and ``\PQc'' categories, $m_{\text{SV}}^{\text{corr}}$ exhibits mass peaks corresponding to \PQb and \PQc hadrons, located around 5\GeV and 1.5\GeV, respectively. In contrast, the ``\Pl'' category shows a smooth distribution. The logarithmic scale is chosen to address the long-tailed distribution caused by the reconstruction precision of $m_{\text{SV}}^{\text{corr}}$, allowing for wider bins in the high-mass region.

Three unconstrained parameters, denoted as $\text{SF}_\PQb$, $\text{SF}_\PQc$, and $\text{SF}_\Pl$, are used to define the normalization of the ``\PQb'', ``\PQc'', and ``\Pl'' categories when passing the selection on the tagger discriminant at a given WP.
The three parameters are simultaneously fitted to data in the regions passing or failing the tagger WP (denoted by the ``pass'' and ``fail'' region).
The total yields of the ``pass'' and ``fail'' regions for each category remain constant.
The fit is delivered separately in three exclusive \pt bins.
Specifically, for the fit performed with a given tagger and WP at a specific \pt bin, for each histogram bin, let the simulated event yields of the three flavour categories in the ``pass'' and ``fail'' regions be denoted as $N_f^{\text{sim,P}}$ and $N_f^{\text{sim,F}}$, respectively, where $f = \PQb,\,\PQc,\,\Pl$. The predicted data yields for the three categories, $N_f^{\text{data,P}}$ and $N_f^{\text{data,F}}$, can be therefore expressed as
\begin{equation}\begin{aligned}\label{eq:tnp_formula}
    N_f^{\text{data,P}} & = \text{SF}_f\, N_f^{\text{sim,P}},\\
    N_f^{\text{data,F}} & = N_f^{\text{sim,F}} + N_f^{\text{sim,P}} - \text{SF}_f\, N_f^{\text{sim,P}}.
\end{aligned}\end{equation}

\begin{figure}[htbp]
    \centering
    \includegraphics[width=0.48\textwidth]{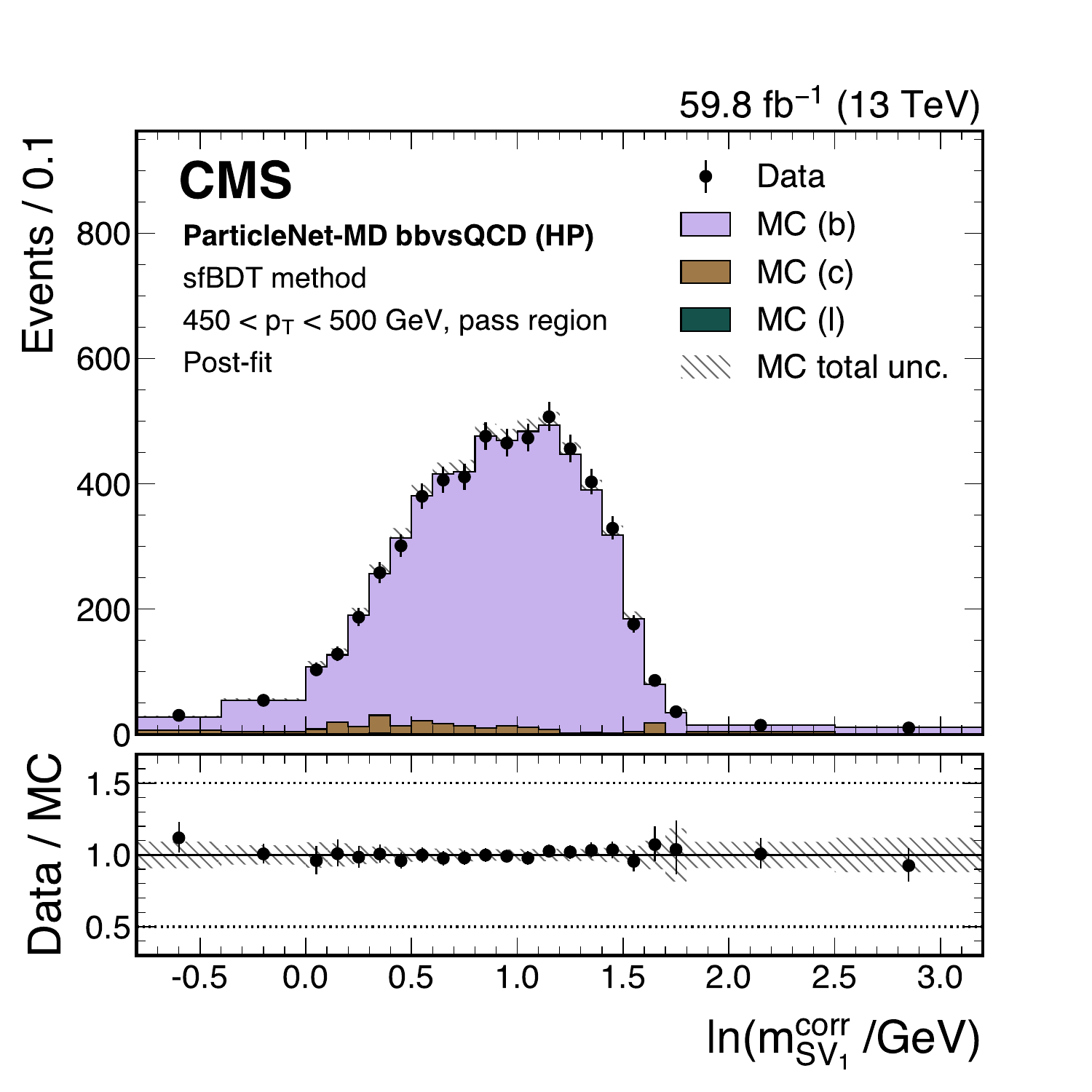}
    \includegraphics[width=0.48\textwidth]{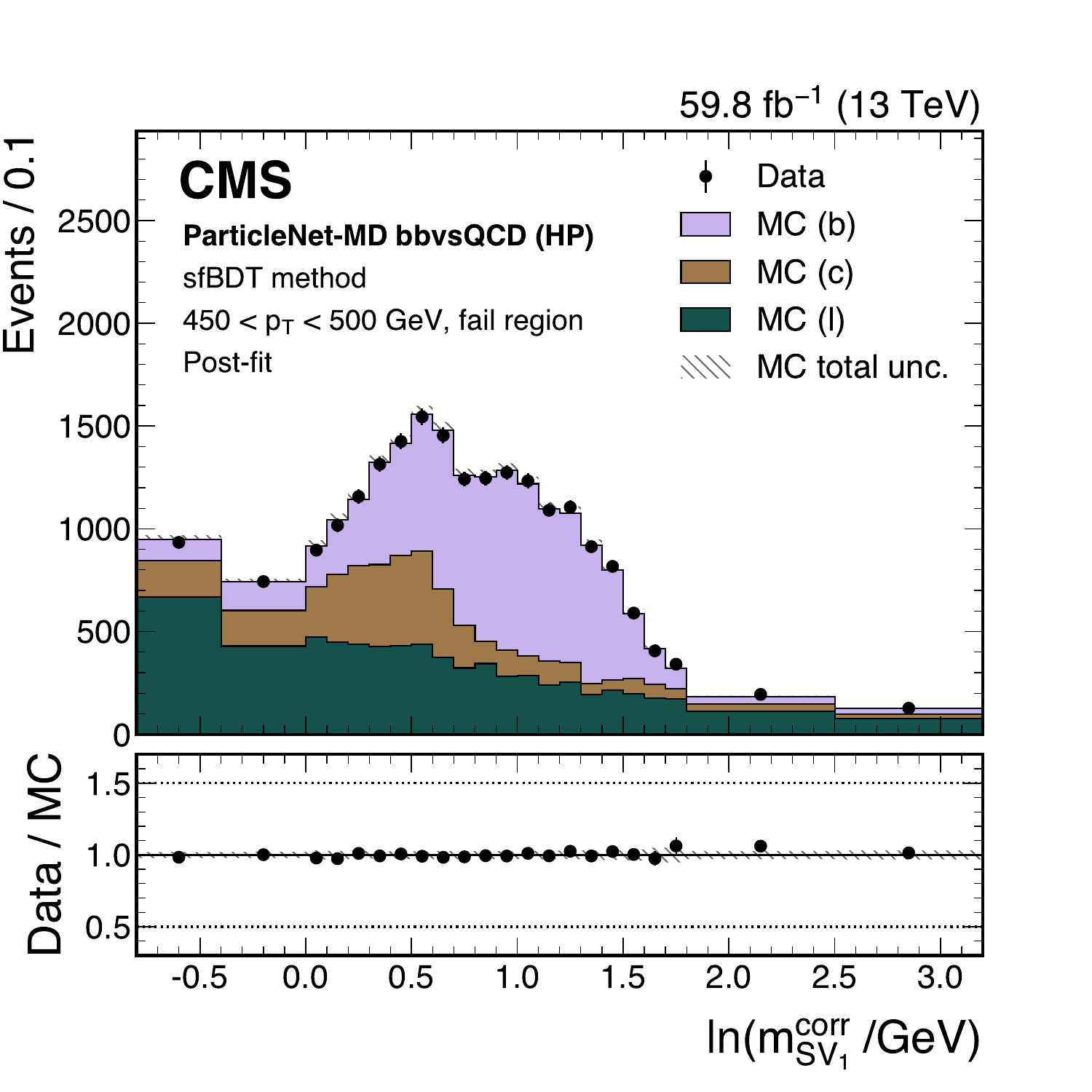}
    \caption{Post-fit distributions from the sfBDT method for events passing (left) and failing (right) the tagger selection, used in the derivation of the scale factor for the ParticleNet-MD \xbb discriminant at the high-purity working point. Error bars represent statistical uncertainties in data, whereas hatched bands denote the total uncertainties in the simulation. The example corresponds to data and simulated events from the 2018 data-taking conditions, in the jet \pt range of $(450,\,500)\GeV$.
    }
    \label{fig:postfit_bb_sfbdt}
\end{figure}

\begin{figure}[htbp]
    \centering
    \includegraphics[width=0.48\textwidth]{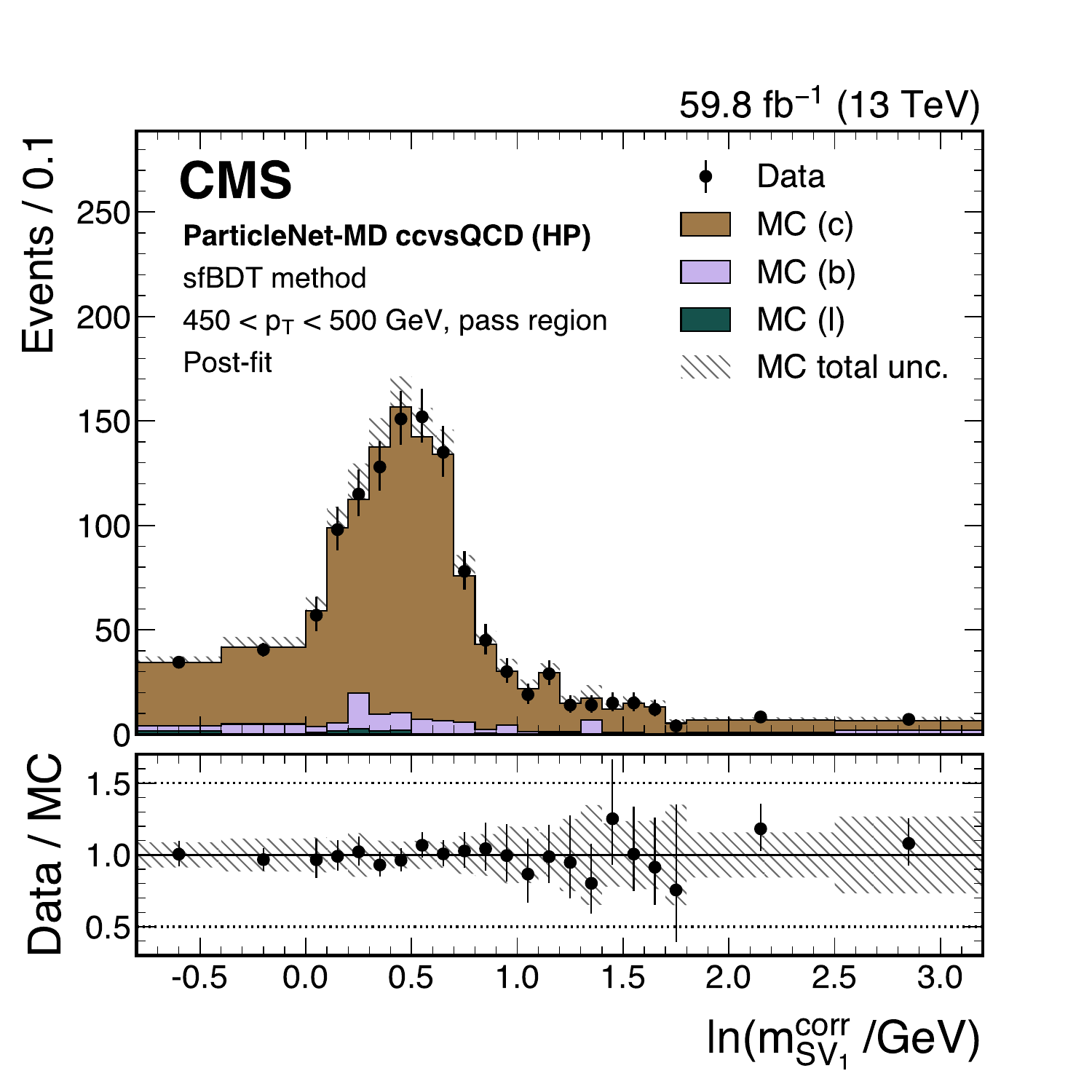}
    \includegraphics[width=0.48\textwidth]{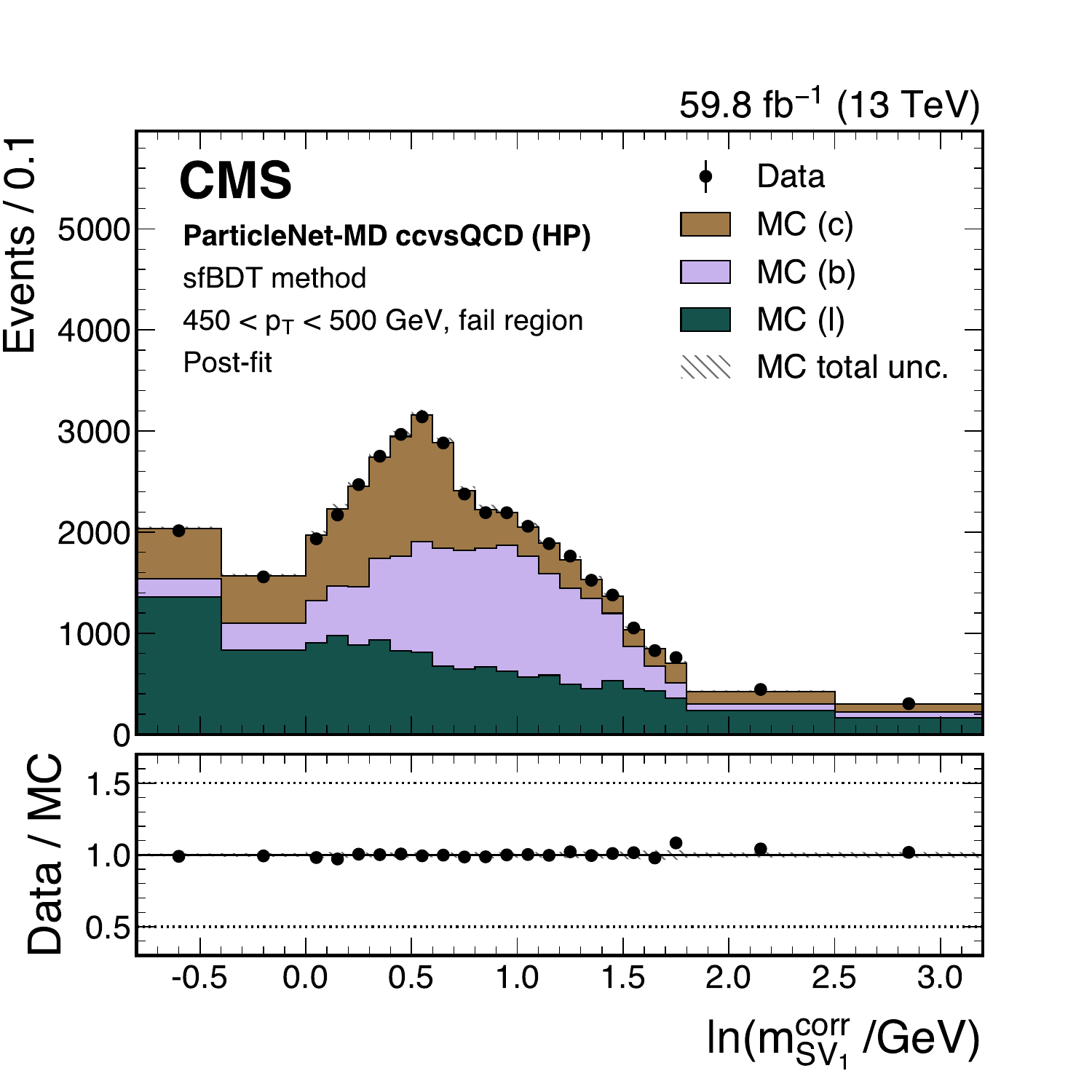}
    \caption{Post-fit distributions from the sfBDT method for events passing (left) and failing (right) the tagger selection, used in the derivation of the scale factor for the ParticleNet-MD \xcc discriminant at the high-purity working point. Error bars represent statistical uncertainties in data, whereas hatched bands denote the total uncertainties in the simulation. The example corresponds to data and simulated events from the 2018 data-taking conditions, in the jet \pt range of $(450,\,500)\GeV$.
    }
    \label{fig:postfit_cc_sfbdt}
\end{figure}

Figures~\ref{fig:postfit_bb_sfbdt} and \ref{fig:postfit_cc_sfbdt} show an example of distributions of data and the corresponding fitted simulated events, in the derivation of SFs of the ParticleNet-MD \xbb and \xcc discriminants, respectively.
The fitted $\text{SF}_\PQb$ or $\text{SF}_\PQc$ are subsequently propagated to derive the final SF for \xbb or \cc jets, respectively. This is achieved through a dedicated post-processing procedure that incorporates additional uncertainties and adjusts the central value, as detailed in Section~\ref{sec:sfbdt_method_uncertainty}.

\subsubsection{Systematic uncertainties and results} \label{sec:sfbdt_method_uncertainty}

A number of systematic uncertainties are considered that affect the shape of the simulated templates used in the fit. They are summarized as follows:

\begin{itemize}

\item \textit{Fractions of \PQb, \PQc, and light-flavour jets}: Three uncertainty sources account for the potential mismodelling of the \PQb-, \PQc-, and light-flavour jet fractions in simulation. Each uncertainty is modelled by varying the yield of the corresponding flavour up and down by 20\%~\cite{CMS:BTVFlvTagger}. Alternative variations have been tested and have a minimal impact on the resulting SF.

\item \textit{Initial- and final-state radiation in parton shower}: The renormalization scale of QCD emissions in the initial-state radiation (ISR) and final-state radiation (FSR) in the parton shower simulation is individually varied up and down by factors of 2 and 0.5.

\item \textit{Jet energy scale}: The uncertainty in the jet energy scale is propagated to the fit template by varying with $\pm1$ standard deviation from its nominal value~\cite{Sirunyan:2019kia,Khachatryan:2016kdb}.

\item \textit{Jet energy resolution}: For the nominal efficiency measurement, the jet energies in the simulation are smeared according to a Gaussian function to accommodate the slightly worse resolution in data. The uncertainty in the jet energy resolution is propagated to the template by varying with $\pm1$ standard deviation of the Gaussian function by its uncertainty~\cite{Sirunyan:2019kia,Khachatryan:2016kdb}.

\item \textit{Integrated luminosity}: The uncertainty in the integrated luminosity is incorporated into the template by uniformly varying the event yields across all samples and regions by 1.2--2.5\% in the 2016--2018 data-taking eras~\cite{CMS:2021xjt,CMS-PAS-LUM-17-004,CMS-PAS-LUM-18-002}.

\item \textit{Pileup reweighting}: The uncertainty in the pileup reweighting procedure is determined by varying the total inelastic cross section used to produce the pileup profile away from the measured central value by 5\%~\cite{Sirunyan:2018nqx}.

\end{itemize}

In addition to these uncertainties contributing to the change of fit templates, two external uncertainty sources are considered.

The first uncertainty assesses the effect of varying the sfBDT selection thresholds.
Each sfBDT selection illustrated in Fig.~\ref{fig:sfbdt} is employed to measure the corresponding SF. Moreover, to evaluate the additional dependency on the sfBDT selection---particularly the effect introduced by varying the signal-to-proxy similarity conditions---separate sfBDT selection thresholds are applied in the ``pass'' and ``fail'' regions, modifying the similarity between the proxy and signal discriminant shapes in an ad-hoc manner. For example, applying a tighter sfBDT selection in the ``pass'' region relative to the ``fail'' region enhances the signal-like characteristics of the proxy jets. This procedure results in 81 distinct selection combinations.
The final SF is obtained by averaging the 81 measured SFs, incorporating both central values and variations. Specifically, each SF is treated as a normally distributed variable, and the combined result is defined by the median and the $\pm 1\,\sigma$ interval of the averaged distribution. As a result, the final SF has a larger uncertainty than that obtained from the individual fits.

The second uncertainty accounts for mismodelling of the sfBDT discriminant score and the fit variable $\ln(m_{\text{SV\textsubscript{1}}}^{\text{corr}}/\GeV)$.
An ``ad-hoc reweighting'' approach is used to evaluate the impact of such mismodelling on the derived SFs.
The SF derivation procedure is repeated in two additional schemes.
In the first scheme, a reweighting of the sfBDT discriminant score is performed such that the total simulated expectation matches data, before any selection on sfBDT is applied.
In the second scheme, a simulation-to-data reweighting is applied on the $\ln(m_{\text{SV\textsubscript{1}}}^{\text{corr}}/\GeV)$ variable after the sfBDT selection and before splitting the template into the ``pass'' and ``fail'' regions.
This results in two additional sets of SFs.
An external uncertainty, determined from the maximum deviation of the central SF values in the new sets with respect to the nominal one, is assigned to the nominal SF.

Table~\ref{tab:sfbdt_syst_breakdown} summarizes the contributions of each source to the final uncertainty in the derived SFs, using ParticleNet-MD \xbb discriminant at the HP WP as a representative example. The presented values are averaged across all derivation points, including all relevant \pt bins and data-taking conditions.
The fit-related uncertainties from individual sources are computed via a breakdown procedure in which nuisance parameters are frozen to their best-fit values sequentially, ordered by descending impact. The contribution from each source is computed by taking the quadrature difference between the total uncertainty with and without the parameter frozen. Uncertainties external to the fit are estimated similarly, by taking the quadrature difference between the total uncertainties obtained with and without the external treatment.
The result indicates that the dominant contributors to the total uncertainty are the two external uncertainty sources and the statistical fluctuations in data and simulation.

\begin{table}[tbp]
\topcaption{Breakdown of the contributions to the total uncertainty in the fitted scale factor (SF) of the ParticleNet-MD \xbb discriminant at the high-purity working point, using the sfBDT method. The numbers are averaged over multiple SF derivation points, including all relevant \pt bins and data-taking eras.}
\label{tab:sfbdt_syst_breakdown}
\centering
{
\begin{tabular}{lcc}
Uncertainty source                               & $\langle \Delta\text{SF} \rangle$  \\
\hline
{Statistical}                                    & 0.063 \\

{Theory}                                         &  \\
\hspace{2em}    Fraction of jet flavours         & 0.039 \\
\hspace{2em}    ISR and FSR in parton shower     & 0.014 \\
{Experimental}                                   &  \\
\hspace{2em}    Effect of varying sfBDT thresholds          & 0.048 \\
\hspace{2em}    Effect of applying ``reweighting schemes''  & 0.091 \\
\hspace{2em}    Jet energy scale and resolution  & 0.008 \\
\hspace{2em}    Integrated luminosity            & 0.001 \\
\hspace{2em}    Pileup reweighting               & 0.009 \\

\end{tabular}
}
\end{table}

The derived SFs for all \xbb and \xcc discriminants are displayed in Figs.~\ref{fig:sf_pnet-b}--\ref{fig:sf_da-c}.
An analysis of the results obtained using the sfBDT method, along with comparisons to other approaches, is provided in Section~\ref{sec:combination}.

\subsection{The \texorpdfstring{$\mu$}{mu}-tagged method} \label{sec:mutagged}

\subsubsection{Method description} \label{sec:mutagged_method_description}

The $\mu$-tagged method calibrates the \xbb (\cc) signal jets using proxy jets from gluon-splitting \bb (\cc) jets that contain a soft muon within their respective jet cones.

Since the hadronized final state initiated from the decay of a bottom (charm) quark has a 20\% (10\%) probability of including an electron or muon \cite{ParticleDataGroup:2024cfk}, the presence of the nonisolated soft leptons provides a good handle to select a pure sample of heavy-flavour jets.
By selecting AK8 jets with \PQb (\PQc) flavours and requiring them to be $\mu$-tagged, the resulting collection is dominated by gluon-splitting \bb (\cc) jets.
This selection ensures a closer resemblance in the phase space between these jets and the signal jets.

For the \xbb (\cc) taggers discussed in Section~\ref{sec:hrtalgo}, the muon information is not explicitly used in the training of the algorithms. Hence, these algorithms are suitable to be calibrated using a subset of \bb (\cc) jets containing soft muons. This essentially imposes a requirement on the kind of tagging algorithm that can be calibrated with this method.
A similar technique is employed in the calibration of AK4 \PQb (\PQc) jet taggers~\cite{CMS:BTVFlvTagger,CMS:BTVCTag}, where soft, non-isolated muons are used as a criterion for selecting \PQb (\PQc) jets for calibration.

In the $\mu$-tagged method, events are selected online by requiring the presence of an AK4 or AK8 jet with $\pt > 300\GeV$ in association with a muon with $\pt > 5\GeV$.
For each event, the leading AK8 jet and the subleading one (if it exists) are selected and required to pass the kinematic preselection of $\pt > 350\GeV$, $\abs{\eta}<2.4$, and $\msd > 40\GeV$.
The simulated $\mu$-enriched QCD multijet events, as described in Section~\ref{sec:datamc}, are used to compare with data, whereas the QCD multijet process simulated with \MGvATNLO is used as an alternative.
Offline, the AK8 jet is required to contain at least one soft muon with $\pt > 5\GeV$ and $\abs{\eta}<2.4$.
To further extract the signal-like \gbb (\cc) jets,
a selection on the $N$-subjettiness ratio, $\tau_{21} < 0.3$, is applied to select two-prong jets.
To correct the $\mu$-enriched QCD modelling to match with data and reduce the discrepancy with the \MGvATNLO-based QCD multijet sample, the QCD multijet sample is reweighted to data after subtracting the \ttbar, single top quark, and \Vjets contributions.
This reweighting is performed in bins of the jet variables $(\pt, \eta, \tau_{21})$, with the leading and subleading jets reweighted separately.

Similar to the sfBDT method in Section~\ref{sec:sfbdt_method_description}, the selected jets are classified into the ``\PQb'', ``\PQc'', and ``\Pl'' flavour categories based on the number of ghost-matched \PQb and \PQc hadrons.
The proxy jets are defined as simulated jets in the ``\PQb'' (``\PQc'') categories for calibration of \xbb (\cc) signal jets, passing the aforementioned selections.

\begin{figure}[htbp]
    \centering
    \includegraphics[width=0.48\textwidth]{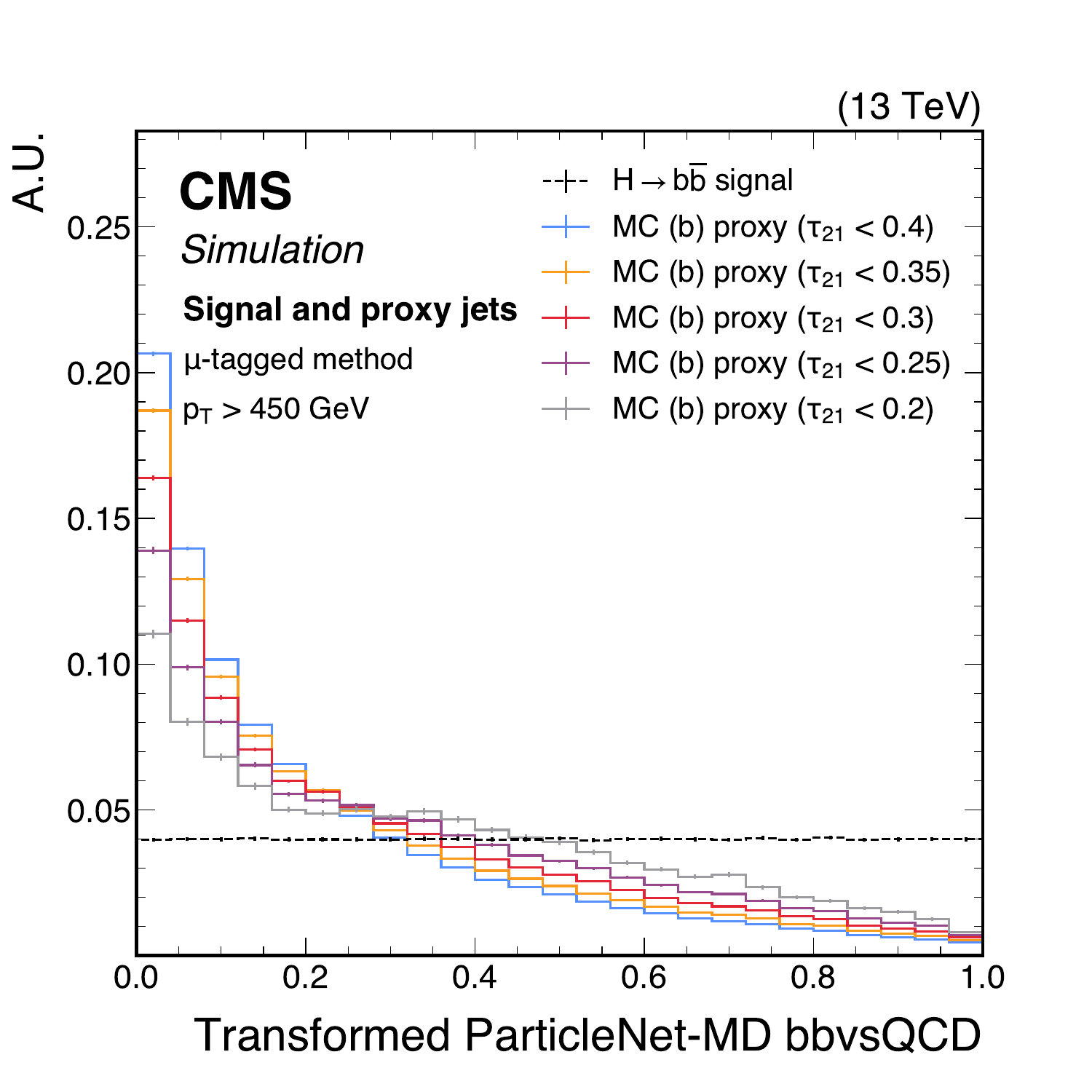}
    \includegraphics[width=0.48\textwidth]{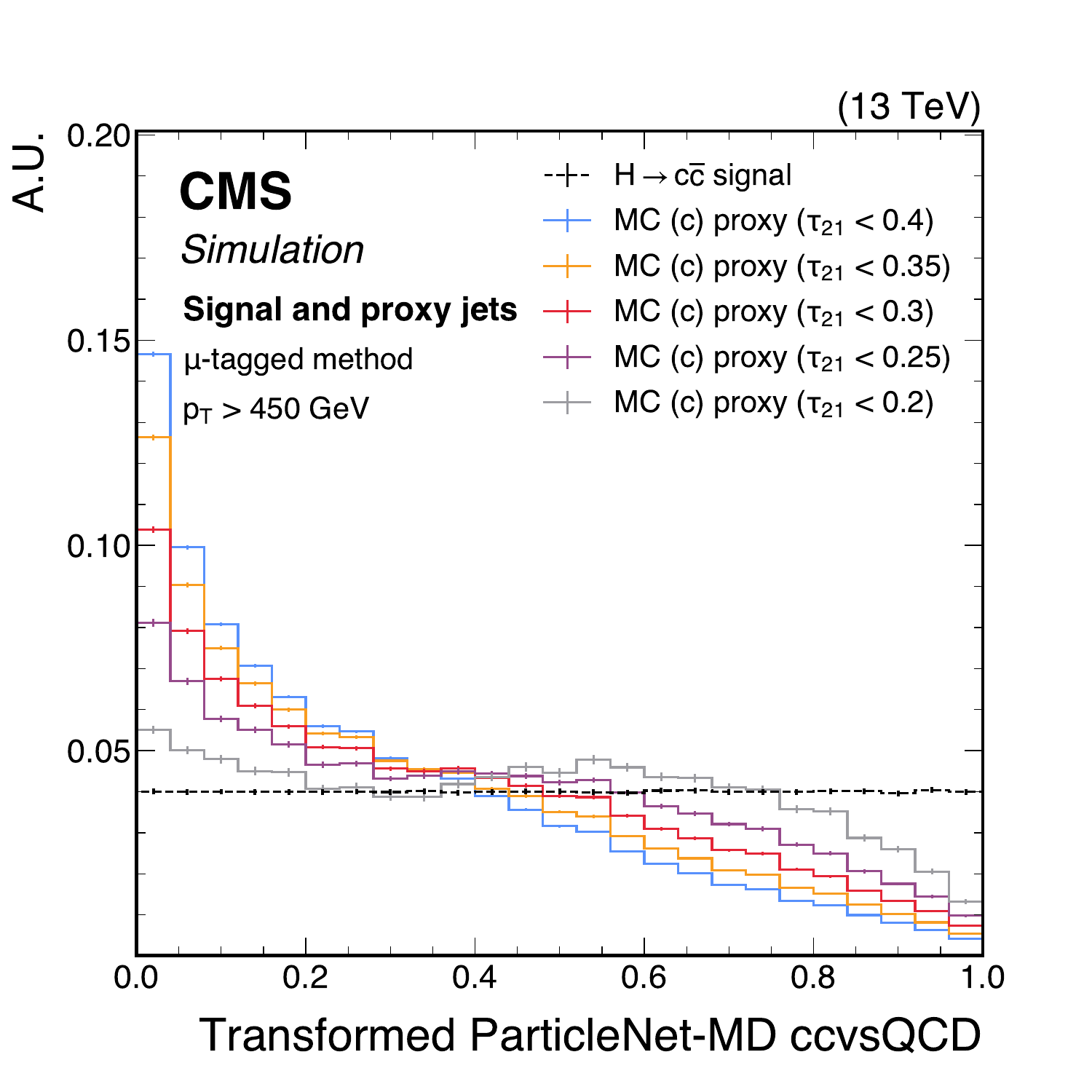}
    \caption{Shapes of the transformed ParticleNet-MD \xbb (left) and \xcc (right) discriminants for SM \hbb (\cc) signal jets and proxy jets selected with different $\tau_{21}$ selection thresholds. The examples correspond to the calibration of the ParticleNet-MD \xbb and \xcc discriminants with the $\mu$-tagged method, using simulated events under 2018 data-taking conditions for jets with $\pt > 450\GeV$. The error bars represent the statistical uncertainties due to the limited number of simulated events.
    }
    \label{fig:sigvsproxy_mutagged}
\end{figure}

The dependence of the resulting SFs on the similarity between proxy and signal jets in the $\mu$-tagged method is evaluated by varying the $\tau_{21}$ selection threshold between 0.2 and 0.4. Figure~\ref{fig:sigvsproxy_mutagged} illustrates the impact of different $\tau_{21}$ thresholds (0.2, 0.25, 0.3, 0.35, and 0.4) on the transformed tagger discriminant for proxy jets. Tighter $\tau_{21}$ selections make the proxy jets more signal-like. The resulting variation in the SF is treated as an additional systematic uncertainty, as detailed in Section~\ref{sec:mutagged_method_uncertainty}.

\begin{figure}[htbp]
    \centering
    \includegraphics[width=0.48\textwidth]{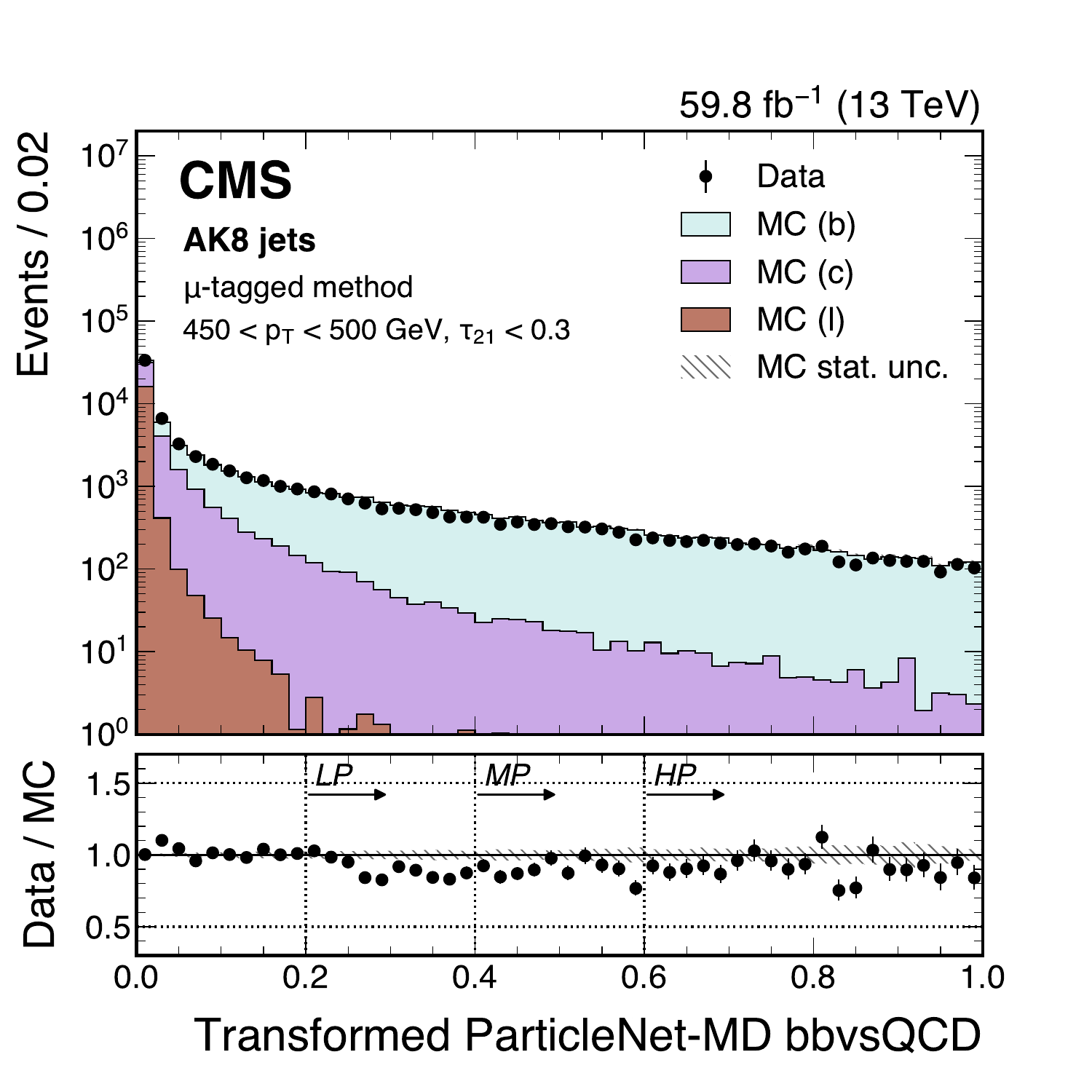}
    \includegraphics[width=0.48\textwidth]{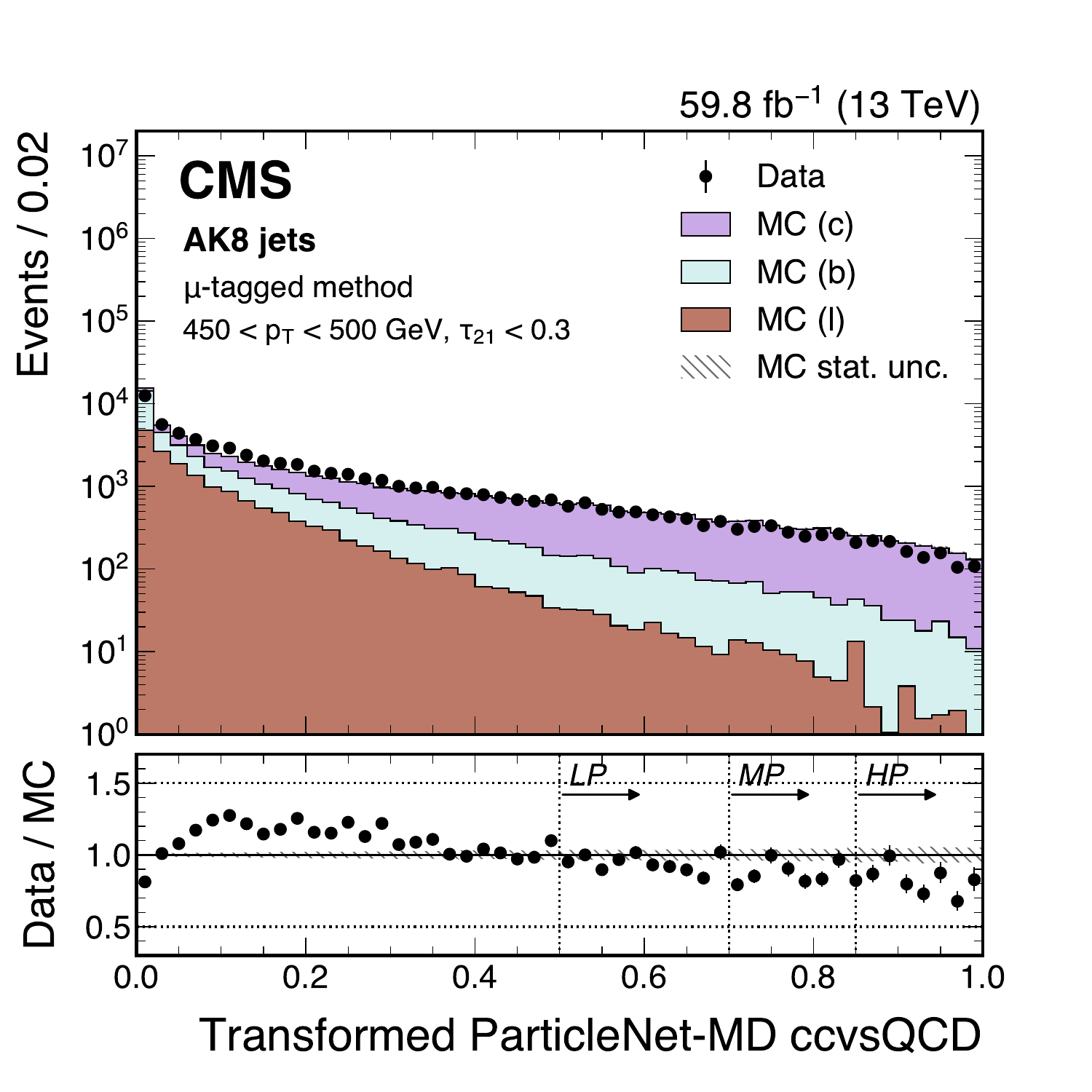}
    \caption{An example of the transformed ParticleNet-MD \xbb (left) and \xcc (right) distribution in data and simulated events, passing the preselection of the $\mu$-tagged method. The high-purity (HP), medium-purity (MP), and low-purity (LP) working points for the left (right) plot correspond to selections of $X>0.6,\,0.4,\,0.2$ $(0.85,\,0.7,\,0.5)$ on the transformed tagger discriminant. The error bars represent the statistical uncertainties in observed data. The lower panels display the ratio of data to simulation, with the hatched bands representing the normalized statistical uncertainty of simulated events for each bin. The distributions are based on data and simulated events with the 2018 data-taking conditions, in the jet \pt range of $(450,\,500)\GeV$.}
    \label{fig:datamc_mutagged_xtagger_pnet}
\end{figure}

\begin{figure}[htbp]
    \centering
    \includegraphics[width=0.48\textwidth]{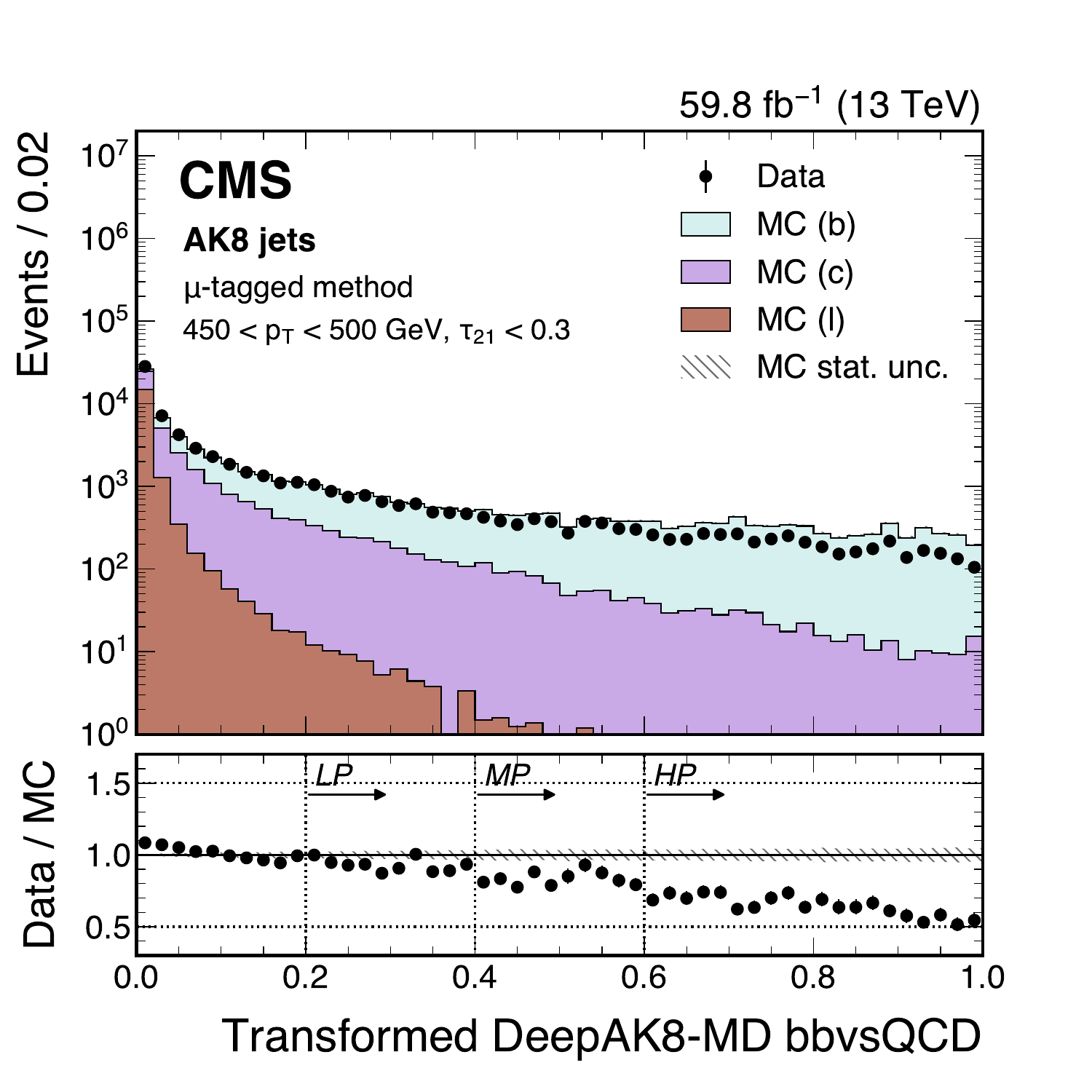}
    \includegraphics[width=0.48\textwidth]{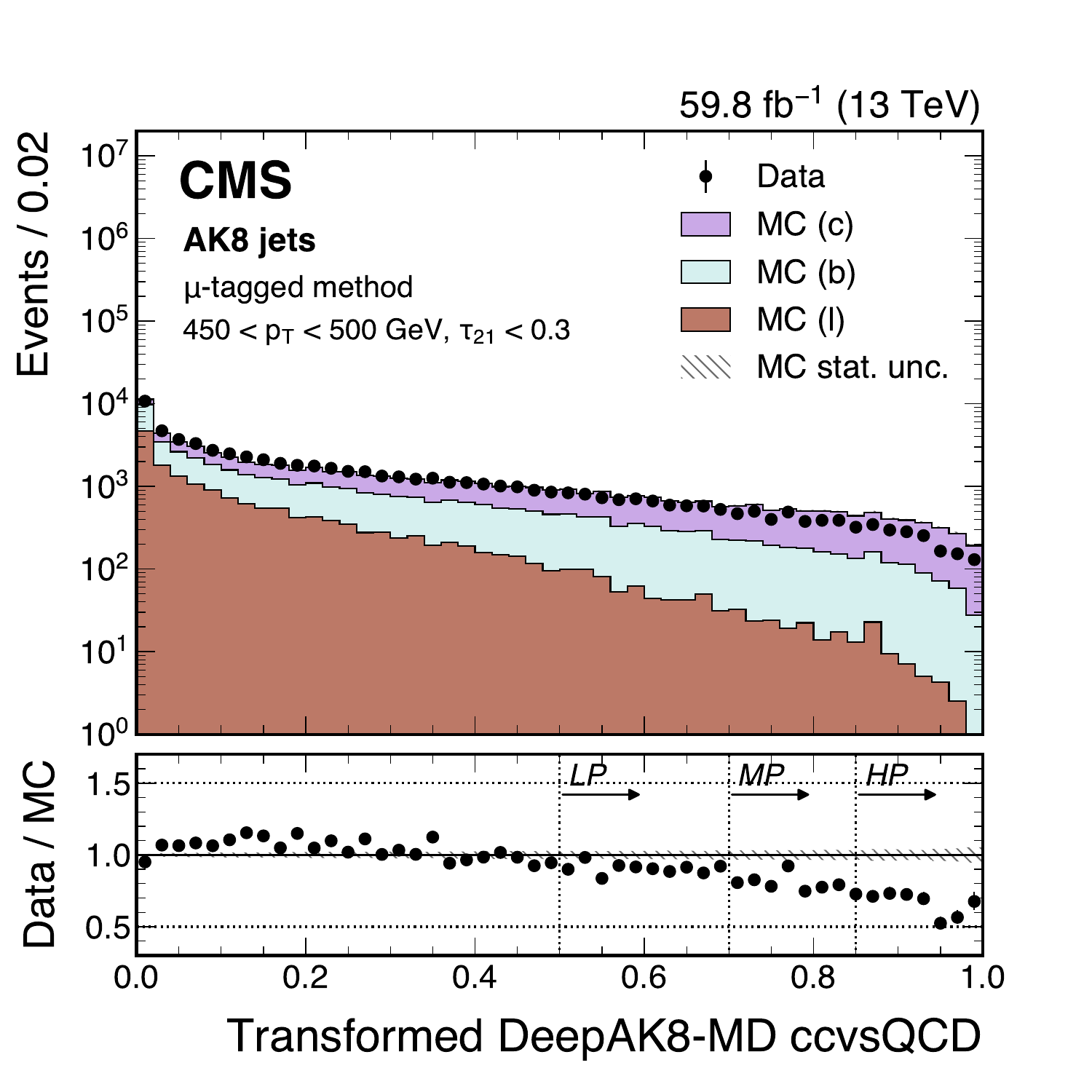}
    \caption{An example of the transformed DeepDoubleX \xbb (left) and \xcc (right) distribution in data and simulated events, passing the preselection of the $\mu$-tagged method. The high-purity (HP), medium-purity (MP), and low-purity (LP) working points for the left (right) plot correspond to selections of $X>0.6,\,0.4,\,0.2$ $(0.85,\,0.7,\,0.5)$ on the transformed tagger discriminant. The error bars represent the statistical uncertainties in observed data. The lower panels display the ratio of data to simulation, with the hatched bands representing the normalized statistical uncertainty of simulated events for each bin. The distributions are based on data and simulated events with the 2018 data-taking conditions, in the jet \pt range of $(450,\,500)\GeV$.}
    \label{fig:datamc_mutagged_xtagger_ddx}
\end{figure}

\begin{figure}[htbp]
    \centering
    \includegraphics[width=0.48\textwidth]{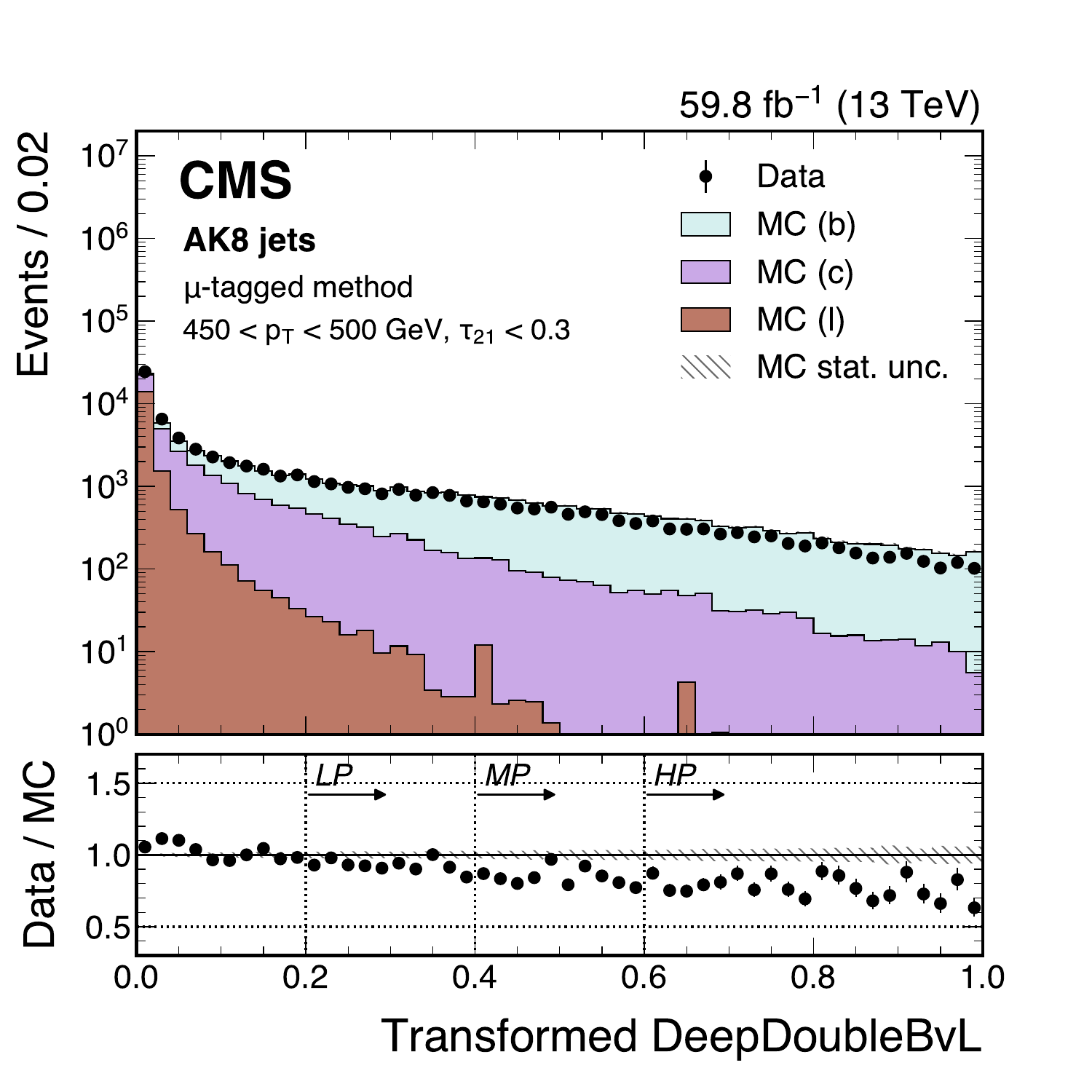}
    \includegraphics[width=0.48\textwidth]{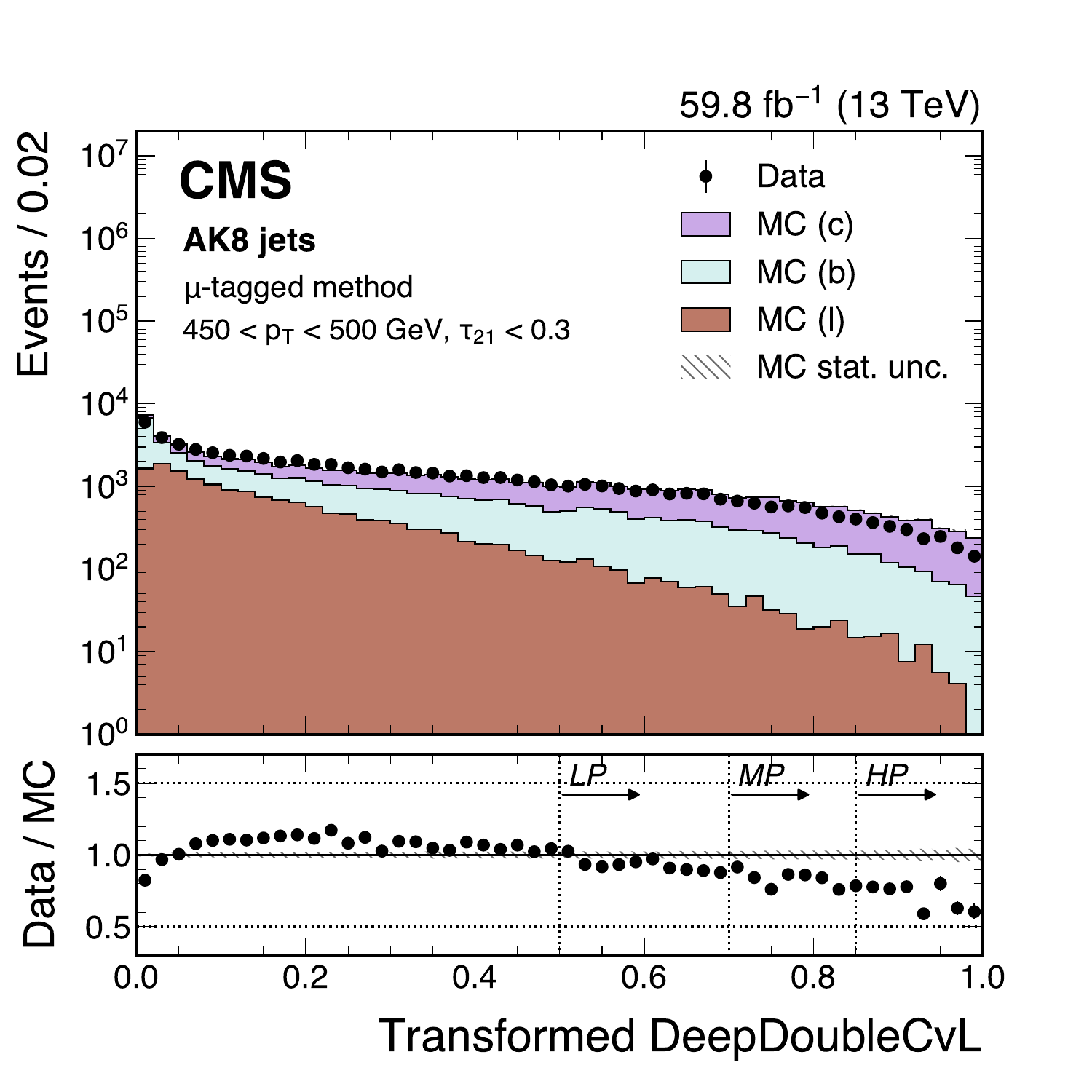}
    \caption{An example of the transformed DeepAK8-MD \xbb (left) and \xcc (right) distribution in data and simulated events, passing the preselection of the $\mu$-tagged method. The high-purity (HP), medium-purity (MP), and low-purity (LP) working points for the left (right) plot correspond to selections of $X>0.6,\,0.4,\,0.2$ $(0.85,\,0.7,\,0.5)$ on the transformed tagger discriminant. The error bars represent the statistical uncertainties in observed data. The lower panels display the ratio of data to simulation, with the hatched bands representing the normalized statistical uncertainty of simulated events for each bin. The distributions are based on data and simulated events with the 2018 data-taking conditions, in the jet \pt range of $(450,\,500)\GeV$.}
    \label{fig:datamc_mutagged_xtagger_da}
\end{figure}

\begin{figure}[htbp]
    \centering
    \includegraphics[width=0.48\textwidth]{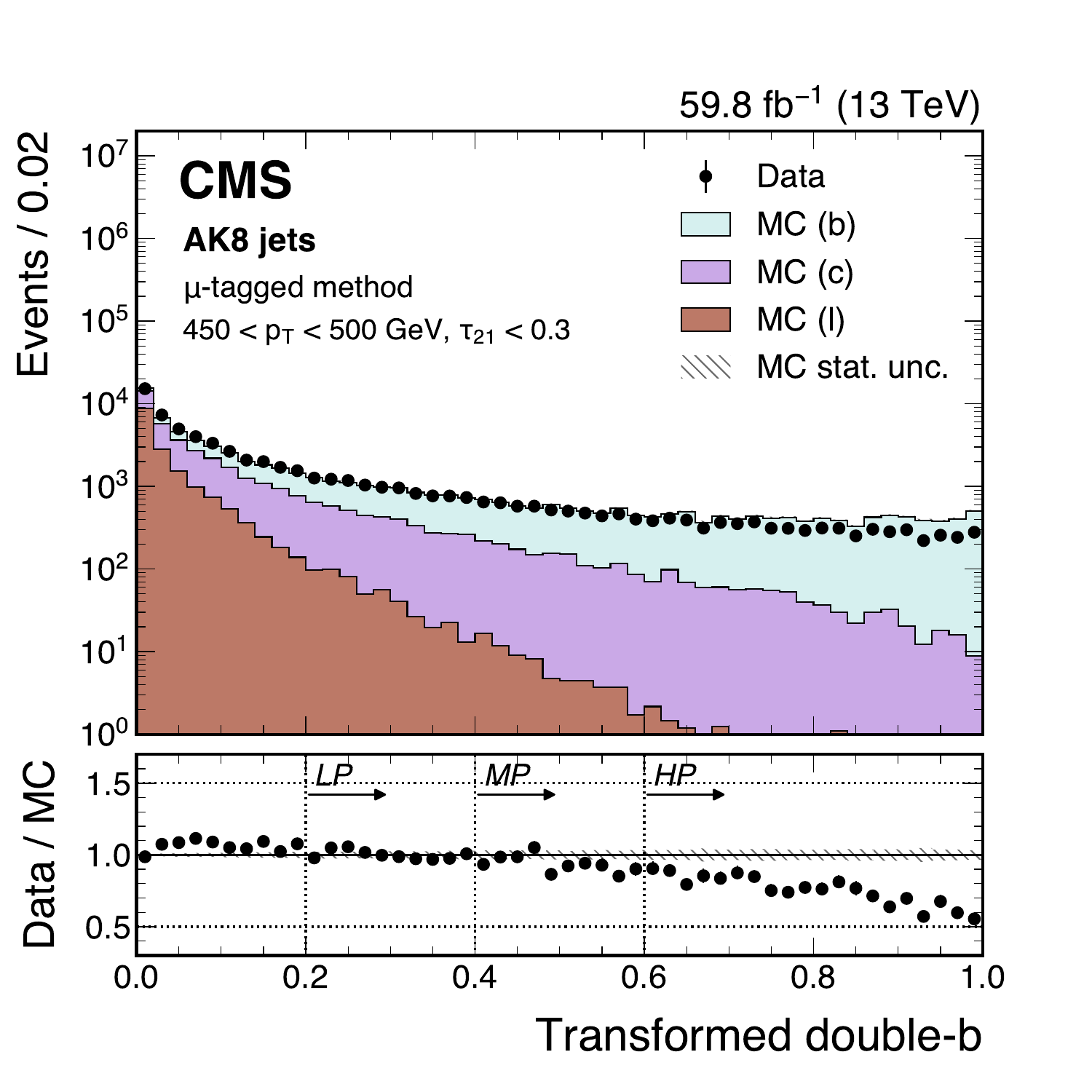}
    \caption{An example of the transformed double-b distribution in data and simulated events, passing the preselection of the $\mu$-tagged method. The high-purity (HP), medium-purity (MP), and low-purity (LP) working points correspond to selections of $X>0.6,\,0.4,\,0.2$ on the transformed tagger discriminant. The error bars represent the statistical uncertainties in observed data. The lower panel displays the ratio of data to simulation, with the hatched bands representing the normalized statistical uncertainty of simulated events for each bin. The distribution is based on data and simulated events with the 2018 data-taking conditions, in the jet \pt range of $(450,\,500)\GeV$.}
    \label{fig:datamc_mutagged_xtagger_doubleb}
\end{figure}

Figures~\ref{fig:datamc_mutagged_xtagger_pnet}--\ref{fig:datamc_mutagged_xtagger_doubleb} show the distributions of the transformed tagger discriminant, passing the preselection above.
The tagger discriminant is transformed to $X\in (0,\,1)$, such that a selection of $X>X_0$ corresponds to the signal jet selection efficiency of $1-X_0$.
The distributions of the tagger discriminants show varying levels of agreement between data and simulation, leading to conclusions consistent with those shown in Figs.~\ref{fig:datamc_xtagger_pnet}--\ref{fig:datamc_xtagger_doubleb} for the sfBDT method.
Furthermore, when comparing these distributions between the $\mu$-tagged method and the sfBDT method for the same tagger and WP, the data-to-simulation ratio in the HP region is slightly smaller for the sfBDT method. Further discussion is provided in Section~\ref{sec:combination}.

To extract the SF for the proxy jet, the $\mu$-tagged method employs a fit procedure analogous to that used in the sfBDT method. Three unconstrained factors, $\text{SF}_\PQb$, $\text{SF}_\PQc$, and $\text{SF}_\Pl$, are assigned to the ``\PQb'', ``\PQc'', and ``\Pl'' categories in simulation, for jets passing a specific WP of tagger discriminant.
The fit is performed on the binned histogram of the variable $\ln(m(\sum\vec{p}_{\text{SV}}^{\,\text{corr}})/\GeV)$, where $m(\sum\vec{p}_{\text{SV}}^{\,\text{corr}})$ denotes the invariant mass of the vector sum of all corrected SV four-momenta, $\vec{p}_{\text{SV}}^{\,\text{corr}}$, associated with the jets.
The corrected SV four-momentum $\vec{p}_{\text{SV}}^{\,\text{corr}}$ is computed from the momenta of tracks associated with the SV and using the corrected SV mass as defined in Eq.~(\ref{eq:svmass}). Alternative fit variables based on SV information have been tested and lead to compatible results.
A simultaneous fit is performed across the ``pass'' and ``fail'' regions of the tagger WP, in three exclusive \pt bins, following the same procedure as used in the sfBDT method.

In addition, a dedicated treatment is put into place in order to account for the degeneracy of the ``\PQb'' and ``\PQc'' flavour 
templates in the ``pass'' region. The background SFs are fixed to unity if the signal and background templates 
are degenerate or if the background contribution is negligible. 
The templates are defined as degenerate if the $\chi^2$ difference between the ``\PQb'' and ``\PQc'' shapes is below one.
This procedure is necessary because the fit of two unconstrained parameters with degenerate templates 
cannot disentangle the effects of each parameter independently, since they are anti-correlated in the fit.
In these cases, the background SF is fixed in the fit and only the signal SF is measured. 
Only 16\% of the fitted points are affected by the degeneracy of signal and background templates and are subject to this special treatment.
In particular, the ``pass'' region of taggers with higher purity, such as the ParticleNet-MD \xbb and \xcc taggers, 
is the most affected by the degeneracy, especially at high \pt and for the HP WP.

\begin{figure}[htbp]
    \centering
    \includegraphics[width=0.48\textwidth]{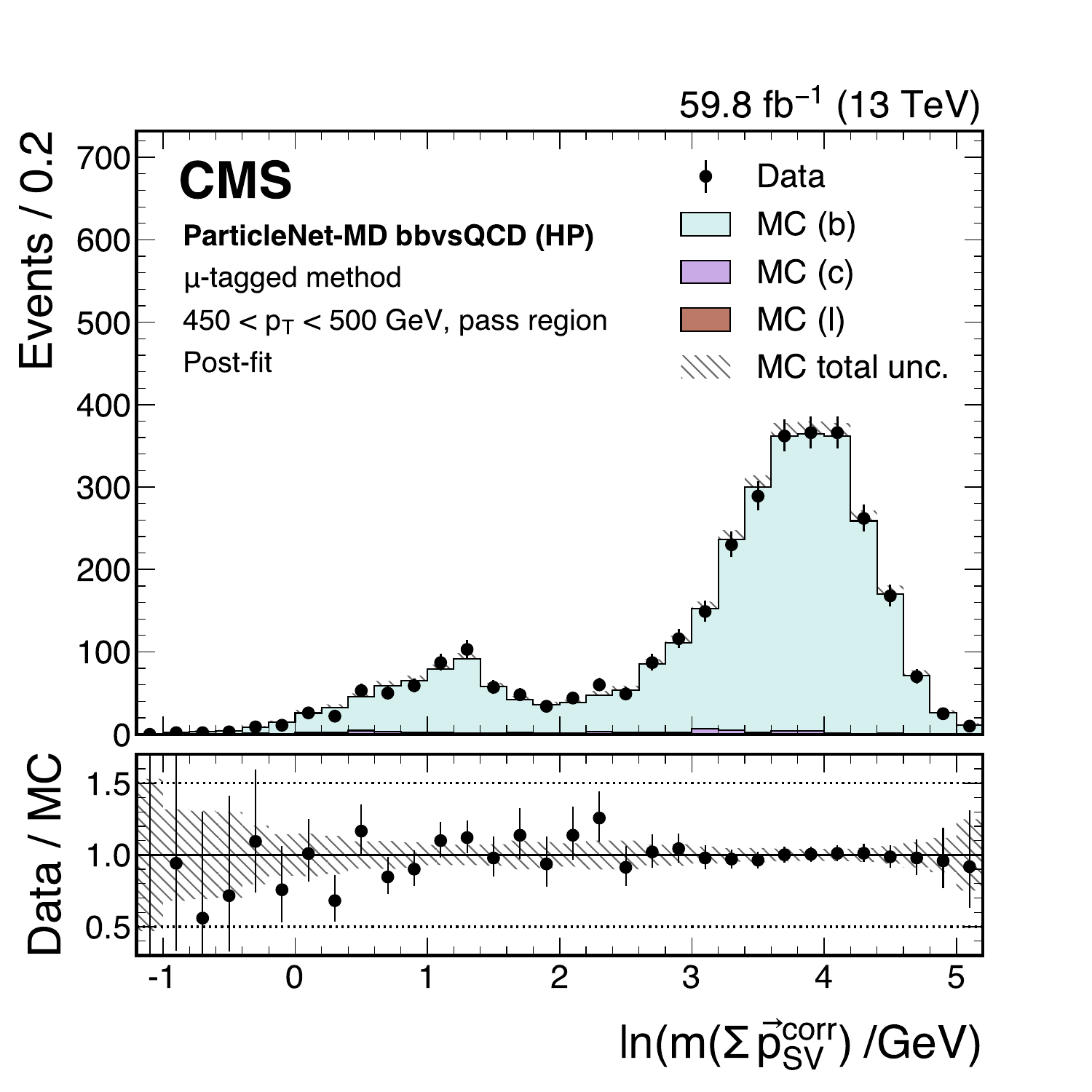}
    \includegraphics[width=0.48\textwidth]{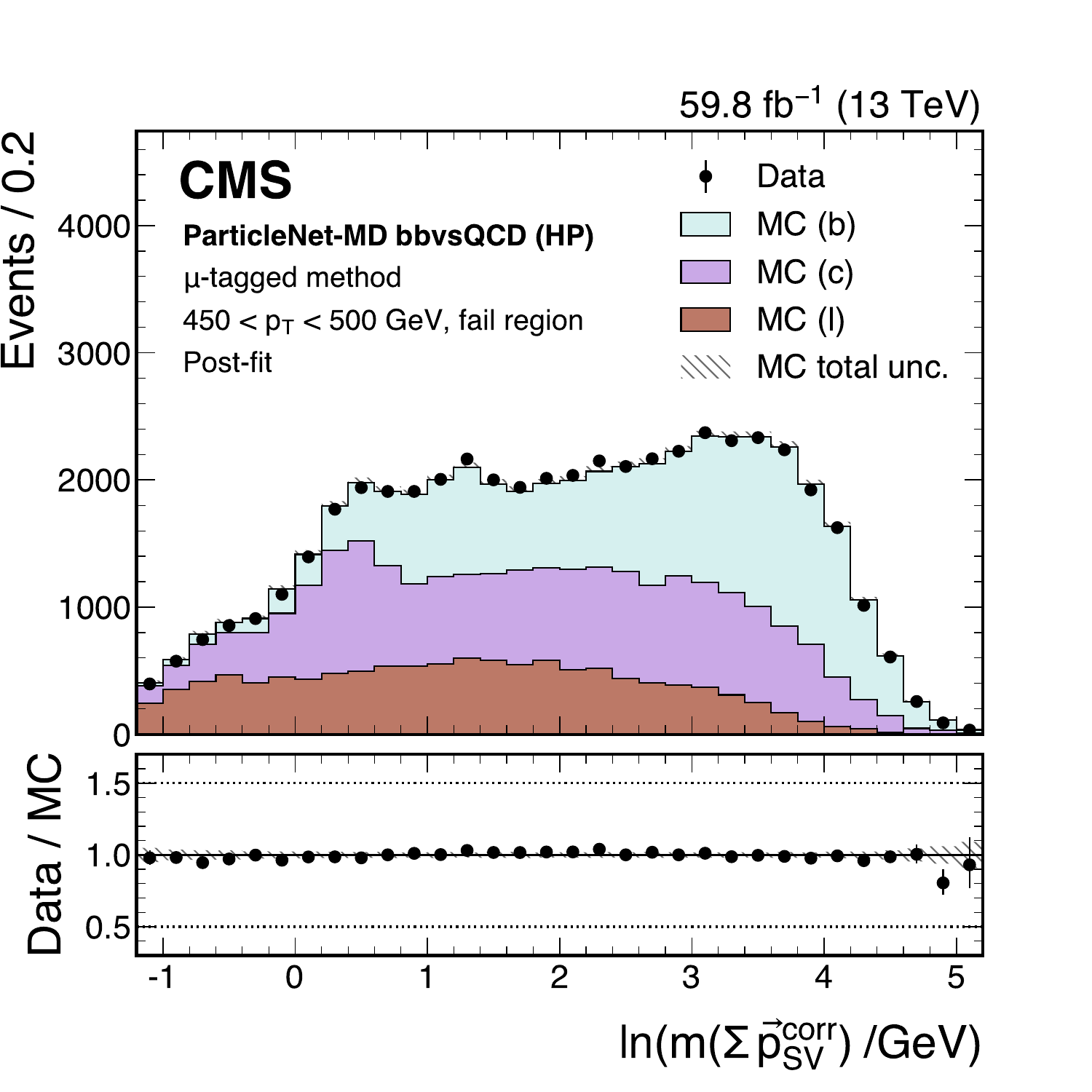}
    \caption{Post-fit distributions from the $\mu$-tagged method for events passing (left) and failing (right) the tagger selection, used in the derivation of the scale factor for the ParticleNet-MD \xbb discriminant at the high-purity working point. Error bars represent statistical uncertainties in data, where hatched bands denote the total uncertainties in the simulation. The example corresponds to data and simulated events from the 2018 data-taking conditions, in the jet \pt range of $(450,\,500)\GeV$.}
    \label{fig:postfit_bb_mutagged}
\end{figure}

\begin{figure}[htbp]
    \centering
    \includegraphics[width=0.48\textwidth]{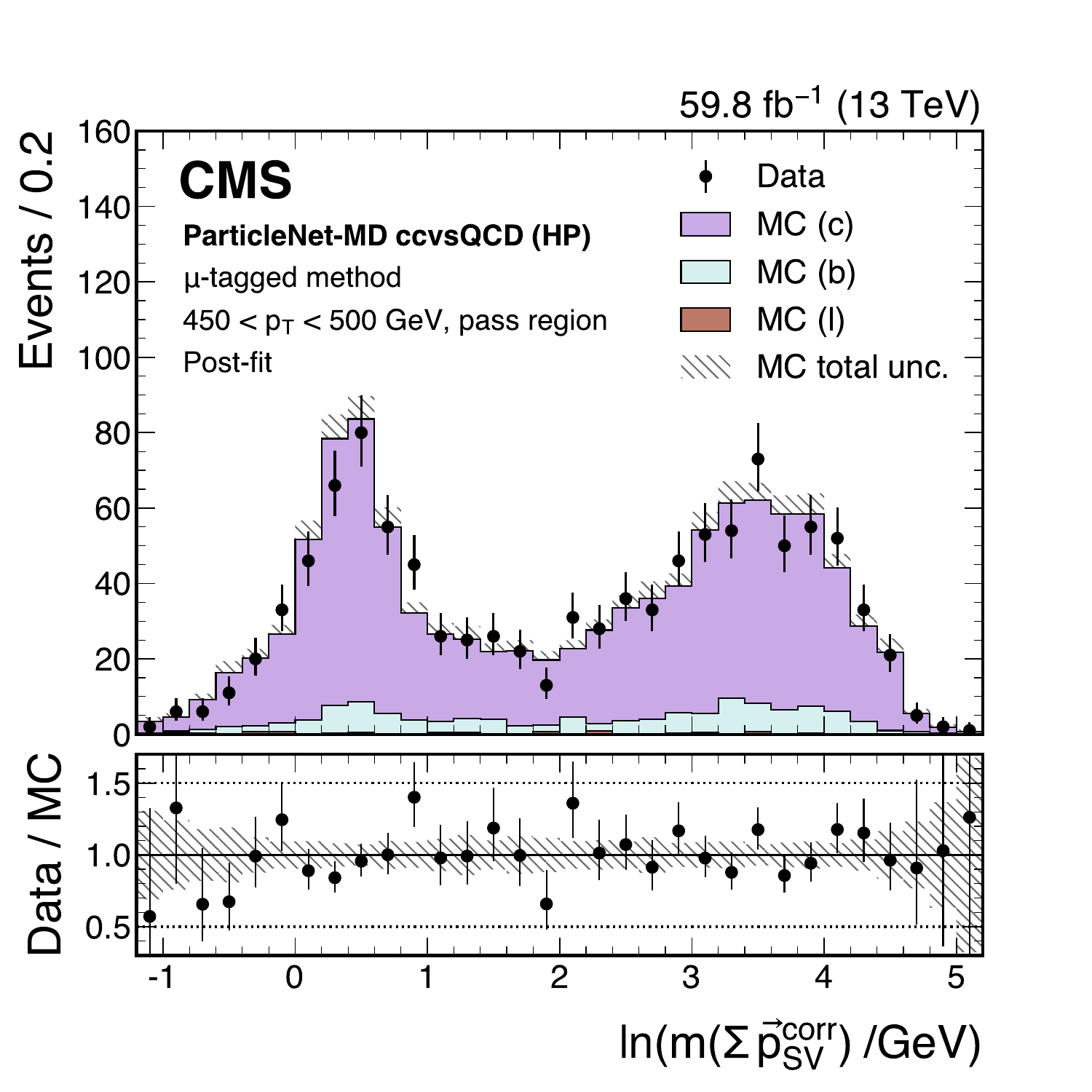}
    \includegraphics[width=0.48\textwidth]{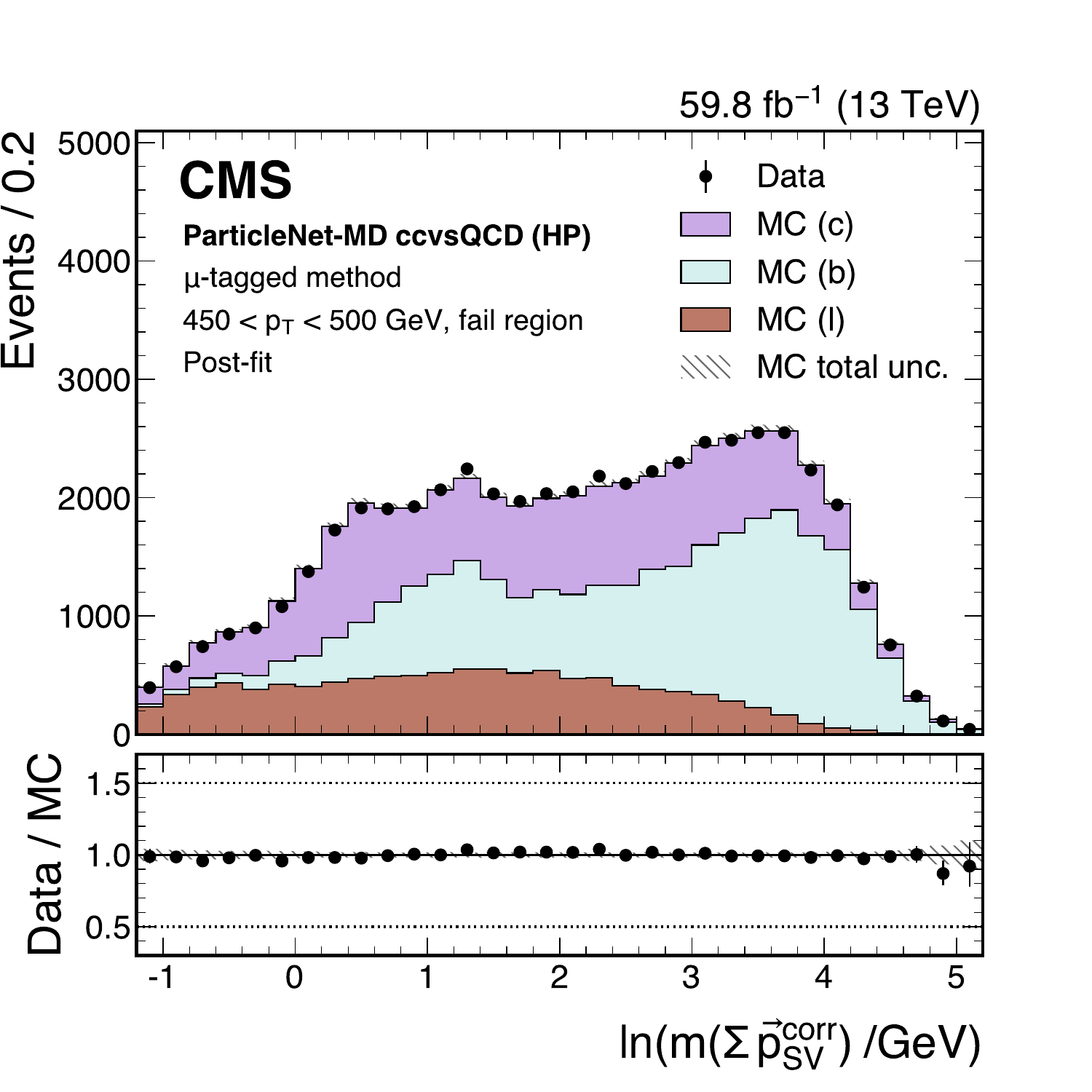}
    \caption{Post-fit distributions from the $\mu$-tagged method for events passing (left) and failing (right) the tagger selection, used in the derivation of the scale factor for the ParticleNet-MD \xcc discriminant at the high-purity working point. Error bars represent statistical uncertainties in data, whereas hatched bands denote the total uncertainties in the simulation. The example corresponds to data and simulated events from the 2018 data-taking conditions, in the jet \pt range of $(450,\,500)\GeV$.}
    \label{fig:postfit_cc_mutagged}
\end{figure}

Figures~\ref{fig:postfit_bb_mutagged} and \ref{fig:postfit_cc_mutagged} show an example of distributions of data and the corresponding fitted simulated events for deriving the SFs of the ParticleNet-MD \xbb and \xcc discriminants.
In the calibration of \xbb (\cc) taggers, the fitted $\text{SF}_{\PQb}$ ($\text{SF}_{\PQc}$) is employed as the central value for the derived SF, and the uncertainty is expanded to incorporate additional uncertainties, as detailed in the following description.

\subsubsection{Systematic uncertainties and results} \label{sec:mutagged_method_uncertainty}

The following systematic uncertainties are included in the fit.

\begin{itemize}
    
\item \textit{Fractions of \PQb, \PQc, and light-flavour jets}: Three uncertainty sources accounting for the fractions of \PQb, \PQc, and light-flavour jets are treated in the same way as described in Section~\ref{sec:sfbdt_method_uncertainty}.

\item \textit{QCD jet modelling}: A systematic uncertainty accounting for the simulation difference between the $\mu$-enriched QCD sample and the \MGvATNLO-based QCD sample is estimated, by adjusting the fit template of each \PQb, \PQc, and light-flavour originating from the former QCD sample to the latter sample.

\item \textit{ISR and FSR in parton shower}: Two uncertainty sources accounting for the ISR and FSR in the parton shower by \PYTHIA are estimated in the same way as described in Section~\ref{sec:sfbdt_method_uncertainty}.

\item \textit{Jet energy scale and resolution}: Two sources of uncertainties, accounting for the jet energy scale and resolution, are propagated to the SF measurement as described in Section~\ref{sec:sfbdt_method_uncertainty}.

\item \textit{Integrated luminosity}: The uncertainty in the integrated luminosity is treated in the same way as described in Section~\ref{sec:sfbdt_method_uncertainty}.

\item \textit{Pileup reweighting}: The uncertainty in the pileup reweighting is treated in the same way as described in Section~\ref{sec:sfbdt_method_uncertainty}.

\end{itemize}

In addition, similar to the sfBDT method, two external uncertainty sources are included in the SF measurements. The first one aims to measure the effect of varying the $\tau_{21} < 0.3$ selection. The threshold is adjusted from 0.4 to 0.2 as a handle to tune the signal and proxy jet similarity. 
The variation observed in the fitted SFs is treated as an additional source of uncertainty.
The second source accounts for the mismodelling of the fit variable $\ln(m(\sum\vec{p}_{\text{SV}}^{\,\text{corr}})/\GeV)$. Prior to measuring the SF, a simulation-to-data reweighting is implemented on the variable within the ``inclusive'' region, which is the combined  ``pass'' and ``fail'' regions.
In degenerate fit points, as defined in Section~\ref{sec:mutagged_method_description}, the impact of the chosen fixed background SF on the signal SF is minor compared to the existing systematic uncertainties. The signal SF measured by fixing the background SF is compatible with the measurement with all SFs freely floating.

The contribution of each uncertainty source is summarized in Table~\ref{tab:mutag_syst_breakdown}, taking the ParticleNet-MD \xbb discriminant at the HP WP as an example. The most significant contribution arises from the external uncertainty associated with the dependence of the SF on the $\tau_{21}$ selection.

\begin{table}[tbp]
\topcaption{Breakdown of the contributions to the total uncertainty in the fitted scale factor (SF) of the ParticleNet-MD \xbb discriminant at the high-purity working point, using the $\mu$-tagged method. The numbers are averaged over multiple SF derivation points, including all relevant \pt bins and data-taking eras.}
\label{tab:mutag_syst_breakdown}
\centering
{
\begin{tabular}{lcc}
Uncertainty source                               & $\langle \Delta\text{SF} \rangle$  \\
\hline
{Statistical}                                    & 0.093 \\

{Theory}                                         &  \\
\hspace{2em}    Fraction of jet flavours         & 0.089 \\
\hspace{2em}    ISR and FSR in parton shower     & 0.027 \\
\hspace{2em}    QCD jet modelling                & 0.014 \\

{Experimental}                                   & \\
\hspace{2em}    Effect of varying $\tau_{21}$ thresholds       & 0.275 \\
\hspace{2em}    Effect of ``simulation-to-data reweighting''   & 0.064 \\
\hspace{2em}    Jet energy scale and resolution  & 0.032 \\
\hspace{2em}    Integrated luminosity            & 0.009 \\
\hspace{2em}    Pileup reweighting               & 0.017 \\

\end{tabular}
}
\end{table}

The derived SFs for all \xbb and \xcc discriminants are displayed in Figs.~\ref{fig:sf_pnet-b}--\ref{fig:sf_da-c}. A detailed analysis of the results is provided in Section~\ref{sec:combination}.

\subsection{The boosted \texorpdfstring{\PZ}{Z} boson method} \label{sec:boostedzbb}

\subsubsection{Method description}

The boosted \PZ boson method calibrates the \xbb signal jets using the proxy jets originating from the decay of a Lorentz-boosted \PZ boson into a \bb pair.
Since the \PZ boson is a massive particle, the boosted \zbb jets are closer in the jet characteristics to the target \xbb jets, compared with \gbb jets. Therefore, no special selection is applied to \zbb proxy jets, contrary to the method based on gluon-splitting proxy jets described in Sections~\ref{sec:sfbdt_method} and \ref{sec:mutagged}.
However, the measurement of \PZ jets comes with a smaller number of events compared with the gluon-splitting jets, and there is a sizeable QCD multijet background.
Hence, the principle of the method is to extract the \PZ boson peak on top of the large nonresonant hadronic background.

In the boosted \PZ boson method, events are selected using a series of online triggers, which impose a combination of requirements on the jet \pt, the jet mass after applying the trimming algorithm~\cite{Krohn:2009th}, or \HT, as detailed in Ref.~\cite{CMS:XY4b}. The trigger efficiency is measured in data using a baseline trigger, which requires a single AK4 jet with $\pt > 260\GeV$, and by applying the offline selection described below. This baseline trigger is a prescaled trigger and has a low threshold, ensuring that it passes all events that also satisfy the offline selection.

The following offline selection criteria are applied.
First, the leading AK8 jet in $\pt$ must satisfy $\pt > 450\GeV$, $\abs{\eta}<2.4$, and $\mpnet > 40\GeV$, where $\mpnet$ is the DNN-based regressed mass, as introduced in Section~\ref{sec:object}.
Then, the subleading AK8 jet, regarded as the recoil jet, is required to pass the selection of $\pt > 200\GeV$ and $\abs{\eta}<2.4$. This condition reduces the background contribution and helps the trigger efficiency without significantly reducing the signal efficiency.
Two vetoing requirements are also imposed to suppress the \ttbar background.
Events with at least one electron or muon with $\pt > 20\GeV$, $\abs{\eta}<2.4$, and satisfying the loosest identification and isolation WP~\cite{Khachatryan:2015hwa,Sirunyan:2018fpa} are vetoed.
Events are also required to have no presence of a \PQb-tagged AK4 jet satisfying $\pt > 30\GeV$ and $\Delta R > 0.8$ with respect to the leading AK8 jet.
After the selections, the leading jet of an event is used to measure the data efficiency of boosted \zbb jets on a given WP of an \xbb tagging discriminant.

For a given WP of a tagger, a fit is performed simultaneously in the regions passing and failing the WP. The parameters of interest are the three unconstrained factors, $\text{SF}_{\PZ, i}$ $(i=1,\,2,\,3)$, assigned to the \Zjets process for three exclusive target $\pt$ bins. Also included is a single factor, $\text{SF}_{\PW}$, assigned to the \Wjets process. SFs represent the ratio of tagging efficiencies between data and simulation.
For the \Zjets process, we only consider the \PZ boson decays to two \PQb or \PQc quarks since the decays to lighter quarks do not contribute significantly to the ``pass'' region. The \zbb and \zcc decay modes are jointly fit as a single process, since their templates are not distinguishable at the {\PZ} boson peak. The \zbb component dominates the boson peak in the ``pass'' region, contributing more than 90\%, so the $\text{SF}_{\PZ, i}$ extracted from the fit can be interpreted as the SF for \zbb jets in each of the three target \pt bins.
Likewise, only the decays with a \PQc quark contribute significantly to \Wjets and are included in the fit.

The fit is performed on a 2D binned histogram on $(\mpnet,\,\pt)$ of the leading jet.
The fit templates are produced both from simulation and from data.
The \Zjets and \Wjets processes are estimated from simulated events.
The QCD multijet background, which is the dominant background source, is modelled with data to achieve better modelling accuracy.
Specifically, for each $(\mpnet,\,\pt)$ bin in the ``fail'' region, which is predominantly composed of QCD multijet events, a free parameter is assigned to represent the QCD multijet background contribution for that bin. During the fitting procedure, these parameters typically converge to values very close to the total data yield minus the small contributions from other processes.

A transfer ratio $R_{\text{P/F}}$, defined as the ratio of the QCD multijet event yields in the ``pass'' and ``fail'' region, is then modelled by a 2D polynomial function in $(\mpnet,\,\pt)$ of order $o$,
\begin{equation}
    R_\text{P/F} = \sum_{p,\,q=0}^{p+q\leq o \; \land \; q<3}k_{p,q}\, (\mpnet)^p (\pt)^q,
\end{equation}
where $k_{p,q}$ are the parameters of the polynomial, determined during the fit. The polynomial order is determined with a Fisher's F-test~\cite{ref:ftest} combined with the chi-square goodness-of-fit test using the saturated model. The selected polynomial orders range from 2 to 4, depending on the tagger, WP, and era.
The feasibility of determining $R_{\text{P/F}}$ in polynomial form relies on the tagging algorithm being decorrelated from the jet mass.
This mass decorrelation prevents the algorithm from introducing peaks at specific masses in the QCD multijet background shape. Consequently, the ratio of distribution shapes in the ``pass'' and ``fail'' regions can be modelled using simple functions.

\begin{figure}[htbp]
    \centering
    \includegraphics[width=0.48\textwidth]{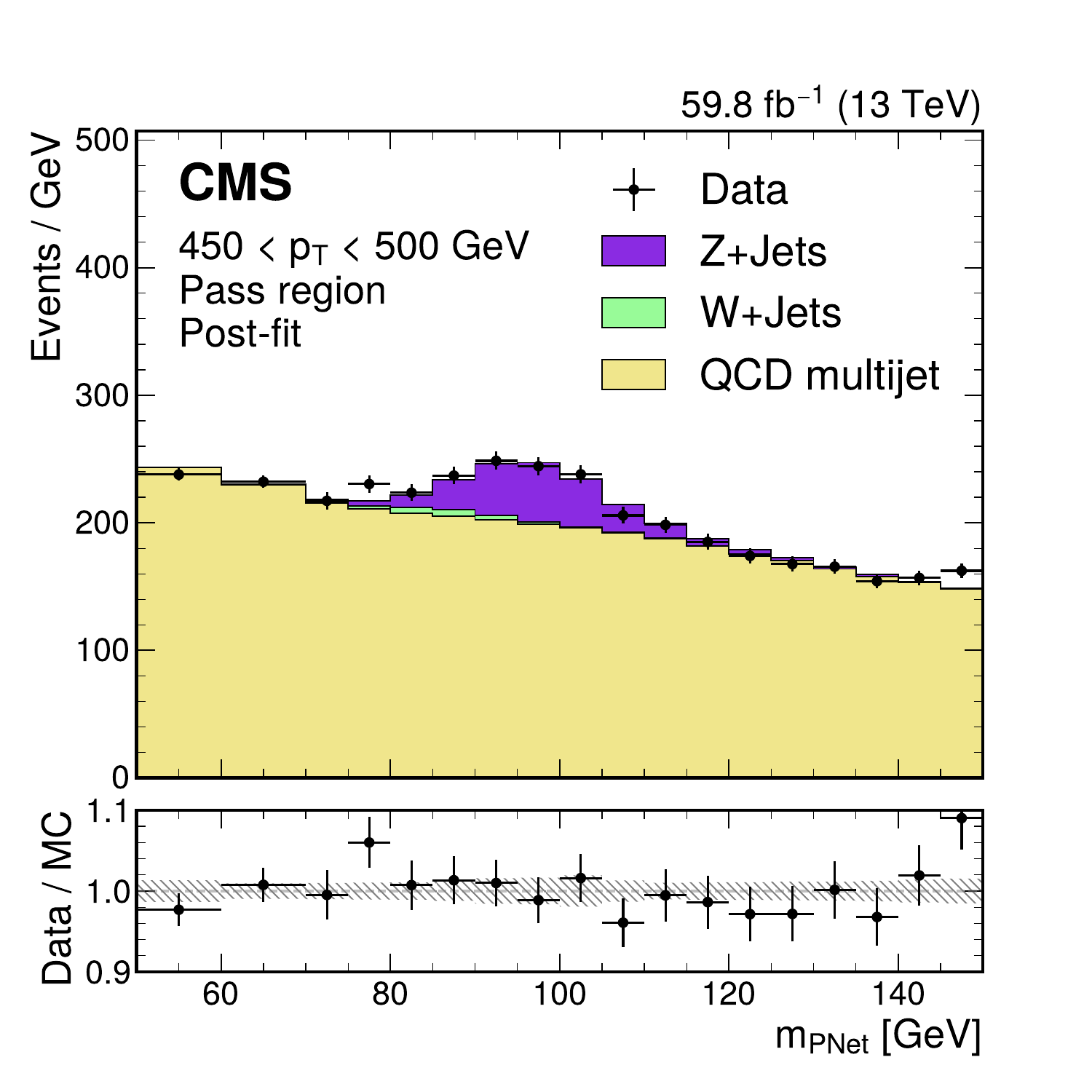}
    \includegraphics[width=0.48\textwidth]{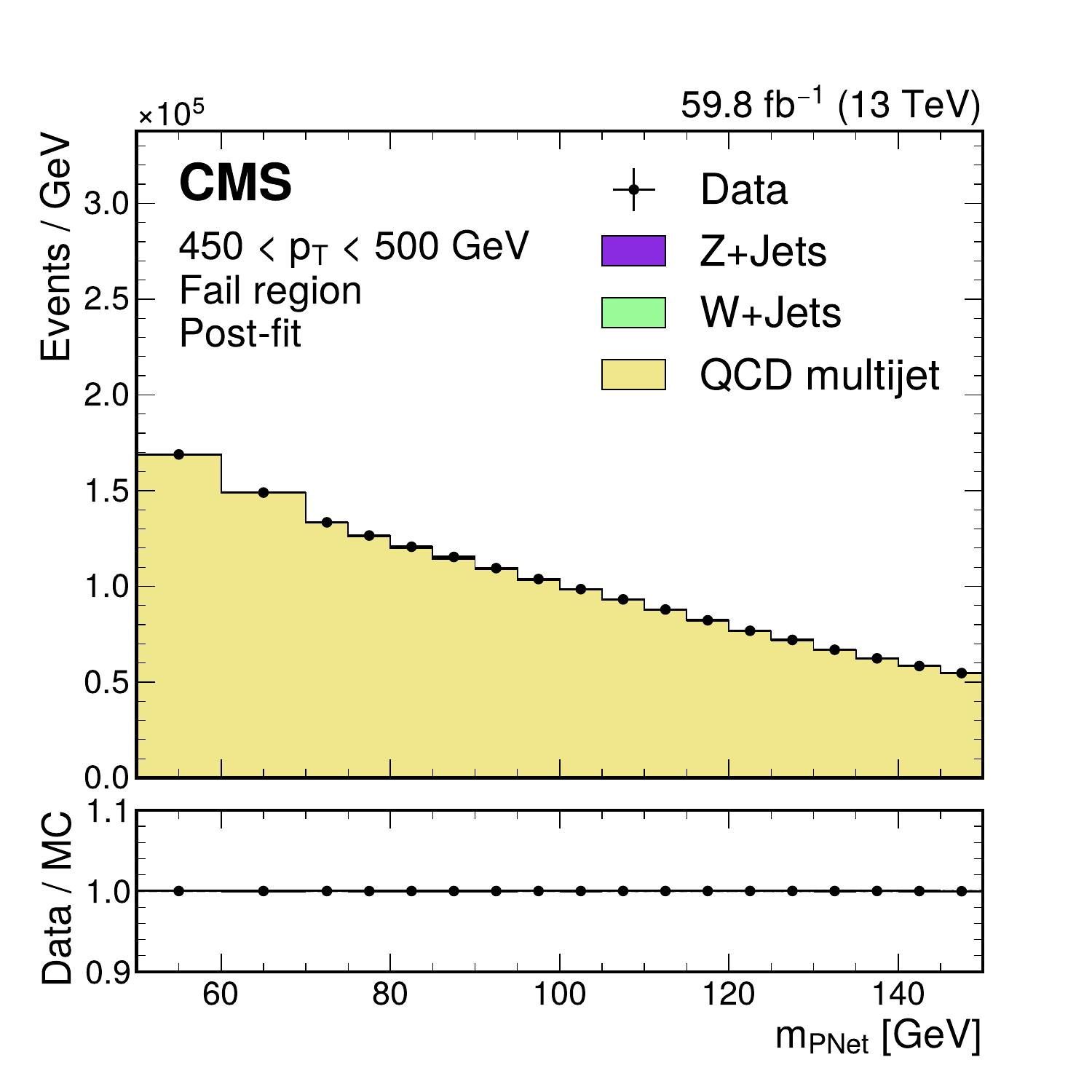}
    \caption{Post-fit distributions from the boosted \PZ boson method for events passing (left) and failing (right) the tagger selection, used in the derivation of the scale factor for the ParticleNet-MD \xbb discriminant at the high-purity working point. The error bars represent the statistical uncertainties in observed data. The lower panels show the pulls defined as $(\text{observed events}\,{-}\,\text{expected events})/\sqrt{\smash[b]{\sigmaobs^{2} + \sigmaexp^{2}}}$, where \sigmaobs and \sigmaexp are the total uncertainties in the observation and the background estimation, respectively. The example corresponds to data and simulated events from the 2018 data-taking conditions, in the jet \pt range of $(450,\,500)\GeV$.}
    \label{fig:postfit_bb_zbb}
\end{figure}

Figure~\ref{fig:postfit_bb_zbb} shows the example post-fit histograms in the ``pass'' and ``fail'' regions, in the derivation of SFs of the ParticleNet-MD \xbb discriminant.

\subsubsection{Systematic uncertainties and results}

The following systematic uncertainties are included in the fit. 
The contribution of each uncertainty source is listed in Table~\ref{tab:zbb_syst_breakdown}, taking ParticleNet-MD \xbb discriminant at the HP WP as an example.

\begin{itemize}

\item \textit{NLO corrections}: The uncertainties in the NLO corrections applied to the \Vjets are taken from Ref.~\cite{Lindert:2017olm}. They account for the renormalization and factorization scale variations and shape uncertainties of the NLO QCD corrections. For the NLO electroweak corrections, uncertainties account for higher-order Sudakov logarithms, hard NNLO emission effects, and the limitations of Sudakov approximation. Details are given in Section~4 of Ref.~\cite{Lindert:2017olm}.

\item \textit{PDF uncertainties}: the uncertainties accounting for the PDF are derived using the PDF4LHC procedure~\cite{Butterworth:2015oua} and the NNPDF3.1 PDF sets.

\item \textit{ISR and FSR in parton shower}: Two uncertainty sources accounting for the ISR and FSR in the parton shower by \PYTHIA are handled in the same way as described in Section~\ref{sec:sfbdt_method_uncertainty}.

\item \textit{Jet mass scale and resolution}: Two sources of uncertainty for the jet mass scale and resolution are propagated to the SF measurement. They account for the simulation-to-data discrepancy in the modelling of \mpnet, and are measured using hadronically decaying, boosted \PW boson jets selected in a dedicated \ttbar-enriched phase space. The uncertainty in the jet mass scale is $<1\%$. The relative uncertainty in the jet mass resolution is around 5\%.

\item \textit{Jet energy scale and resolution}: Two sources of uncertainties for the jet energy scale and resolution are propagated to the SF measurement as described in Section~\ref{sec:sfbdt_method_uncertainty}.

\item \textit{Trigger efficiency}: The overall uncertainty is calculated by taking the statistical uncertainty of the trigger efficiency measurement, less than 1\%, to which an additional 1\% is included. The latter corresponds to the jet energy scale uncertainties of the efficiency measurement.

\item \textit{Integrated luminosity}: The uncertainty in the integrated luminosity for different years ranges from 1.2\% to 2.5\%.~\cite{CMS:2021xjt,CMS-PAS-LUM-17-004,CMS-PAS-LUM-18-002}.

\item \textit{Pileup reweighting}: The uncertainty in the pileup reweighting is treated in the same way as described in Section~\ref{sec:sfbdt_method_uncertainty}.

\end{itemize}

\begin{table}[tbp]
\topcaption{Breakdown of the contributions to the total uncertainty in the fitted scale factor (SF) of the ParticleNet-MD \xbb discriminant at the high-purity working point, using the boosted \PZ boson method. The numbers are averaged over multiple SF derivation points, including all relevant \pt bins and data-taking eras.}
\label{tab:zbb_syst_breakdown}
\centering
{
\begin{tabular}{lcc}
Uncertainty source                               & $\langle \Delta\text{SF} \rangle$  \\
\hline
{Statistical}                                    & 0.354 \\

{Theory}                                         &  \\
\hspace{2em}    ISR and FSR in parton shower     & 0.081 \\
\hspace{2em}    NLO corrections                  & 0.074 \\
\hspace{2em}    PDF uncertainties                & 0.019 \\

{Experimental}                                   &  \\
\hspace{2em}    Jet mass scale and resolution    & 0.033 \\
\hspace{2em}    Jet energy scale and resolution  & 0.078 \\
\hspace{2em}    Trigger effiency                 & 0.020 \\
\hspace{2em}    Integrated luminosity            & 0.036 \\
\hspace{2em}    Pileup reweighting               & 0.007 \\

\end{tabular}
}
\end{table}

The boosted \PZ boson method is used to measure the SFs of the ParticleNet-MD \xbb discriminants at all WPs, the DeepDoubleX \xbb discriminants at the HP and MP WPs, and the double-b tagger at the HP WP. For the unmeasured WPs of the three discriminants, the method cannot converge with reasonable uncertainties because the \Zjets contribution is negligible compared with the large QCD multijet yield in the ``pass'' region. Additionally, the DeepAK8-MD \xbb discriminant is not calibrated with this method since jet \mpnet distributions for the QCD multijet background differ significantly between the ``pass'' and ``fail'' regions. This is due to some residual mass correlation of this discriminant. The derived SFs are displayed in Figs.~\ref{fig:sf_pnet-b}--\ref{fig:sf_da-c}. 
It is also worth noting that the method has limitations in calibrating \xcc discriminants, given the notable contribution of \zbb jets after applying a \xcc discriminant WP selection. Therefore, the method is only used for the \xbb discriminants.

\begin{figure}[htbp]
    \centering
    \includegraphics[width=0.60\textwidth]{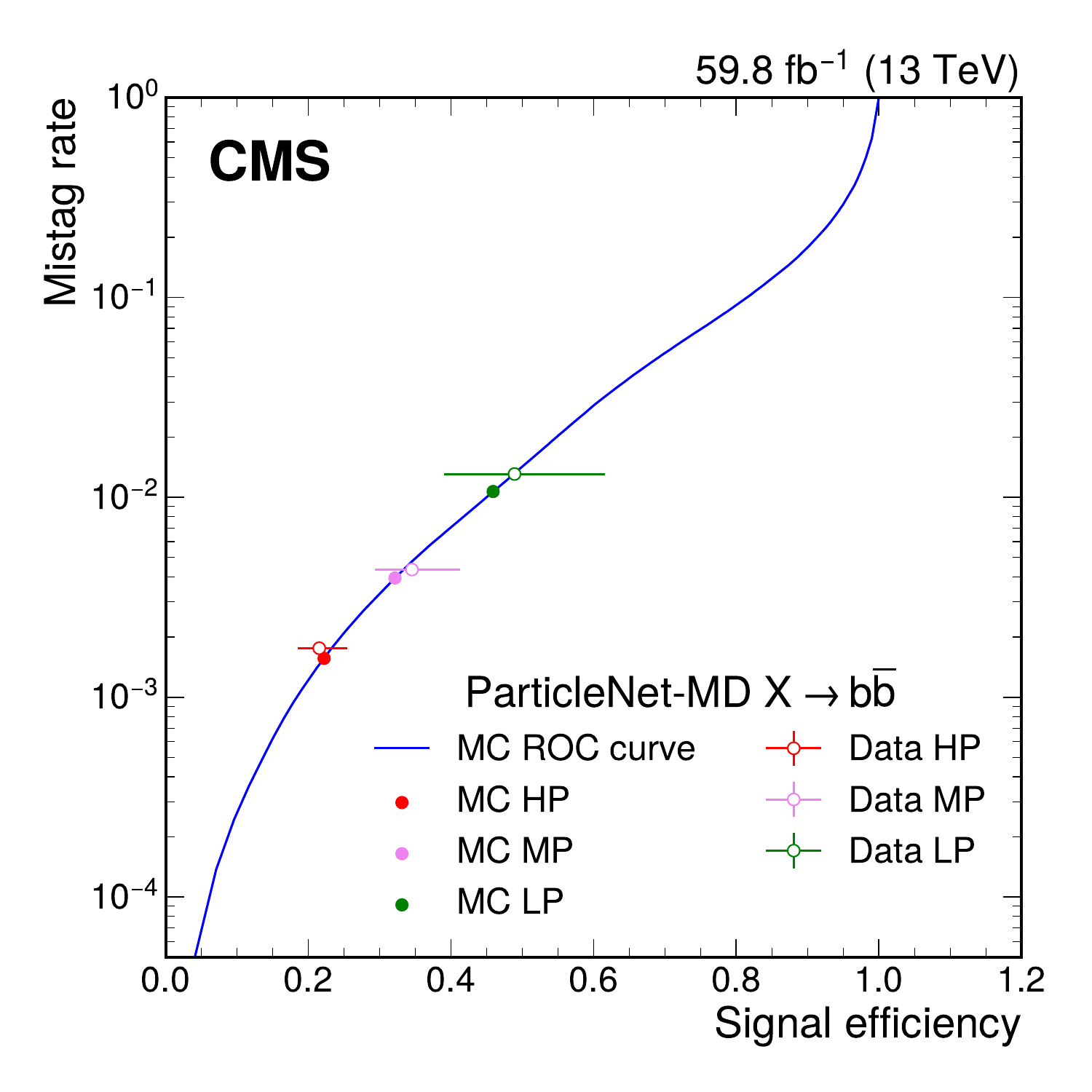}
    \caption{Receiver operating characteristic (ROC) curve of the ParticleNet-MD \xbb discriminant obtained from simulation (blue), under 2018 data-taking conditions with $\pt > 450\GeV$. The high-purity (HP), medium-purity (MP), and low-purity (LP) working points are indicated by filled circles for simulation and hollow circles for data. The error bars represent the statistical uncertainties in observed data. }
    \label{fig:zbb_roc}
\end{figure}

Since the method extracts the yields of both the \zbb signal and the QCD multijet background, we can measure the signal efficiency versus the mistag rate.
Figure~\ref{fig:zbb_roc} shows the ROC curve of the ParticleNet-MD \xbb discriminant, obtained in simulation and the three WPs measured in data.
The uncertainty in the measured mistag rate is much lower than the uncertainty in the measured signal efficiency due to the large QCD background in both the ``pass'' and ``fail'' regions. The uncertainty in the measured signal efficiency is larger for the LP WP because of the lowering of the signal-to-background ratio in the ``pass'' region.

\subsection{Combination of measured scale factors}\label{sec:combination}

In previous sections, individual SFs for \xbb and \xcc tagging efficiencies, measured from the sfBDT method, the $\mu$-tagged method, and the boosted \PZ boson method, have been presented.
The sfBDT and $\mu$-tagged methods are employed to derive the full set of tagging efficiency SFs, whereas the boosted \PZ boson method provides measurements for a subset of the \xbb tagging SF derivation points.
In this section, a combination of the available measurements is performed for the SFs at each derivation point.
The combination is a weighted average taking into account the full covariance matrix for the uncertainties using the best linear unbiased estimator (BLUE) method~\cite{BLUE}. As adopted in the combination of AK4 SFs for \PQb and \PQc jets~\cite{CMS:BTVFlvTagger}, the BLUE method is extended to fit all the jet \pt bins simultaneously, providing a more comprehensive treatment of bin-to-bin correlations for the systematic uncertainties~\cite{ExtBLUE}.

\begin{figure}[htbp]
    \centering
    \includegraphics[width=0.32\textwidth]{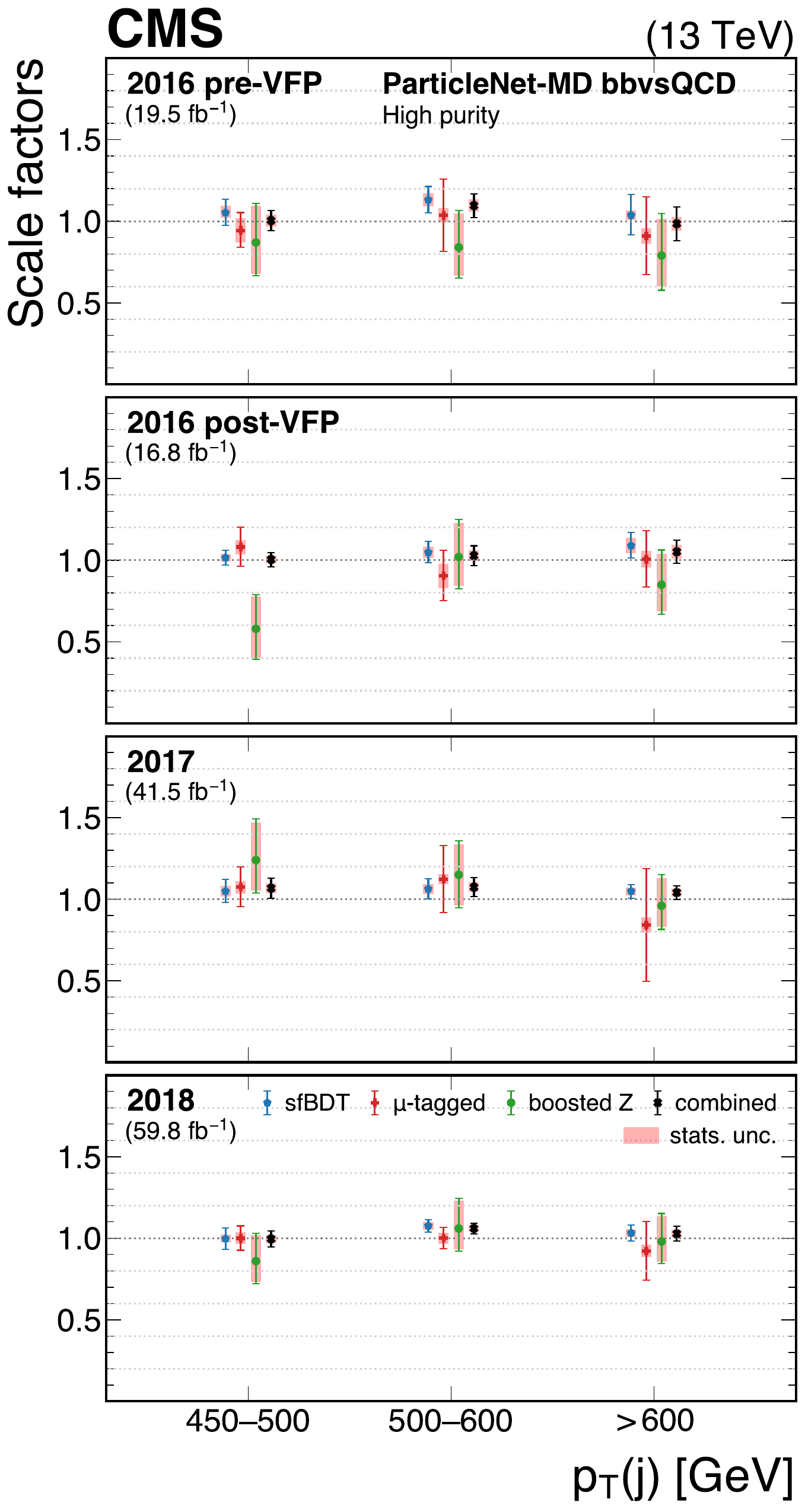}
    \includegraphics[width=0.32\textwidth]{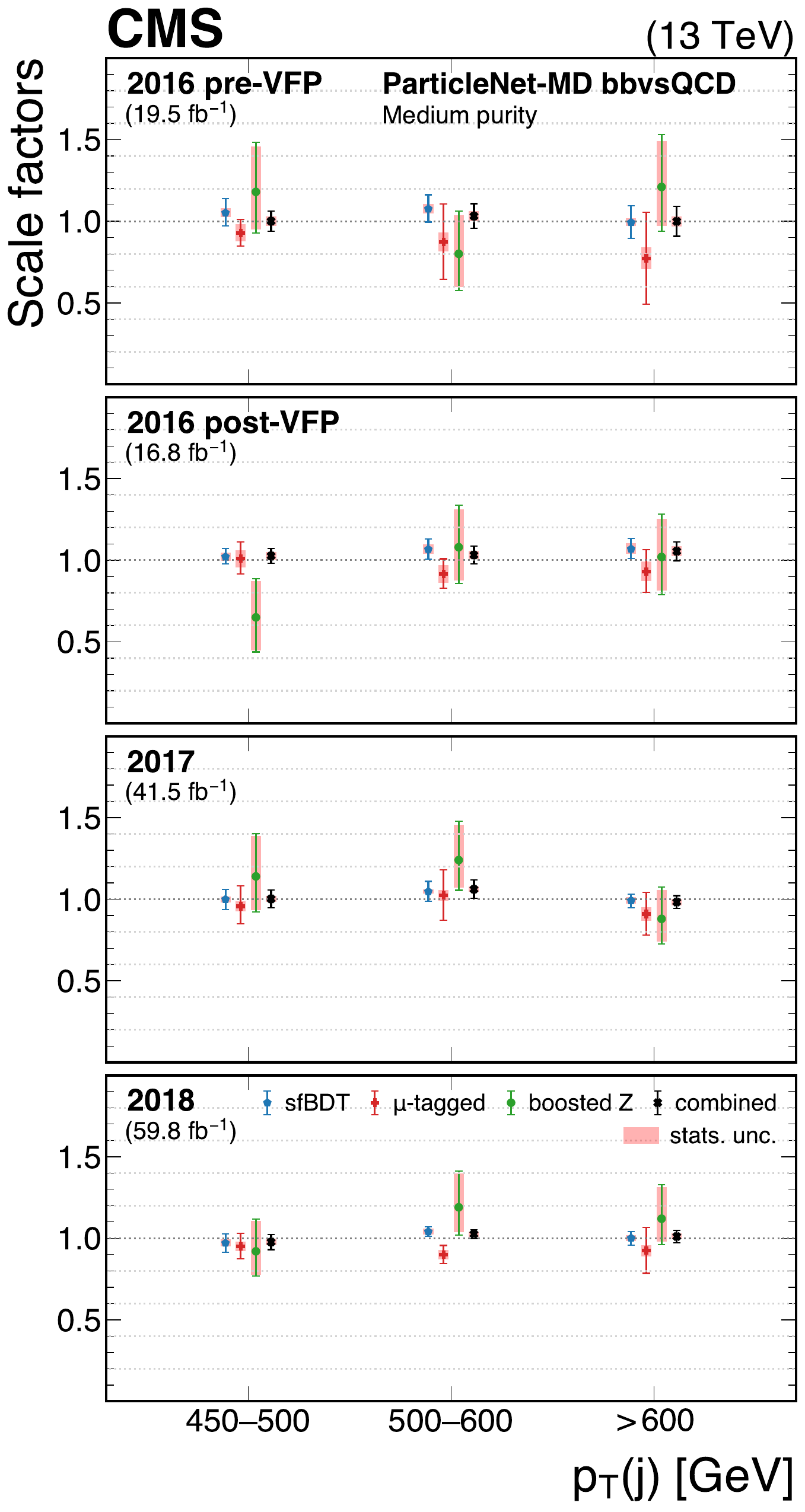}
    \includegraphics[width=0.32\textwidth]{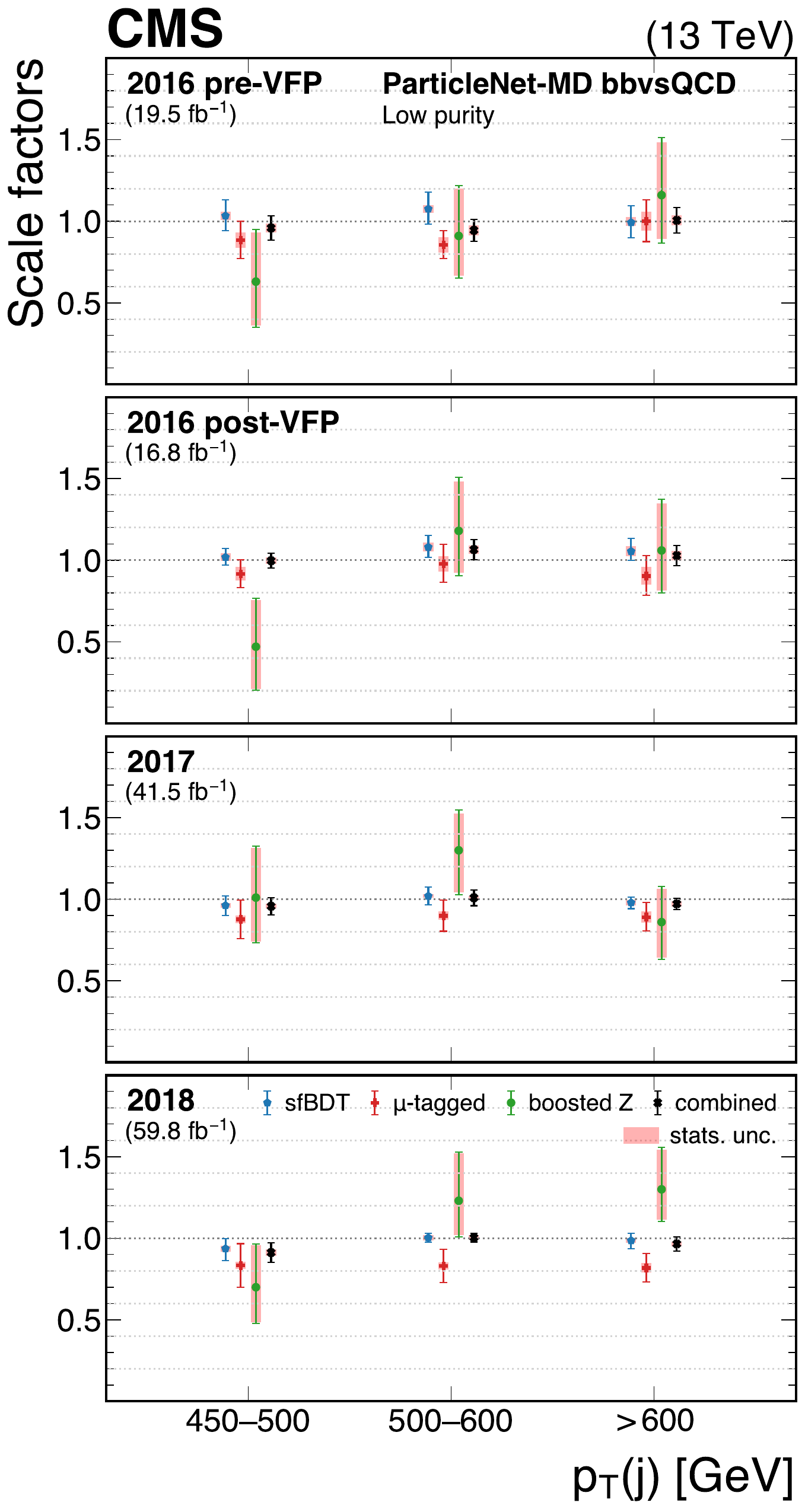}
    \caption{The measured scale factors of the ParticleNet-MD \xbb discriminant in the high-purity (left), medium-purity (middle), and low-purity (right) working points. Three methods are presented in the measurements: the sfBDT method, the $\mu$-tagged method, and the boosted \PZ boson method. The combined measurements from available methods are also shown.}
    \label{fig:sf_pnet-b}
\end{figure}

\begin{figure}[htbp]
    \centering
    \includegraphics[width=0.32\textwidth]{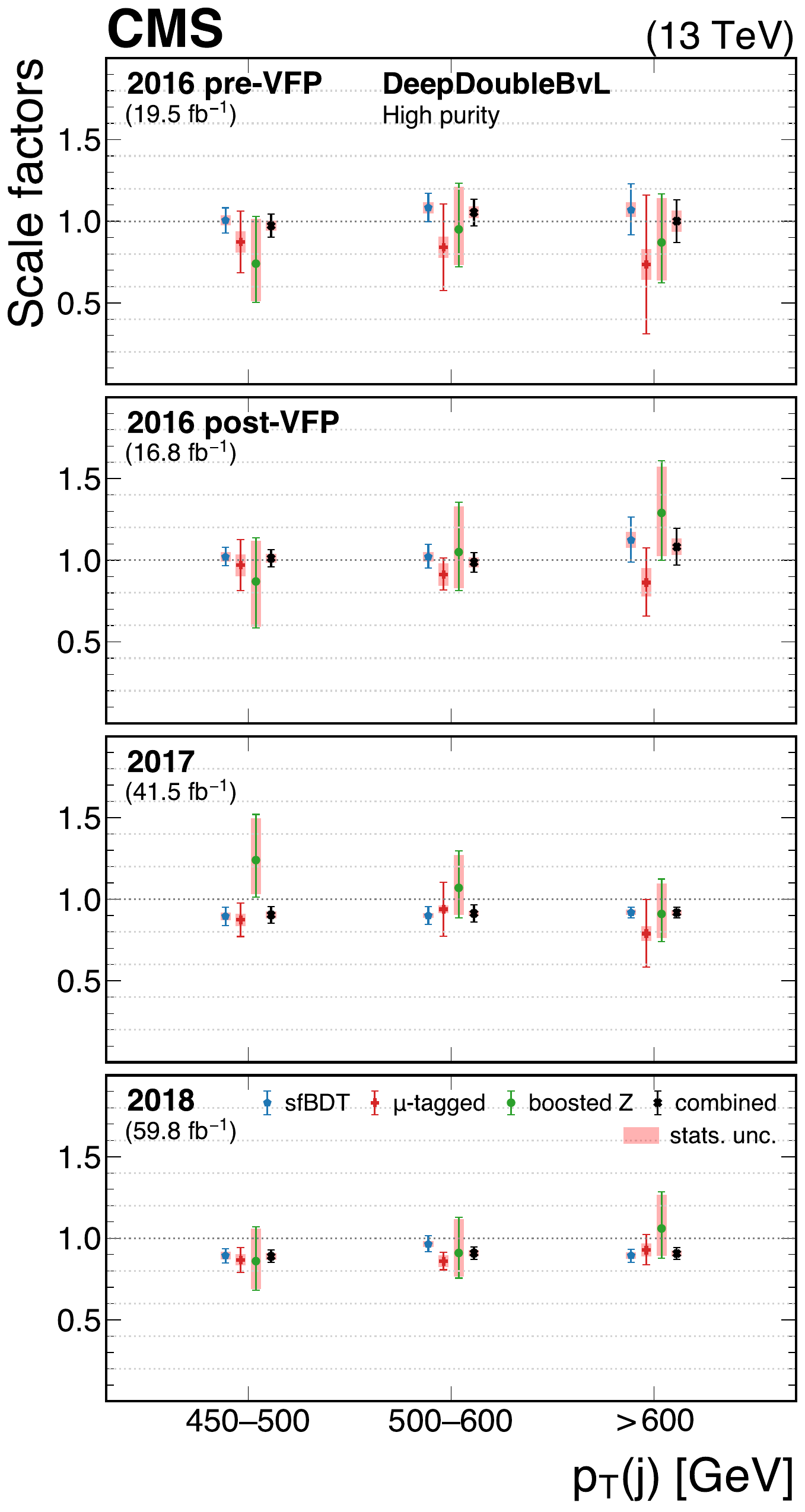}
    \includegraphics[width=0.32\textwidth]{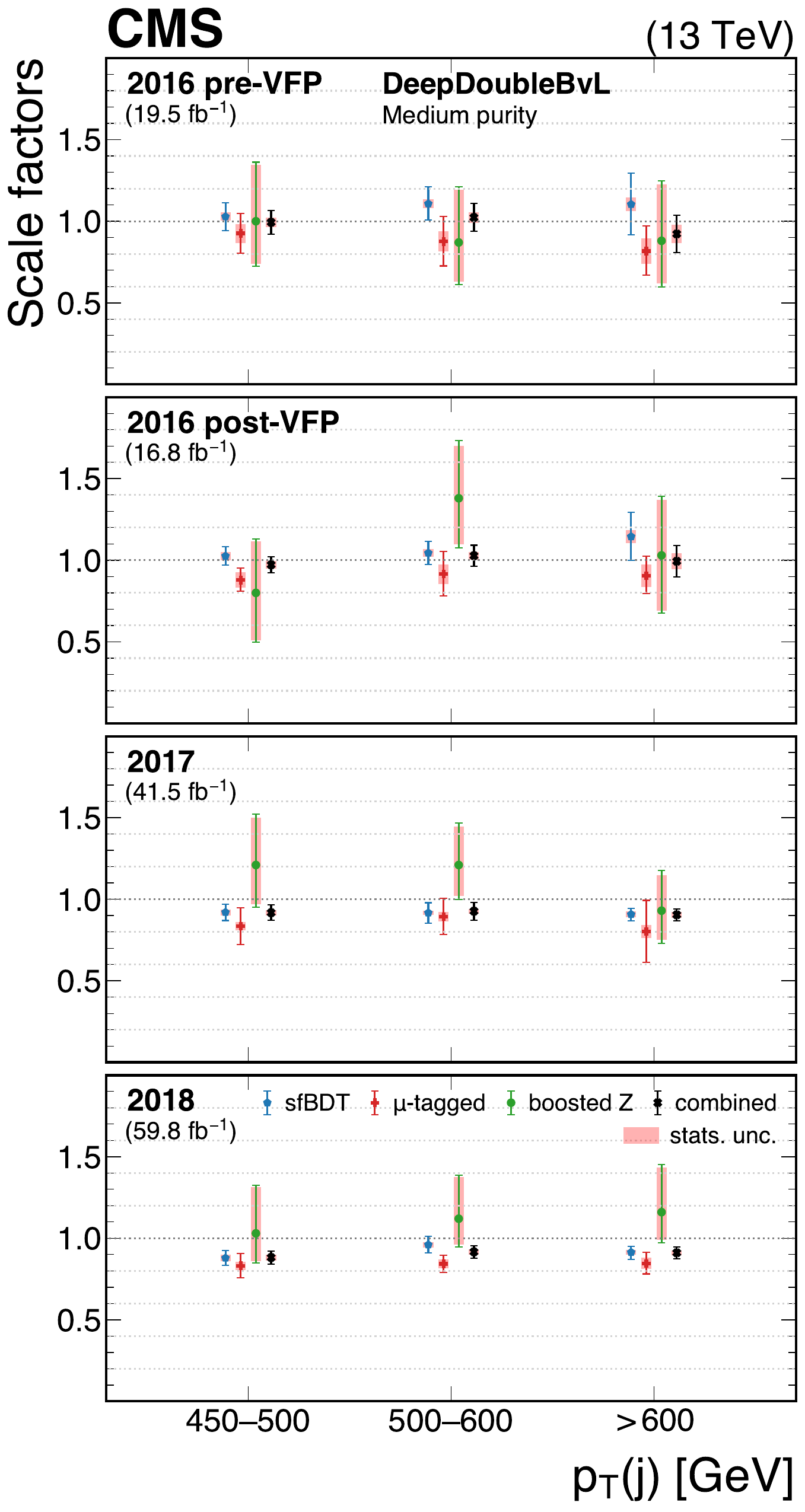}
    \includegraphics[width=0.32\textwidth]{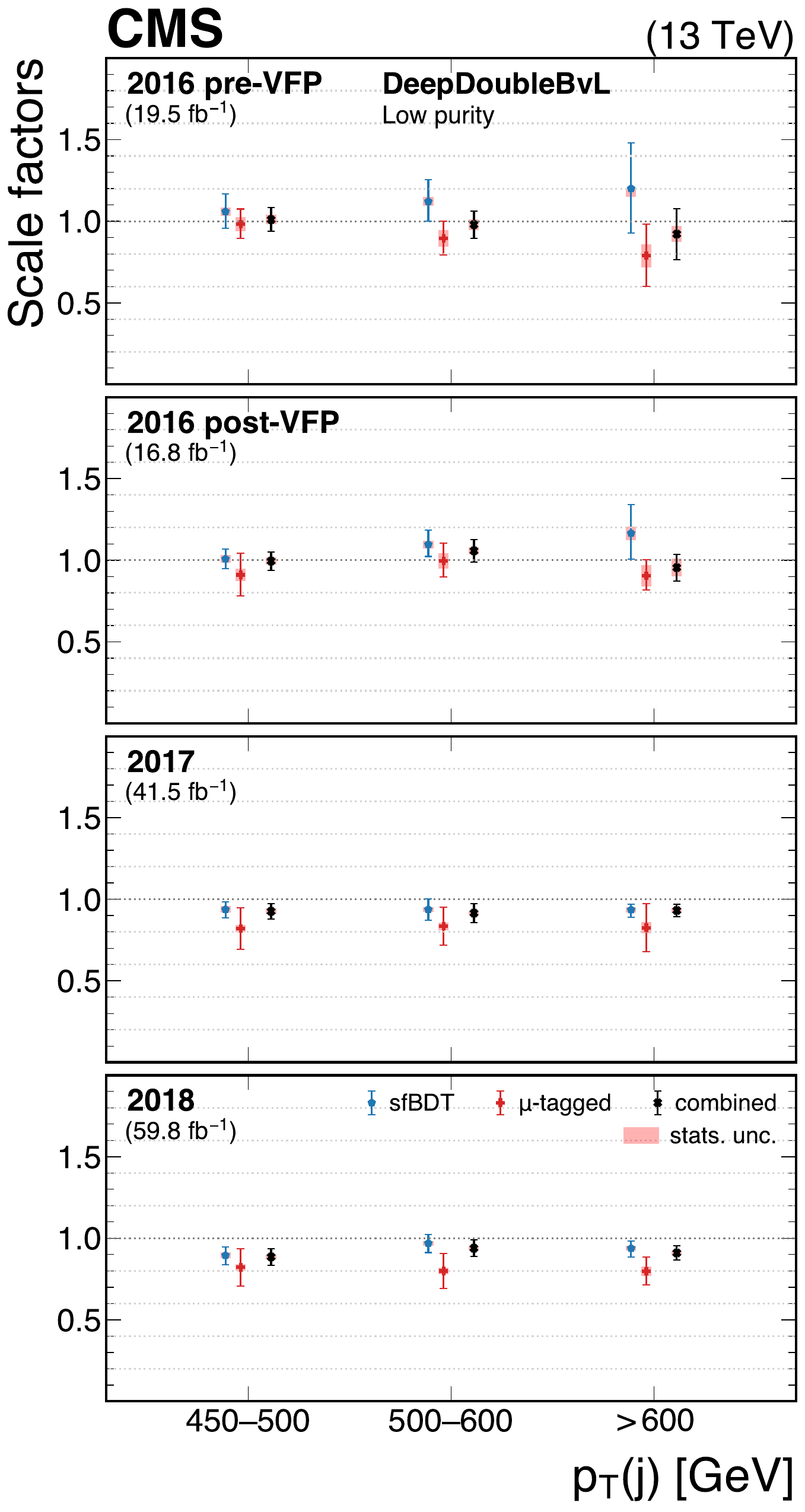}
    \caption{The measured scale factors of the DeepDoubleX \xbb discriminant in the high-purity (left), medium-purity (middle), and low-purity (right) working points. Three methods are presented in the measurements: the sfBDT method, the $\mu$-tagged method, and the boosted \PZ boson method. The combined measurements from available methods are also shown.}
    \label{fig:sf_ddx-b}
\end{figure}

\begin{figure}[htbp]
    \centering
    \includegraphics[width=0.32\textwidth]{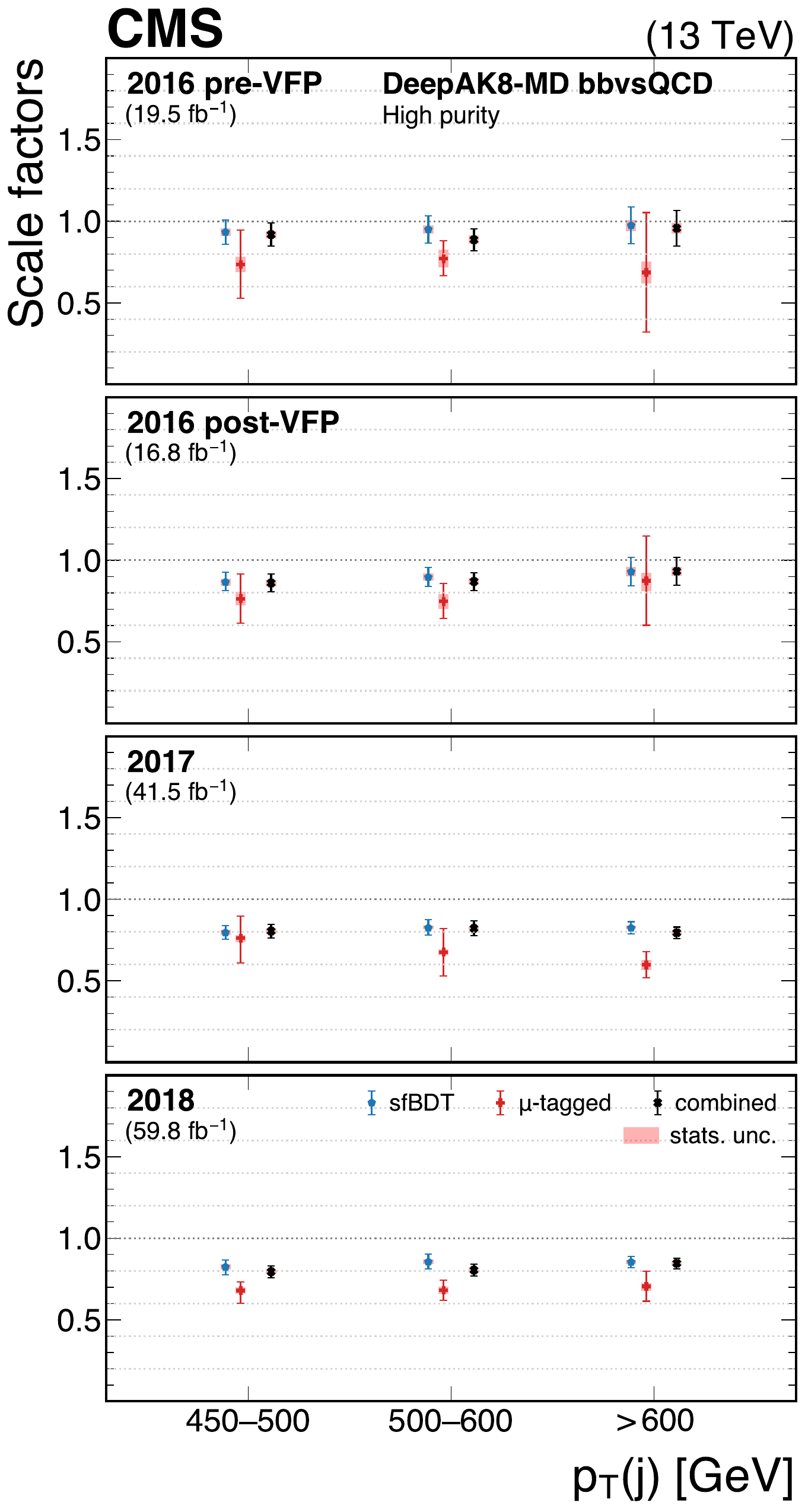}
    \includegraphics[width=0.32\textwidth]{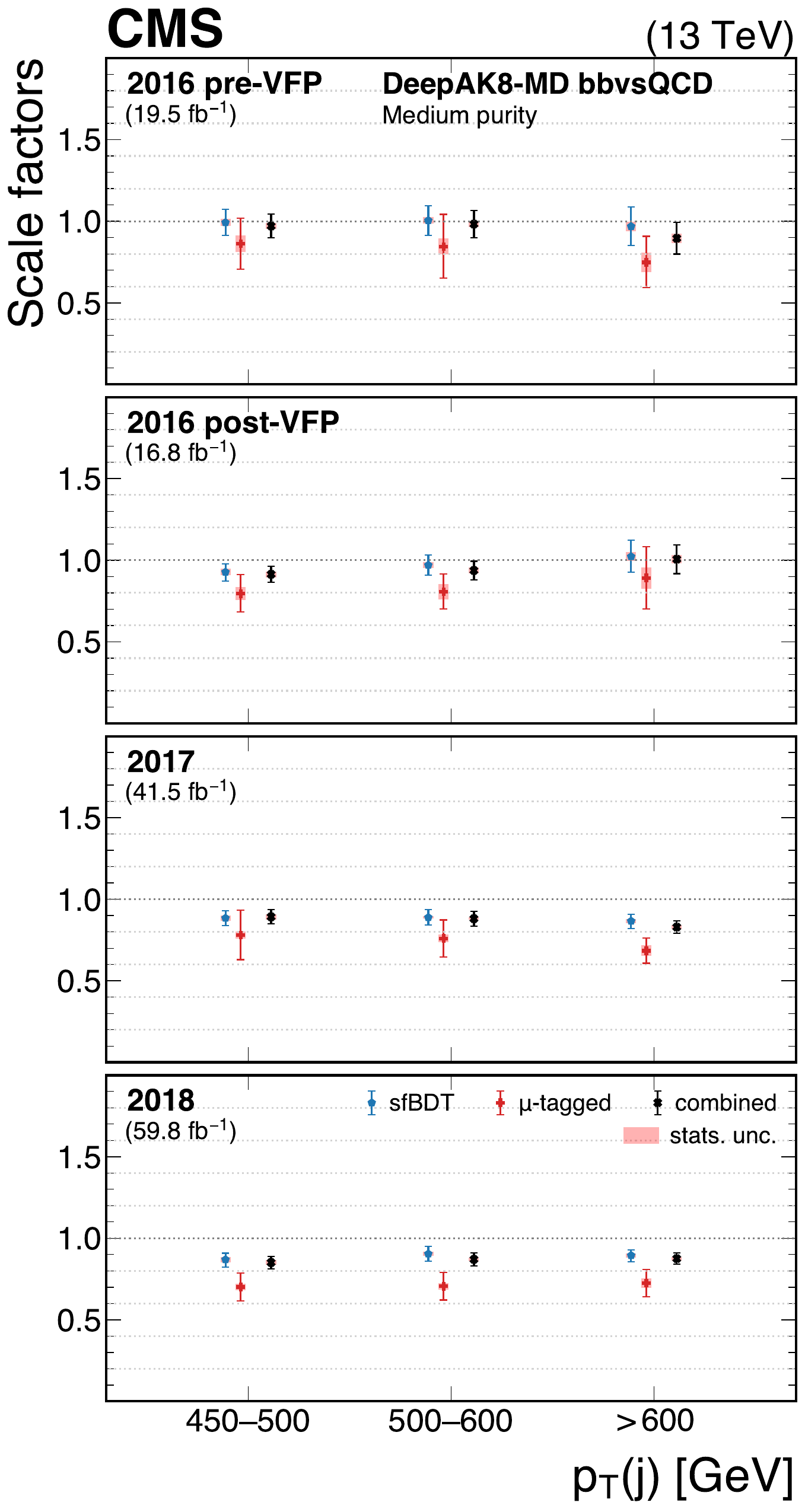}
    \includegraphics[width=0.32\textwidth]{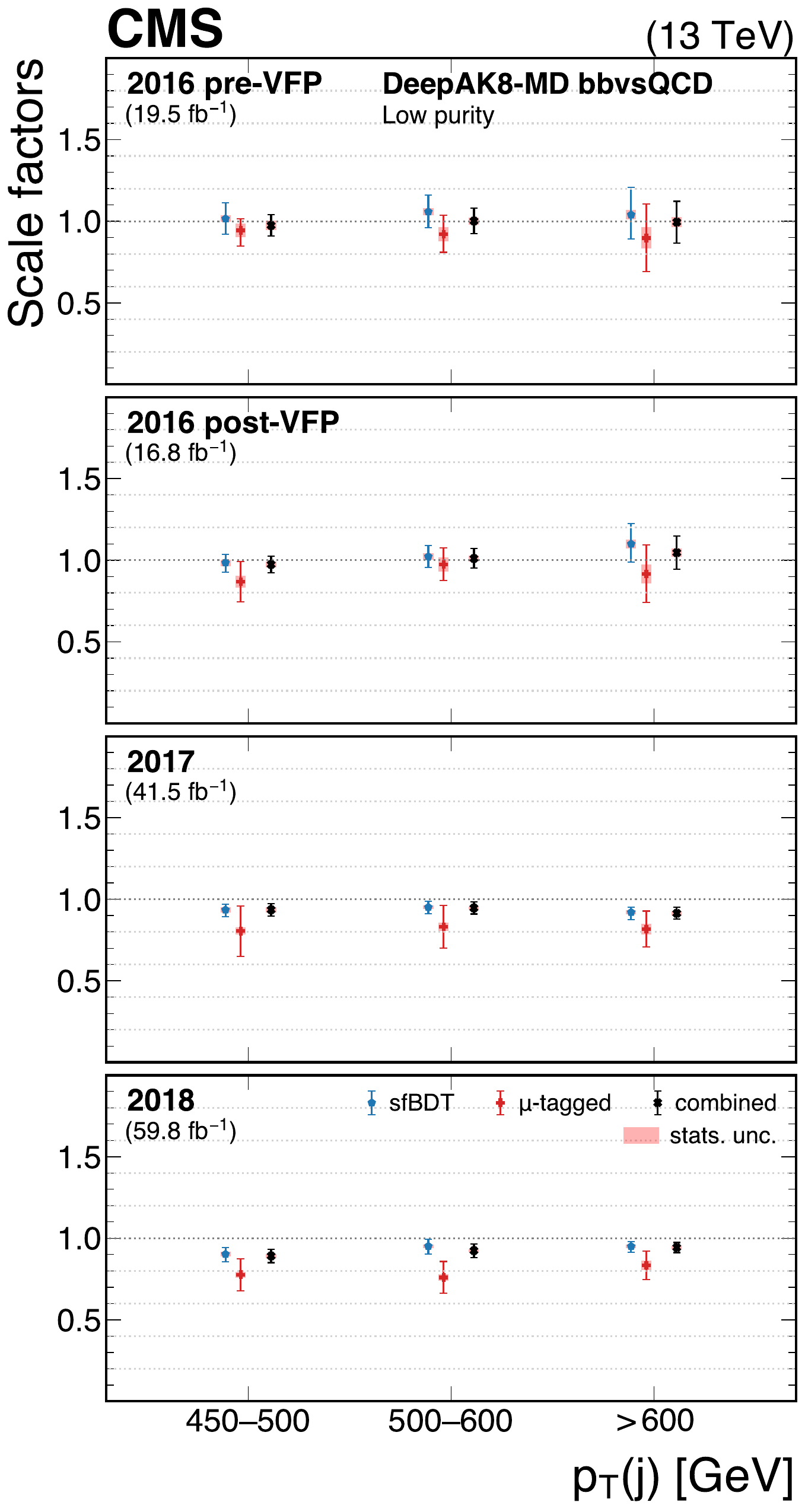}
    \caption{The measured scale factors of the DeepAK8-MD \xbb discriminant in the high-purity (left), medium-purity (middle), and low-purity (right) working points. Two methods are presented in the measurements: the sfBDT method and the $\mu$-tagged method. The combined measurements from available methods are also shown.}
    \label{fig:sf_da-b}
\end{figure}

\begin{figure}[htbp]
    \centering
    \includegraphics[width=0.32\textwidth]{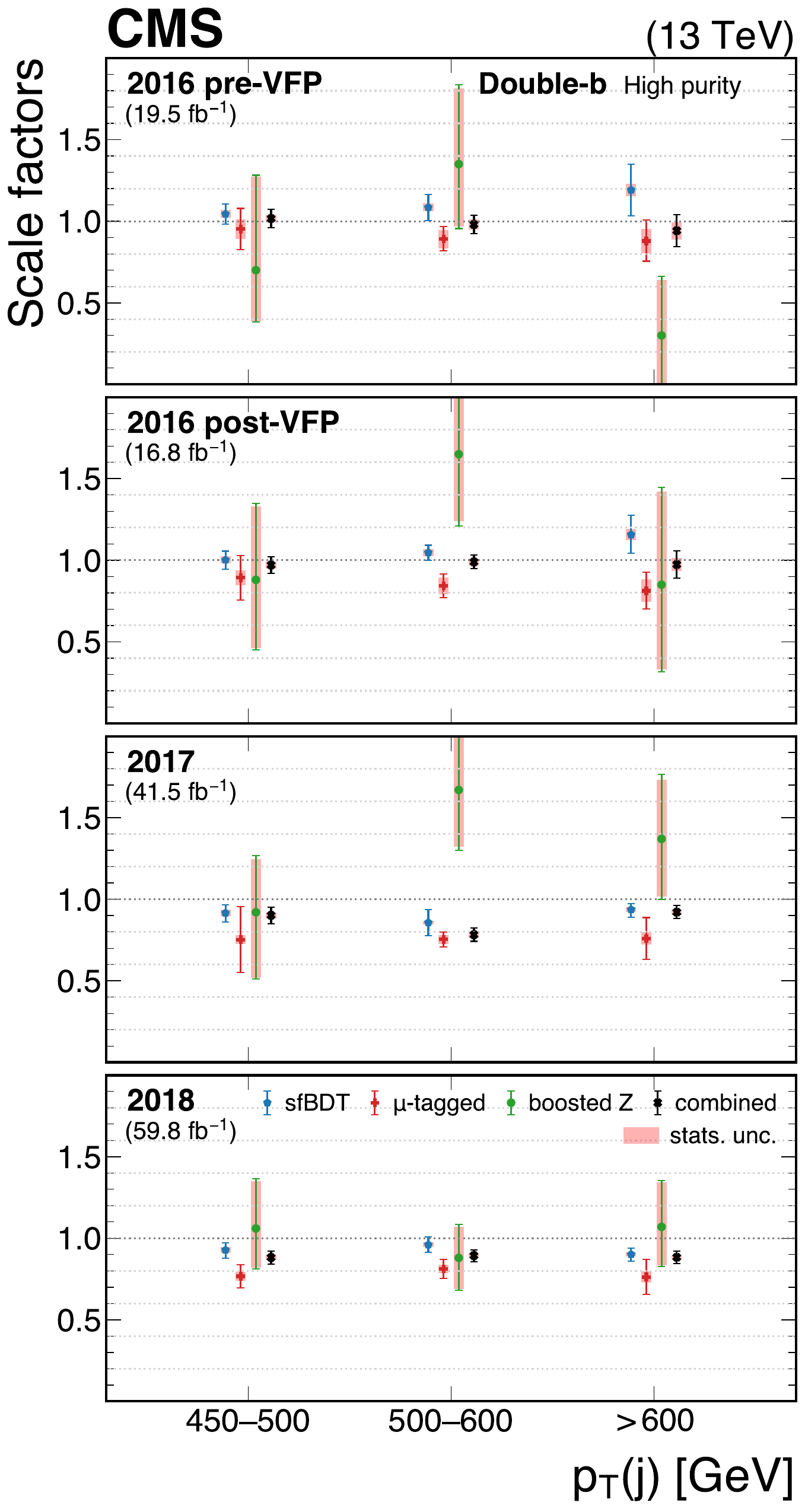}
    \includegraphics[width=0.32\textwidth]{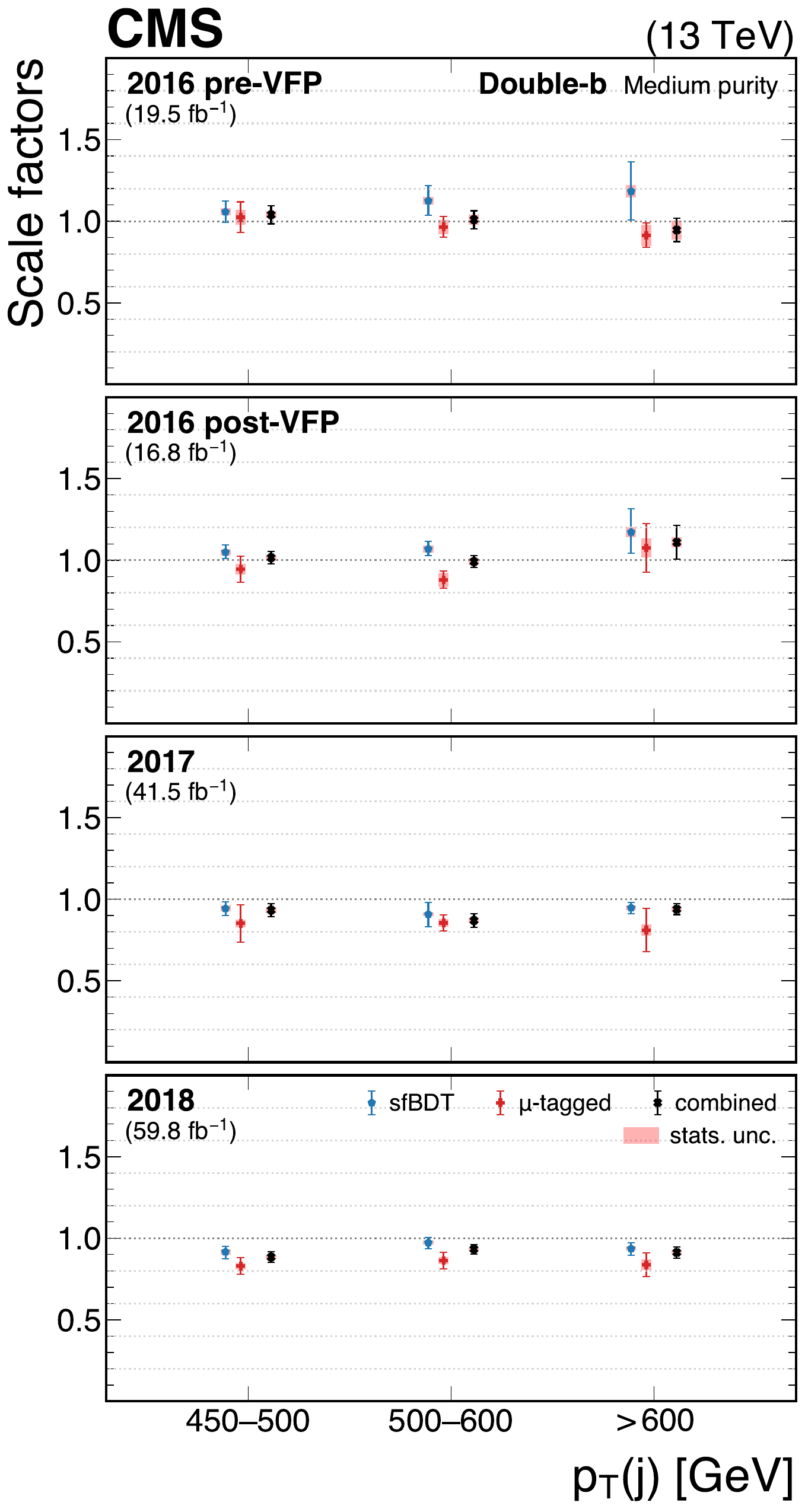}
    \includegraphics[width=0.32\textwidth]{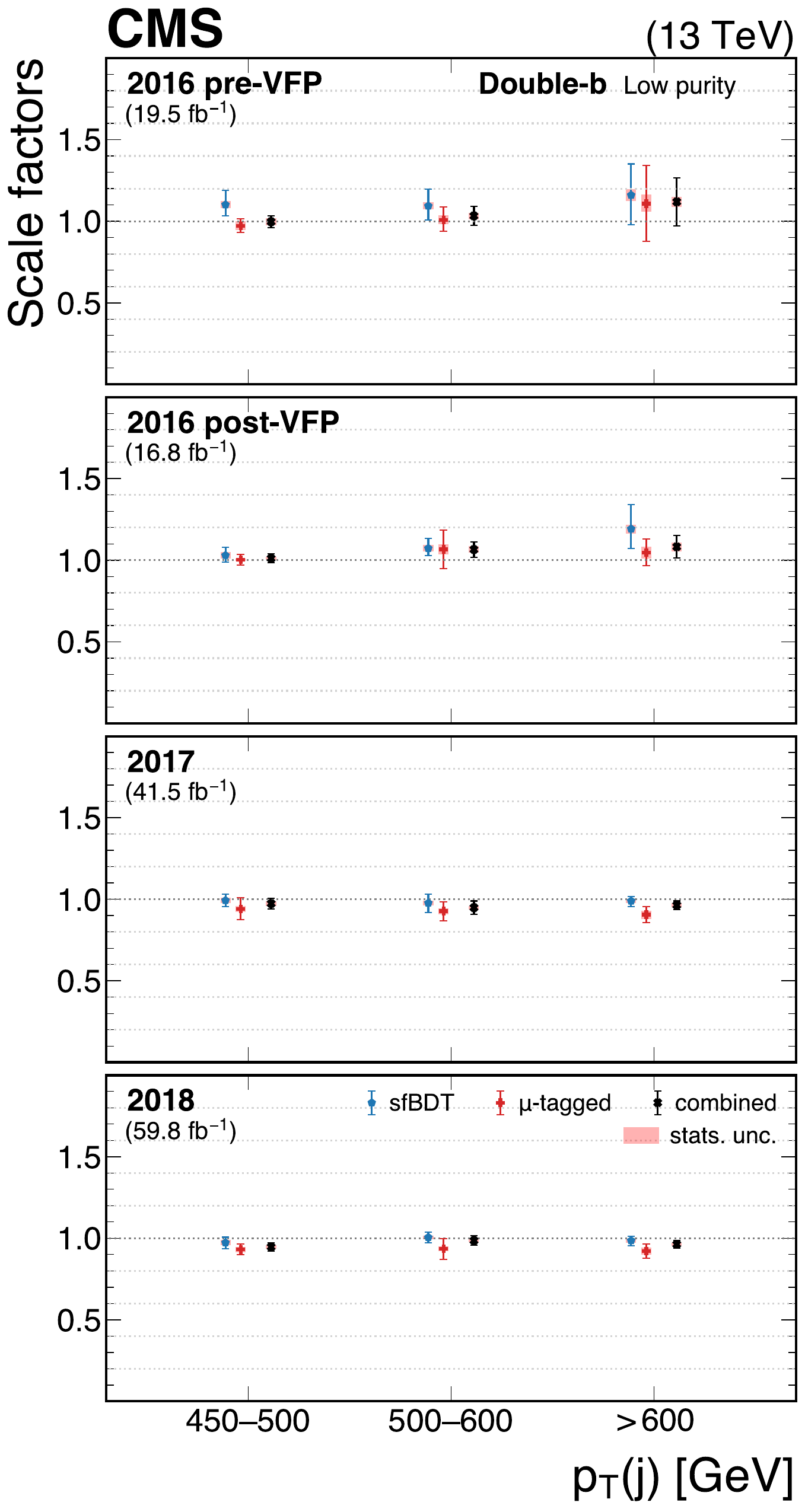}
    \caption{The measured scale factors of the double-b \xbb discriminant in the high-purity (left), medium-purity (middle), and low-purity (right) working points. Three methods are presented in the measurements: the sfBDT method, the $\mu$-tagged method, and the boosted \PZ boson method. The combined measurements from available methods are also shown.}
    \label{fig:sf_double-b}
\end{figure}

\begin{figure}[htbp]
    \centering
    \includegraphics[width=0.32\textwidth]{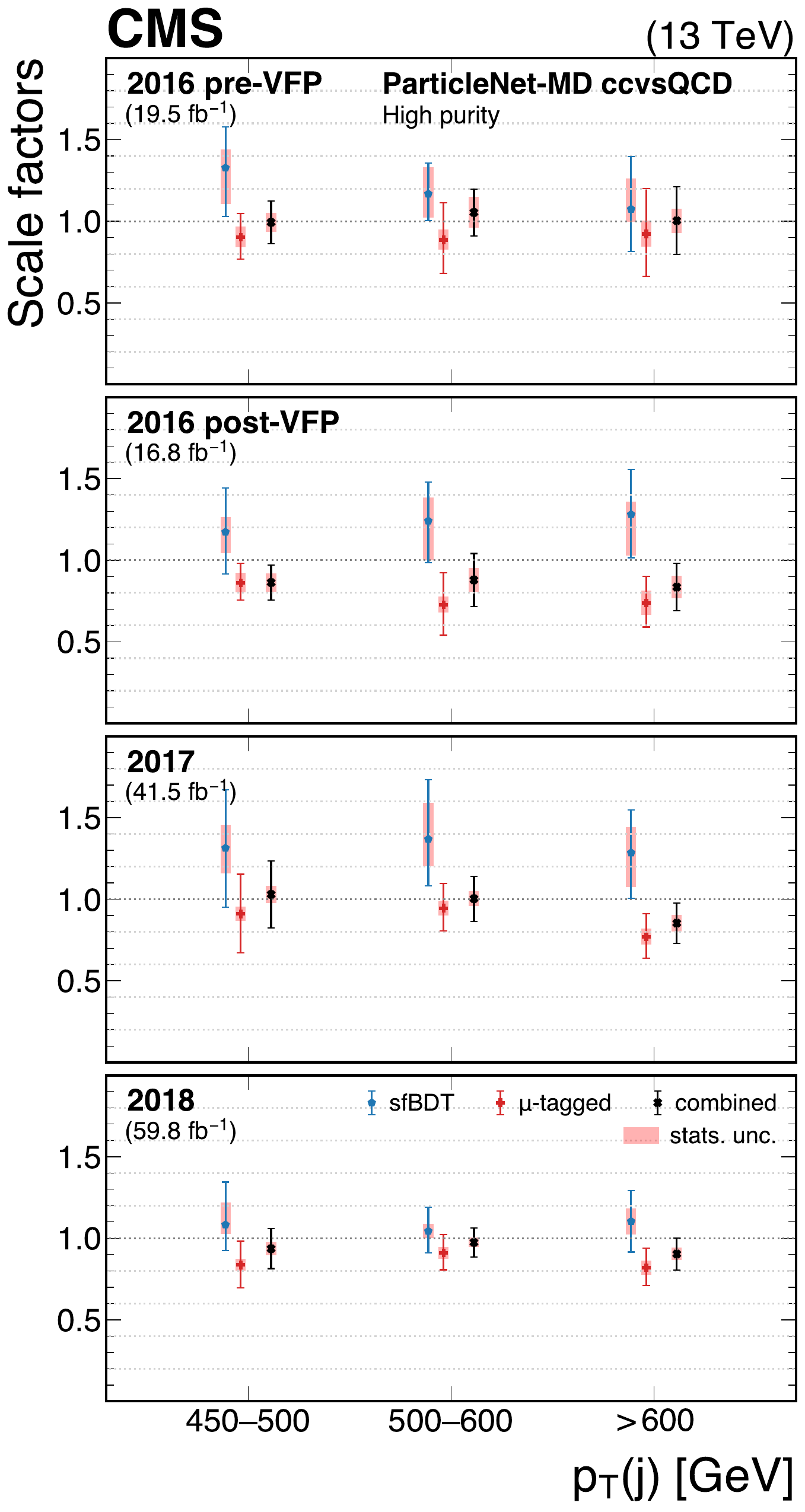}
    \includegraphics[width=0.32\textwidth]{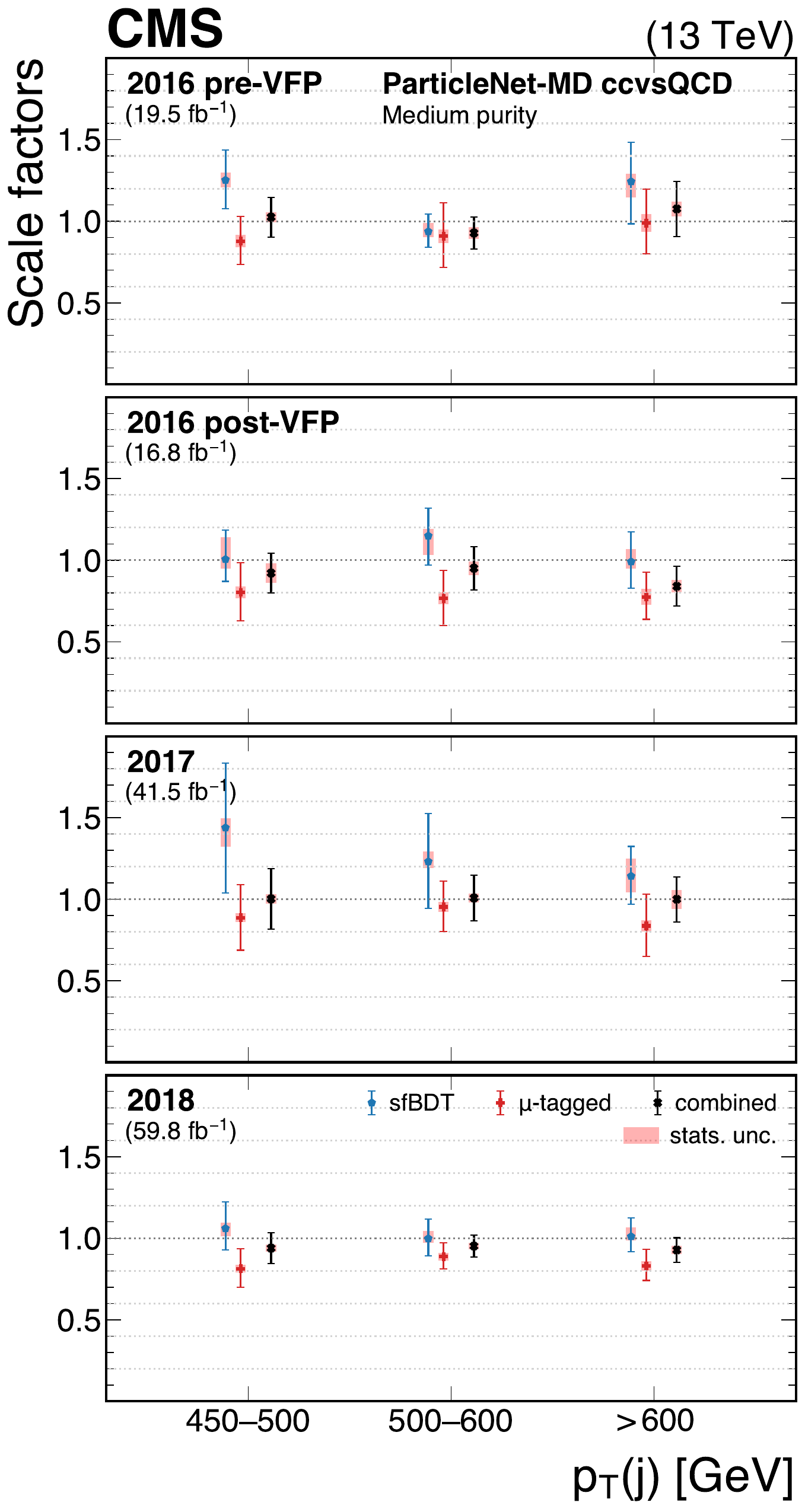}
    \includegraphics[width=0.32\textwidth]{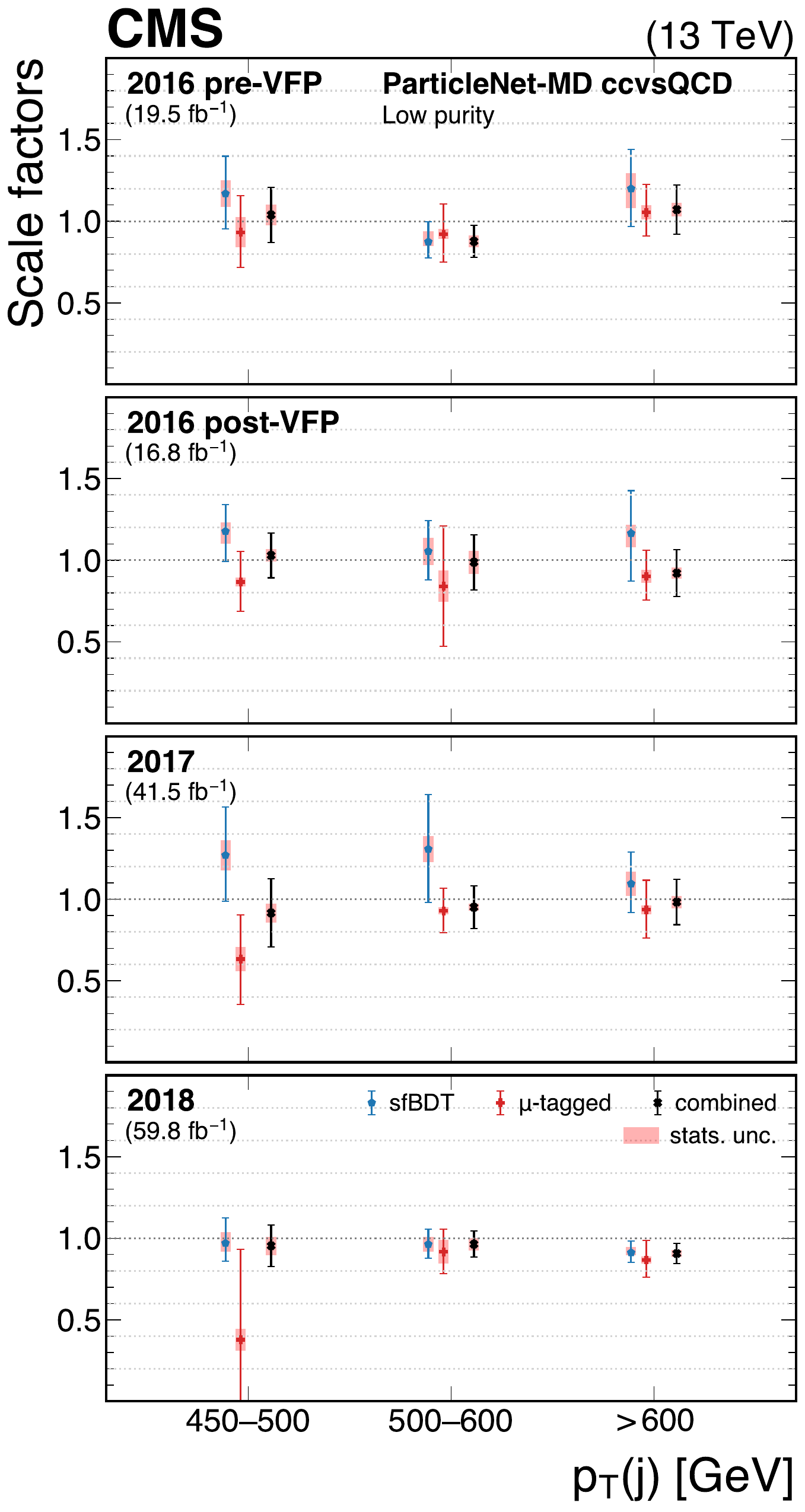}
    \caption{The measured scale factors of the ParticleNet-MD \xcc discriminant in the high-purity (left), medium-purity (middle), and low-purity (right) working points. Two methods are presented in the measurements: the sfBDT method and the $\mu$-tagged method. The combined measurements from available methods are also shown.}
    \label{fig:sf_pnet-c}
\end{figure}

\begin{figure}[htbp]
    \centering
    \includegraphics[width=0.32\textwidth]{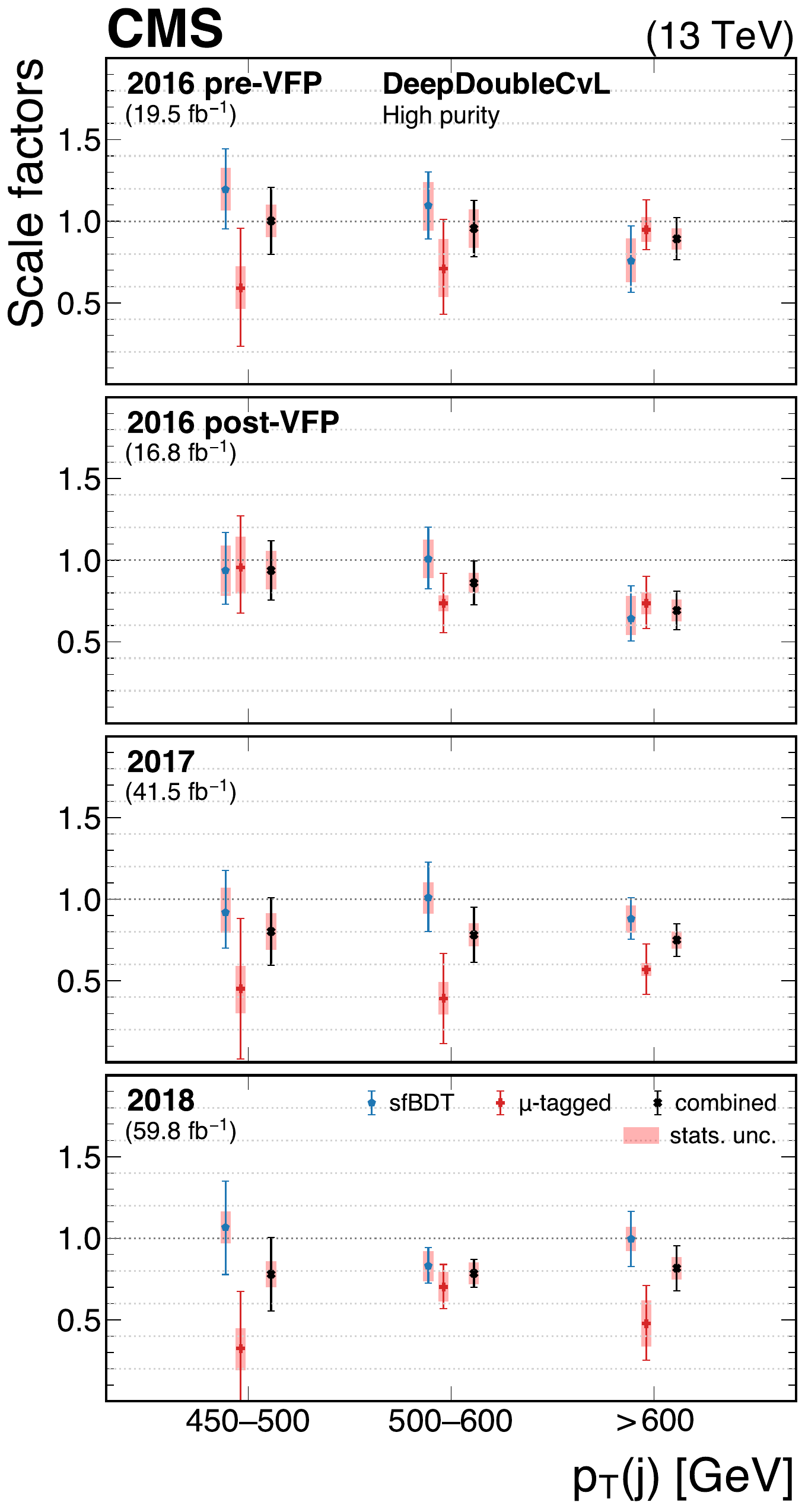}
    \includegraphics[width=0.32\textwidth]{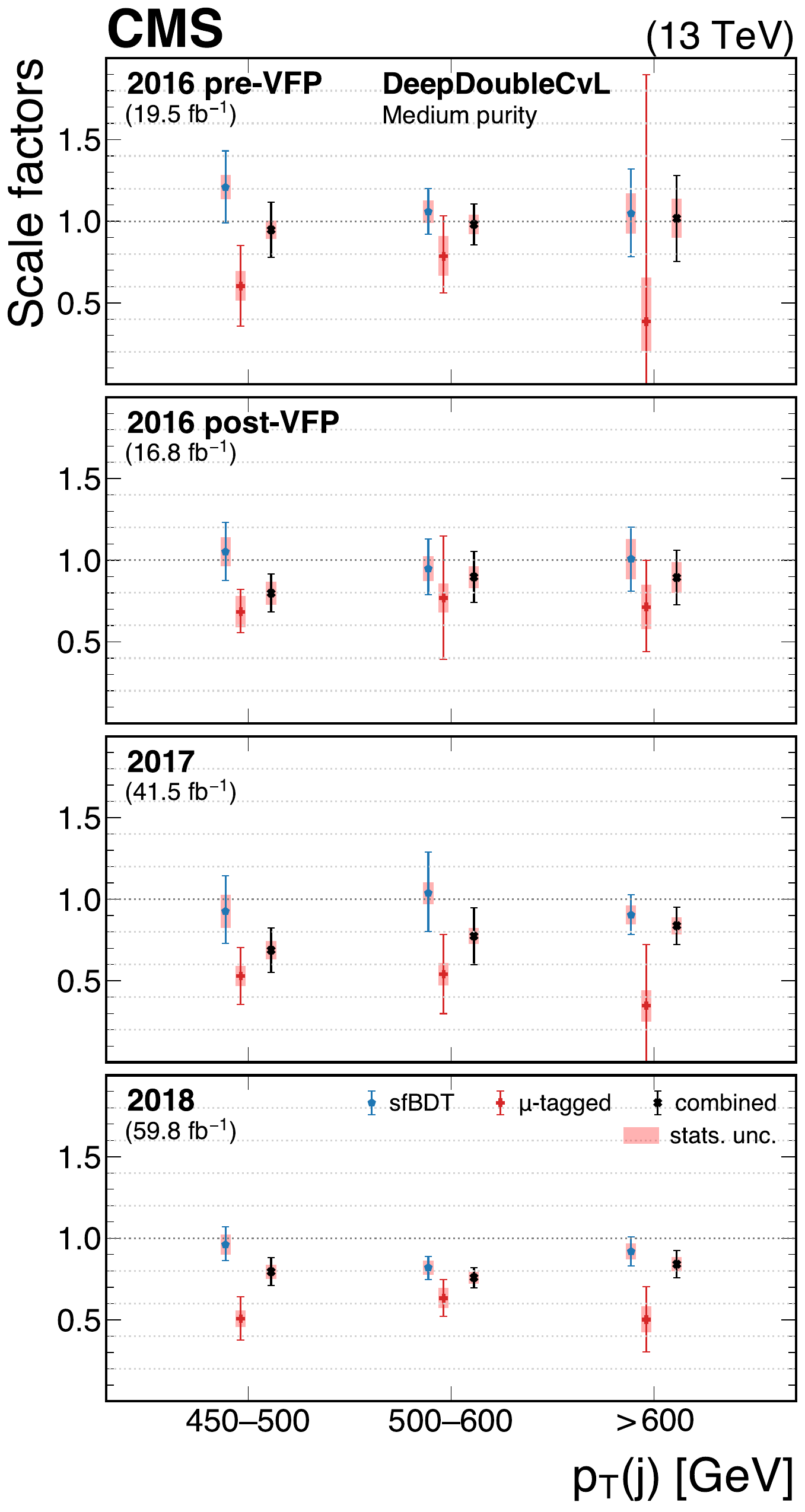}
    \includegraphics[width=0.32\textwidth]{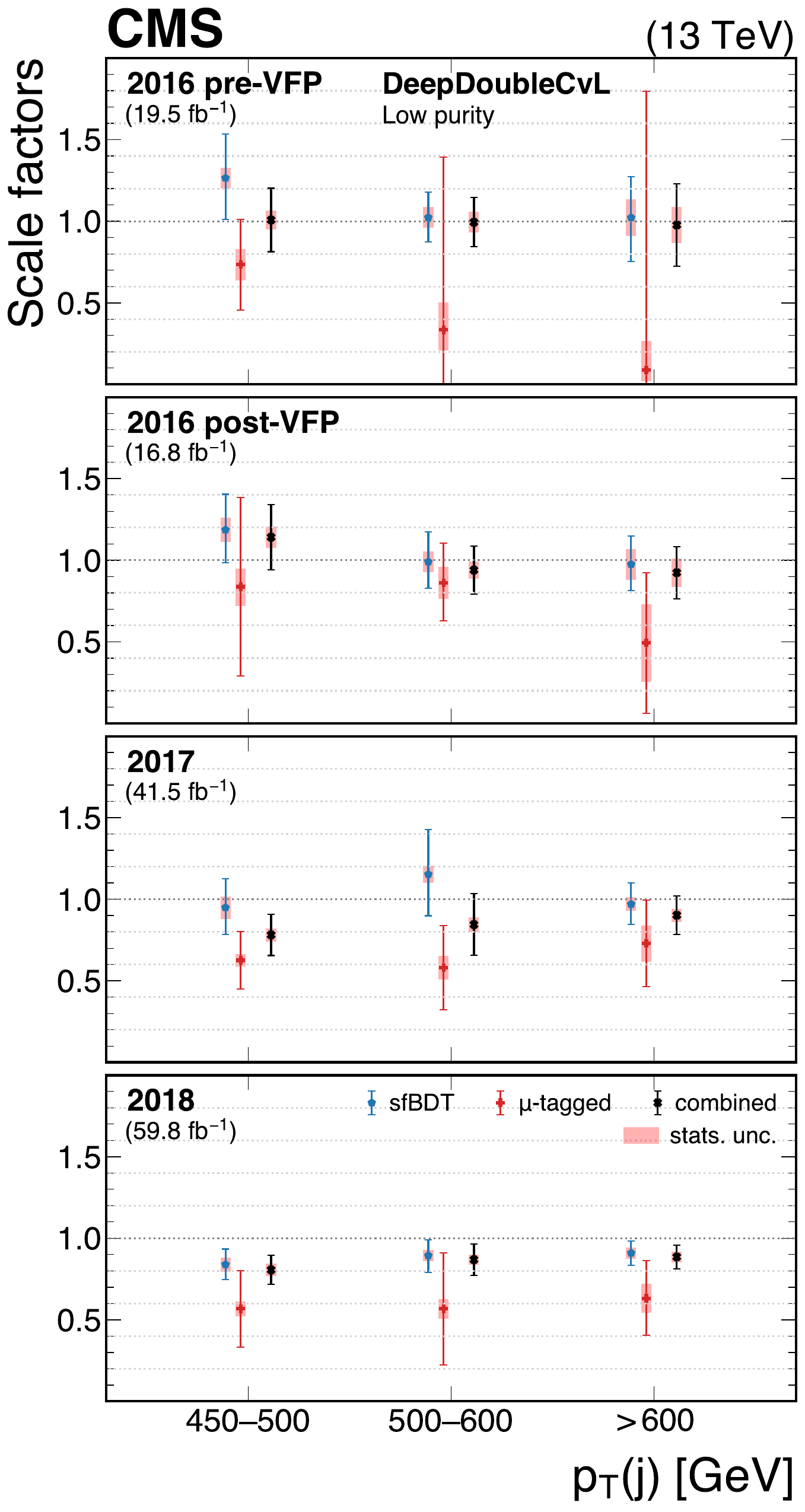}
    \caption{The measured scale factors of the DeepDoubleX \xcc discriminant in the high-purity (left), medium-purity (middle), and low-purity (right) working points. Two methods are presented in the measurements: the sfBDT method and the $\mu$-tagged method. The combined measurements from available methods are also shown.}
    \label{fig:sf_ddx-c}
\end{figure}

\begin{figure}[htbp]
    \centering
    \includegraphics[width=0.32\textwidth]{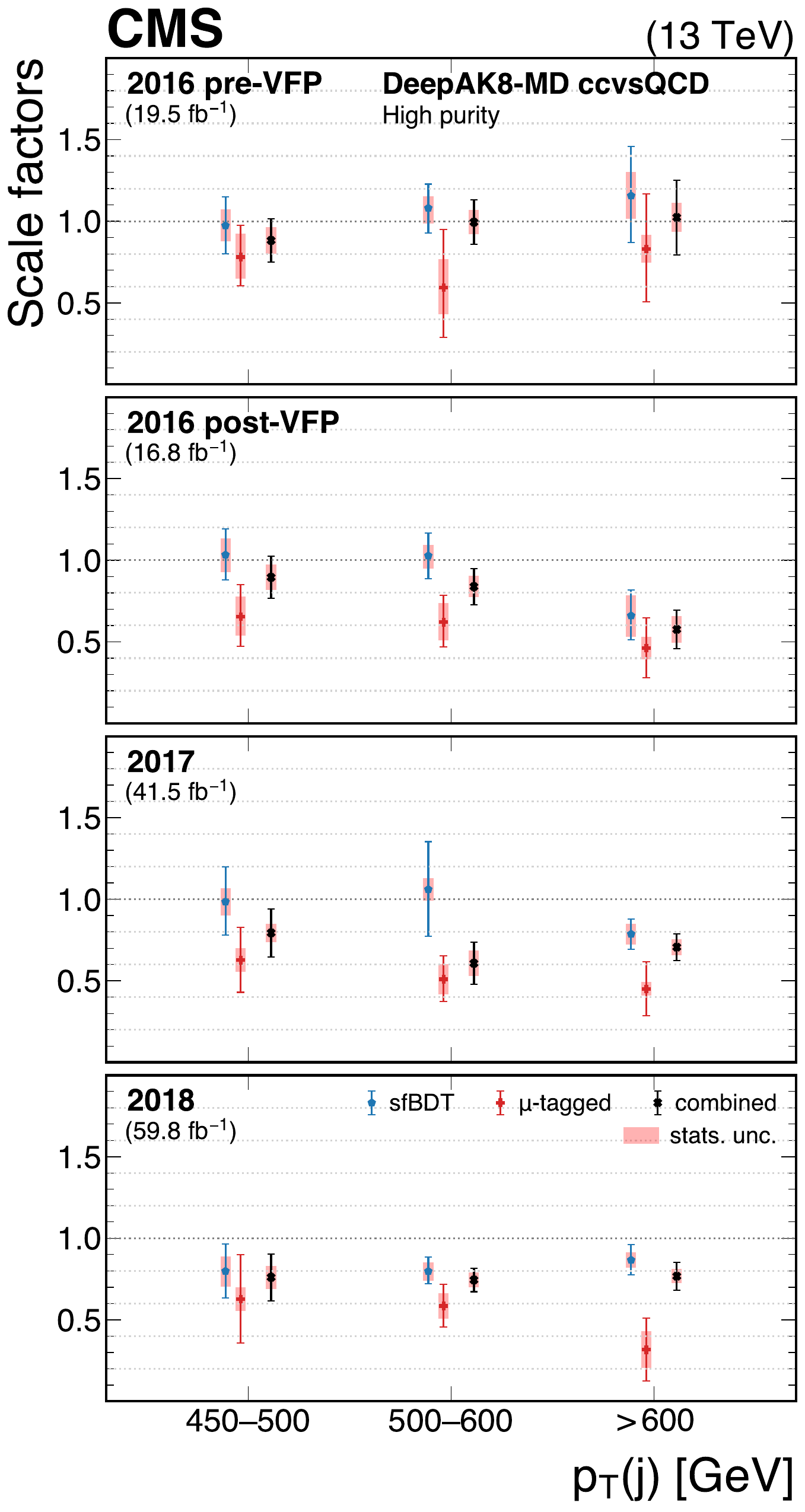}
    \includegraphics[width=0.32\textwidth]{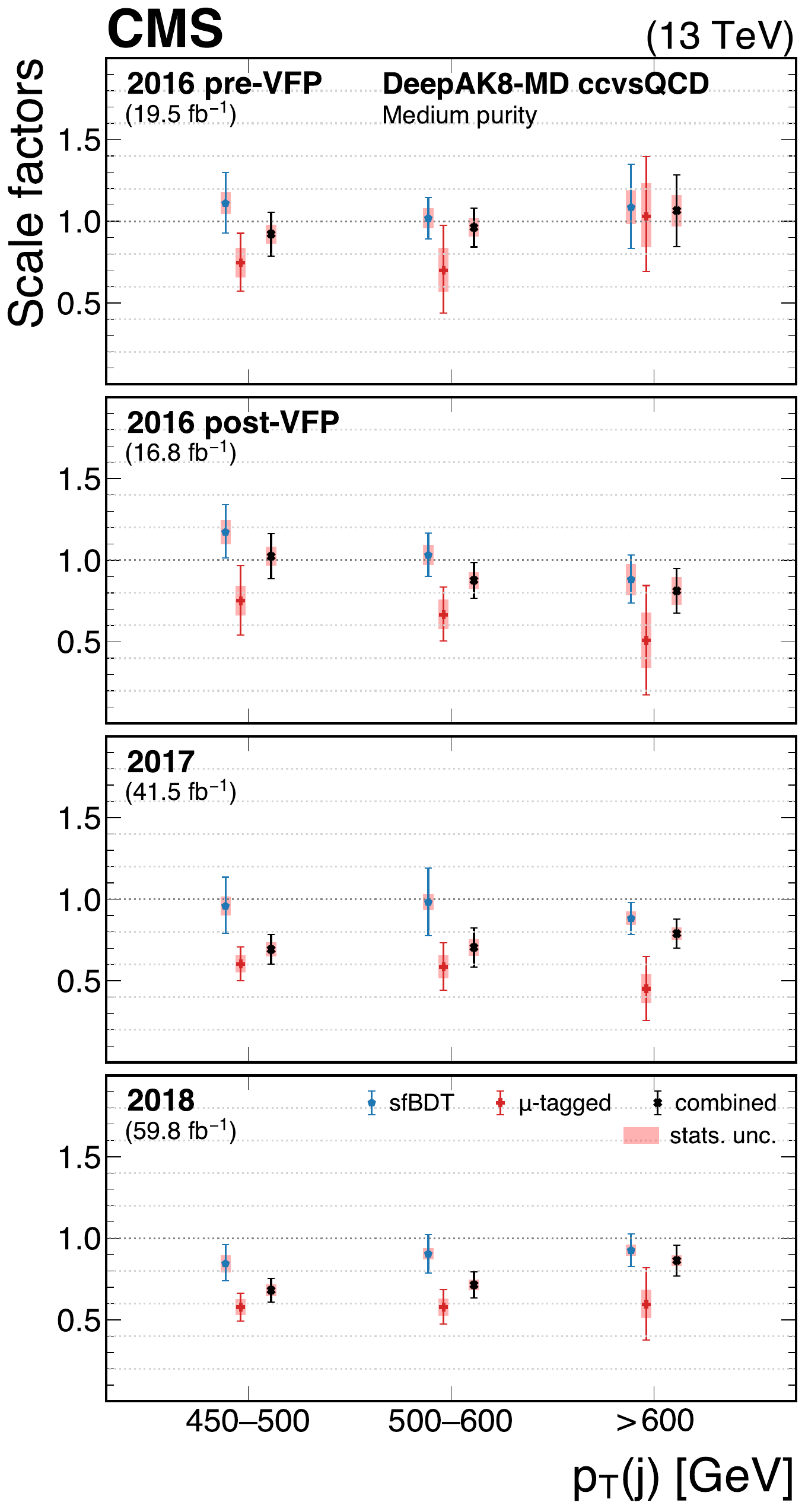}
    \includegraphics[width=0.32\textwidth]{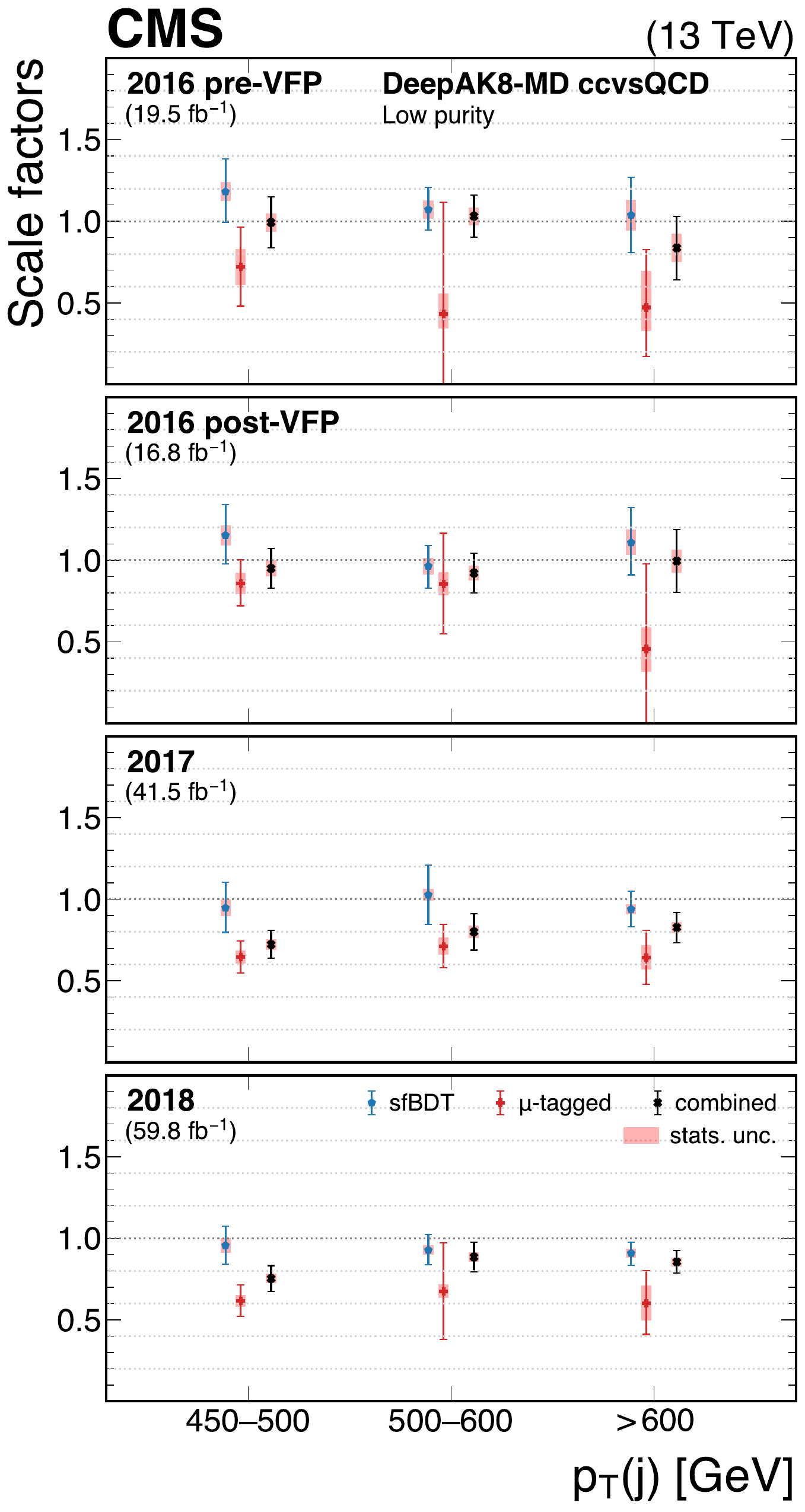}
    \caption{The measured scale factors of the DeepAK8-MD \xcc discriminant in the high-purity (left), medium-purity (middle), and low-purity (right) working points. Two methods are presented in the measurements: the sfBDT method and the $\mu$-tagged method. The combined measurements from available methods are also shown.}
    \label{fig:sf_da-c}
\end{figure}

For the combination of AK8 SFs, the common systematic uncertainties shared by the three measurements are treated as fully correlated. These systematic uncertainties include the ISR and FSR uncertainty in parton shower, the jet energy scale and resolution, the integrated luminosity uncertainty, and the uncertainty from pileup reweighting.
The sfBDT method and the $\mu$-tagged method, both based on QCD proxy jets, include uncertainties in the fraction of the \PQb, \PQc, and light-flavour jets. These three systematic uncertainty sources are considered correlated between the two methods.
Other uncertainty sources that are specific to an individual measurement are treated as fully uncorrelated.
Since the phase space of the proxy definitions of the three methods is largely orthogonal, the statistical uncertainty in data is also considered fully uncorrelated.
The result of the combination is also shown in Figs.~\ref{fig:sf_pnet-b}--\ref{fig:sf_da-c}.
The derived SFs are presented on the basis of the tagger discriminant at certain WPs. Each plot summarizes the SF results under the four data-taking eras for three exclusive \pt bins, obtained from the measurements from two or three methods.
For each SF derivation point, available individual SFs from either two or three measurements are combined via the BLUE method.


The derived SFs from the two or three methods, as well as the combined SF, lead to several findings. When comparing multiple tagger discriminants, better agreement between data and simulated events is observed with the ParticleNet-MD and DeepDoubleX discriminants, whereas the DeepAK8-MD discriminant generally yields SFs that are systematically lower than unity. This observation is supported by the data and simulation distributions depicted in Figs.~\ref{fig:datamc_xtagger_pnet}--\ref{fig:datamc_xtagger_doubleb} and \ref{fig:datamc_mutagged_xtagger_pnet}--\ref{fig:datamc_mutagged_xtagger_doubleb}.

For all three methods, the uncertainties in the SFs are dominated either by method-specific systematic sources or by statistical uncertainties, which are uncorrelated across the different methods.
When considering all methods collectively, consistent results are found among the available approaches within their respective uncertainties. The sfBDT and $\mu$-tagged methods exhibit smaller uncertainties compared with the boosted \PZ boson method, which extracts the signal from a predominantly QCD multijet background and is largely constrained by statistical limitations.
Furthermore, the sfBDT method generally yields larger SFs than the $\mu$-tagged method. This difference can be attributed, in part, to systematic effects arising from the distinct phase space probed by the two methods. The observed trend in SF divergence is supported by the comparison of Figs.~\ref{fig:datamc_mutagged_xtagger_pnet}--\ref{fig:datamc_mutagged_xtagger_doubleb} with \ref{fig:datamc_xtagger_pnet}--\ref{fig:datamc_xtagger_doubleb}, based on the data-to-simulation ratio in the high-discriminant-score region. Given that the sfBDT and $\mu$-tagged methods perform calibration in two distinct regions of phase space, their combination can mitigate the systematic bias in scale factors introduced by each specific method.

The results presented in this paper are compared with earlier corresponding measurements by the CMS Collaboration using Run~2 data.
For the double-b tagger, the extracted efficiency SFs are consistent with those reported in Ref.~\cite{CMS:BTVFlvTagger} for 2016 data, with central SF values being close to unity.
The uncertainty reported in this work is larger, primarily due to the inclusion of new external uncertainty sources considered in both the sfBDT and the $\mu$-tagged methods. These uncertainties are introduced to account for the calibration of taggers with improved discrimination power for \hbb against \gbb jets. The estimation strategies for these uncertainties are relatively conservative and contribute significantly to the overall uncertainty in the SFs.
When compared with the ParticleNet-MD \xcc and DeepDoubleX \xcc SFs presented in Refs.~\cite{CMS:VHcc,CMS-DP-2022-005,CMS:ggHcc} derived for all conditions in Run~2, reasonable agreement is observed regarding the central SFs, the SF dependence on the year of data taking, and the level of uncertainties.
Overall, the methods summarized in this paper provide a unified framework for deriving SFs across all detector conditions and tagger discriminants, including those with substantial discrimination power between the signal and the QCD multijet background.

\section{Summary} \label{sec:summary}

This paper presents the performance of heavy-flavour \xbb and \xcc jet tagging algorithms in the boosted topology, with a focus on the performance of various taggers in simulation and the calibration of tagging efficiencies using data collected by the CMS detector during the 2016--2018 data-taking period (LHC Run~2).
With the boosted topology gaining increasing relevance in physics searches during Run~2, the development of dedicated jet-tagging techniques and robust calibration methods for taggers on data has become increasingly important.

In this paper, we first provide a complete review and a comparison of \xbb and \xcc tagging algorithms, which were developed by the CMS Collaboration for analyzing Run~2 data and have been used for various physics measurements.
These algorithms include the ParticleNet-MD, DeepDoubleX, DeepAK8-MD, and the double-b tagging algorithms.
Three methods for evaluating the performance of the algorithms on data, in terms of deriving the scale factors to correct the selection efficiency of simulated \xbb and \xcc jets, are presented in detail.
The three methods define the proxy jets based on
(1)~a novel phase space selected from gluon-splitting \bb and \cc jets via a dedicated boosted decision tree discriminant;
(2)~gluon-splitting \bb and \cc jets containing a soft muon, with an auxiliary selection on the $N$-subjettiness variable;
and (3)~boosted \zbb jets for representing the \xbb signal jet.
The phase space of the selected proxy jets is largely orthogonal across the methods, which enables a meaningful comparison of their calibration results.
Scale factors and their uncertainties are derived for all working points of the seven tagging discriminants developed for \xbb and \xcc tagging. These scale factors are presented both individually and in a combined form, obtained using the best linear unbiased estimator method.

A reasonable agreement is found when comparing the results with previous CMS studies, which calibrated some of the discriminants studied in this work, either partially or under full Run~2 conditions.
Additionally, the scale factors presented by the three methods remain consistent within the uncertainty range. Their combination provides the highest measurement precision for the scale factor while also reducing the systematic biases inherent to each individual method.
The tagging algorithms and calibration approaches documented in this paper serve as a comprehensive summary and are considered as benchmarks for the techniques adopted by the CMS Collaboration during Run~2. These outcomes will facilitate further in-depth studies and wider experimental explorations of the boosted phase space with heavy-flavour tagging in the future.

\begin{acknowledgments}
We congratulate our colleagues in the CERN accelerator departments for the excellent performance of the LHC and thank the technical and administrative staffs at CERN and at other CMS institutes for their contributions to the success of the CMS effort. In addition, we gratefully acknowledge the computing centres and personnel of the Worldwide LHC Computing Grid and other centres for delivering so effectively the computing infrastructure essential to our analyses. Finally, we acknowledge the enduring support for the construction and operation of the LHC, the CMS detector, and the supporting computing infrastructure provided by the following funding agencies: SC (Armenia), BMBWF and FWF (Austria); FNRS and FWO (Belgium); CNPq, CAPES, FAPERJ, FAPERGS, and FAPESP (Brazil); MES and BNSF (Bulgaria); CERN; CAS, MoST, and NSFC (China); MINCIENCIAS (Colombia); MSES and CSF (Croatia); RIF (Cyprus); SENESCYT (Ecuador); ERC PRG, TARISTU24-TK10 and MoER TK202 (Estonia); Academy of Finland, MEC, and HIP (Finland); CEA and CNRS/IN2P3 (France); SRNSF (Georgia); BMFTR, DFG, and HGF (Germany); GSRI (Greece); NKFIH (Hungary); DAE and DST (India); IPM (Iran); SFI (Ireland); INFN (Italy); MSIT and NRF (Republic of Korea); MES (Latvia); LMTLT (Lithuania); MOE and UM (Malaysia); BUAP, CINVESTAV, CONACYT, LNS, SEP, and UASLP-FAI (Mexico); MOS (Montenegro); MBIE (New Zealand); PAEC (Pakistan); MES, NSC, and NAWA (Poland); FCT (Portugal); MESTD (Serbia); MICIU/AEI and PCTI (Spain); MOSTR (Sri Lanka); Swiss Funding Agencies (Switzerland); MST (Taipei); MHESI (Thailand); TUBITAK and TENMAK (T\"{u}rkiye); NASU (Ukraine); STFC (United Kingdom); DOE and NSF (USA).

\hyphenation{Rachada-pisek} Individuals have received support from the Marie-Curie programme and the European Research Council and Horizon 2020 Grant, contract Nos.\ 675440, 724704, 752730, 758316, 765710, 824093, 101115353, 101002207, 101001205, and COST Action CA16108 (European Union); the Leventis Foundation; the Alfred P.\ Sloan Foundation; the Alexander von Humboldt Foundation; the Science Committee, project no. 22rl-037 (Armenia); the Fonds pour la Formation \`a la Recherche dans l'Industrie et dans l'Agriculture (FRIA) and Fonds voor Wetenschappelijk Onderzoek contract No. 1228724N (Belgium); the Beijing Municipal Science \& Technology Commission, No. Z191100007219010, the Fundamental Research Funds for the Central Universities, the Ministry of Science and Technology of China under Grant No. 2023YFA1605804, the Natural Science Foundation of China under Grant No. 12061141002, 12535004, and USTC Research Funds of the Double First-Class Initiative No.\ YD2030002017 (China); the Ministry of Education, Youth and Sports (MEYS) of the Czech Republic; the Shota Rustaveli National Science Foundation, grant FR-22-985 (Georgia); the Deutsche Forschungsgemeinschaft (DFG), among others, under Germany's Excellence Strategy -- EXC 2121 ``Quantum Universe" -- 390833306, and under project number 400140256 - GRK2497; the Hellenic Foundation for Research and Innovation (HFRI), Project Number 2288 (Greece); the Hungarian Academy of Sciences, the New National Excellence Program - \'UNKP, the NKFIH research grants K 131991, K 133046, K 138136, K 143460, K 143477, K 146913, K 146914, K 147048, 2020-2.2.1-ED-2021-00181, TKP2021-NKTA-64, and 2021-4.1.2-NEMZ\_KI-2024-00036 (Hungary); the Council of Science and Industrial Research, India; ICSC -- National Research Centre for High Performance Computing, Big Data and Quantum Computing, FAIR -- Future Artificial Intelligence Research, and CUP I53D23001070006 (Mission 4 Component 1), funded by the NextGenerationEU program (Italy); the Latvian Council of Science; the Ministry of Education and Science, project no. 2022/WK/14, and the National Science Center, contracts Opus 2021/41/B/ST2/01369, 2021/43/B/ST2/01552, 2023/49/B/ST2/03273, and the NAWA contract BPN/PPO/2021/1/00011 (Poland); the Funda\c{c}\~ao para a Ci\^encia e a Tecnologia, grant CEECIND/01334/2018 (Portugal); the National Priorities Research Program by Qatar National Research Fund; MICIU/AEI/10.13039/501100011033, ERDF/EU, "European Union NextGenerationEU/PRTR", and Programa Severo Ochoa del Principado de Asturias (Spain); the Chulalongkorn Academic into Its 2nd Century Project Advancement Project, the National Science, Research and Innovation Fund program IND\_FF\_68\_369\_2300\_097, and the Program Management Unit for Human Resources \& Institutional Development, Research and Innovation, grant B39G680009 (Thailand); the Kavli Foundation; the Nvidia Corporation; the SuperMicro Corporation; the Welch Foundation, contract C-1845; and the Weston Havens Foundation (USA).
\end{acknowledgments}

\bibliography{auto_generated}
\cleardoublepage \appendix\section{The CMS Collaboration \label{app:collab}}\begin{sloppypar}\hyphenpenalty=5000\widowpenalty=500\clubpenalty=5000
\cmsinstitute{Yerevan Physics Institute, Yerevan, Armenia}
{\tolerance=6000
A.~Hayrapetyan, A.~Tumasyan\cmsAuthorMark{1}\cmsorcid{0009-0000-0684-6742}
\par}
\cmsinstitute{Institut f\"{u}r Hochenergiephysik, Vienna, Austria}
{\tolerance=6000
W.~Adam\cmsorcid{0000-0001-9099-4341}, J.W.~Andrejkovic, L.~Benato\cmsorcid{0000-0001-5135-7489}, T.~Bergauer\cmsorcid{0000-0002-5786-0293}, S.~Chatterjee\cmsorcid{0000-0003-2660-0349}, K.~Damanakis\cmsorcid{0000-0001-5389-2872}, M.~Dragicevic\cmsorcid{0000-0003-1967-6783}, P.S.~Hussain\cmsorcid{0000-0002-4825-5278}, M.~Jeitler\cmsAuthorMark{2}\cmsorcid{0000-0002-5141-9560}, N.~Krammer\cmsorcid{0000-0002-0548-0985}, A.~Li\cmsorcid{0000-0002-4547-116X}, D.~Liko\cmsorcid{0000-0002-3380-473X}, I.~Mikulec\cmsorcid{0000-0003-0385-2746}, J.~Schieck\cmsAuthorMark{2}\cmsorcid{0000-0002-1058-8093}, R.~Sch\"{o}fbeck\cmsAuthorMark{2}\cmsorcid{0000-0002-2332-8784}, D.~Schwarz\cmsorcid{0000-0002-3821-7331}, M.~Sonawane\cmsorcid{0000-0003-0510-7010}, W.~Waltenberger\cmsorcid{0000-0002-6215-7228}, C.-E.~Wulz\cmsAuthorMark{2}\cmsorcid{0000-0001-9226-5812}
% ...
\par}
\vskip\cmsinstskip
\dag:~Deceased\\
$^{1}$Also at Yerevan State University, Yerevan, Armenia\\
$^{2}$Also at TU Wien, Vienna, Austria\\
% ...
\end{sloppypar}
\end{document}